\documentclass[fleqn,usenatbib]{mnras}

\usepackage[T1]{fontenc}

\DeclareRobustCommand{\VAN}[3]{#2}
\let\VANthebibliography\thebibliography
\def\thebibliography{\DeclareRobustCommand{\VAN}[3]{##3}\VANthebibliography}

\usepackage{graphicx}	
\usepackage{amsmath}	
\usepackage{amssymb}	

\title[The AGNIFS survey]{The AGNIFS survey: distribution and excitation of the hot molecular and ionised gas in the inner kpc of nearby AGN hosts}

\author[R. A. Riffel et al.]{R. A. Riffel,$^{1}$\thanks{E-mail: rogemar@ufsm.br (RAR)}
T. Storchi-Bergmann,$^{2}$
R. Riffel,$^{2}$
M. Bianchin,$^{1}$
N. L. Zakamska,$^{3}$
\newauthor
D. Ruschel-Dutra,$^{4}$
A. J. Sch\"onell,$^{5}$
D. J. Rosario,$^{6}$
A. Rodriguez-Ardila,$^{7}$
T. C. Fischer,$^{8}$
\newauthor
R. I. Davies,$^{9}$
N. Z. Dametto,$^{10}$
L. G. Dahmer-Hahn,$^{7}$
D. M. Crenshaw,$^{11}$
L. Burtscher,$^{12}$
\newauthor
M. C. Bentz$^{11}$\\
$^{1}$Departamento de F\'isica, Centro de Ci\^encias Naturais e Exatas, Universidade Federal de Santa Maria, 97105-900, Santa Maria, RS, Brazil\\
$^{2}$Instituto de F\'isica, Universidade Federal do Rio Grande do Sul, Av. Bento Gon\c calves 9500, 91501-970 Porto Alegre, RS, Brazil \\
$^{3}$Department of Physics \& Astronomy, Johns Hopkins University, Bloomberg Center, 3400 N. Charles St, Baltimore, MD 21218, USA\\
$^{4}$Departamento de F\'isica, Universidade Federal de Santa Catarina, P.O. Box 476, 88040-900, Florian\'opolis, SC, Brazil \\
$^{5}$ Instituto Federal de Educa\c c\~ao, Ci\^encia e Tecnologia Farroupilha, BR287, km 360, Estrada do Chapad\~ao, 97760-000, 
Jaguari - RS, Brazil\\
$^{6}$Centre for Extragalactic Astronomy, Department of Physics, Durham University, South Road, DH1 3LE, Durham, UK\\
$^{7}$Laborat\'orio Nacional de Astrof\'isica. Rua dos Estados Unidos, 154, CEP 37504-364 Itajub\'a, MG, Brazil \\
$^{8}$AURA for ESA, Space Telescope Science Institute, 3700 San Martin Drive, Baltimore, MD 21218\\
$^{9}$Max-Planck-Institut f\"ur Extraterrestrische Physik, Postfach 1312, 85741, Garching, Germany\\
$^{10}$Centro de Astronom\'ia (CITEVA), Universidad de Antofagasta, Avenida Angamos 601, Antofagasta, Chile\\
$^{11}$Department of Physics and Astronomy, Georgia State University, 25 Park Place, Suite 605, Atlanta, GA 30303, USA\\
$^{12}$Leiden Observatory, PO Box 9513, 2300 RA, Leiden, The Netherlands\\
}

\date{Accepted XXX. Received YYY; in original form ZZZ}

\pubyear{2020}

\begin{document}
\label{firstpage}
\pagerange{\pageref{firstpage}--\pageref{lastpage}}
\maketitle

\begin{abstract}
We use the Gemini NIFS instrument to map the H$_2\,2.1218\,\mu$m and Br$\gamma$ flux distributions in the inner 0.04--2\,kpc of a sample of 36 nearby active galaxies ($0.001\lesssim z\lesssim0.056$)
at spatial resolutions from 4 to 250\,pc. We find extended emission 
in 34 galaxies. In $\sim$55\% of them, the emission in both lines is most extended along the galaxy major axis, while in the other 45\% the extent follows a distinct orientation.  
The emission of H$_2$ is less concentrated than that of Br$\gamma$, presenting a radius that contains half of the flux 60\,\% greater, on average. The H$_2$ emission is driven by thermal processes -- X-ray heating and shocks -- at most locations for all galaxies, where $0.4<\rm H_2/Br\gamma<6$. For regions where H$_2$/Br$\gamma>6$  (seen in 40\% of the galaxies), shocks are the main H$_2$ excitation mechanism, while in regions with H$_2$/Br$\gamma<0.4$ (25\% of the sample) the H$_2$ emission is produced by fluorescence. The only difference we found between type 1 and type 2 AGN was in the nuclear emission-line equivalent widths, that are smaller in type 1 than in type 2 due to a larger contribution to the continuum from the hot dusty torus in the former.
The gas masses in the inner 125\,pc radius are in the range $10^1-10^4$ M$_\odot$ for the hot H$_2$ and $10^3-10^6$ M$_\odot$ for the ionised gas and would be enough to power the AGN in our sample for $10^5-10^8$\,yr at their current accretion rates.
\end{abstract}




\section{Introduction}

The presence of a gas reservoir in the inner few tens of parsecs of galaxies is a necessary requirement to trigger an Active Galactic Nucleus (AGN) and/or a nuclear starburst. Understanding the origin of the gas emission at these scales is critical to investigate the role of AGN and star formation (SF) feedback in galaxy evolution. The cold molecular gas is the raw fuel of star formation in the central region of galaxies and AGN, while the ionised gas is usually observed as a consequence of star formation and nuclear activity. The ionised gas is easier to trace, since the strongest emission lines are observed in the optical region. These emission lines are good tracers of many galactic components, such as disks, outflows and AGN \citep[e.g.][]{Ricci+14}. However, there are no strong emission lines of molecular gas in the optical. For this reason, most studies of molecular hydrogen distribution and kinematics usually use other molecules as its tracer, such as the CO emission at sub-millimeter wavelengths. These studies have found that there are two main classes of galaxies regarding their molecular gas distribution: a starburst one, where the molecular gas distribution is very compact with short consumption times (10$^7$--10$^8$\,yr), and a quiescent one, with the gas distributed in extended disks and longer consumption times ($\sim$10$^9$\,yr) \citep{Daddi+10, Genzel+10, Sargent+14, Silverman+15}. In order to properly compare ionised and molecular gas distributions, however, it is ideal that they are observed in the same wavelength range and with the same spatial resolution.

In the near-infrared (hereafter near-IR) both gas phases, hot molecular and ionised, can be observed simultaneously. In particular, the K-band spectra of galaxies include ro-vibrational emission lines from the H$_2$, tracing the hot molecular gas that represents only a small fraction of the total molecular gas, and the hydrogen recombination line Br$\gamma$, a tracer of the ionised gas \citep[e.g][]{rogerio06}. Physically motivated models \citep{hopkins10} show that the relevant feeding processes occur within the inner kiloparsec, which can only be resolved in nearby galaxies. Near-IR integral field spectroscopy (IFS) on 8--10\,m telescopes is a unique tool to investigate the distribution and kinematics of the molecular and ionised gas at spatial resolutions of 10--100\,pc providing observational constraints to better understand the AGN feeding and feedback processes. Such constraints are fundamental ingredients of theoretical models and numerical simulations of galaxy evolution aimed to understand the co-evolution of AGN and their host galaxies \citep{kormendy13,heckman14,harrison17,harrison18,sb19}.

The near-IR emission of molecular and ionised gas have been investigated over the years 
using both long slit spectroscopy \citep[e.g.][]{ardila04,ardila05,rogerio06,rogerio13,lamperti17} and spatially resolved observations \citep[e.g.][]{rogemarN4051,rogemar_N1275,davies09,hicks09,sbN4151Exc,mazzalay13,schawachter13,barbosa14,durre14,Schonell14,SchonellN5548,durre18,Diniz15,fischer17,Husemann19,Shimizu19, rosario19}. In most cases, the hot molecular and ionised gas show distinct flux distributions and kinematics, with the H$_2$ more restricted to the galaxy plane following circular rotation and the ionised gas showing more collimated emission and contribution of non-circular motions. However, in few cases,  hot molecular outflows are also observed \citep[e.g.][]{davies14,fischer17,Gnilka20,may17,may20,rogemar_N1275}. These results, combined with the fact that the ionised gas is usually associated with higher temperatures when compared to the molecular gas, has led \citet{sbN4151Exc} to suggest that the molecular gas is a better tracer of AGN feeding, whereas the ionised gas is a better tracer of AGN feedback. While the studies above have addressed the origin, morphology and amount of molecular gas in individual sources, it is now necessary to trace a more complete picture of the subject from a statistical point of view. This will allow us to detect common points and differences in order to advance on the knowledge of the origin and black hole feeding mechanisms in AGN and galaxies overall.

A few studies using near-IR IFS on larger samples have also been carried out, as for example those of the AGNIFS (AGN Integral Field Spectroscopy) survey \citep{rogemar_stellar,rogemar_sample,schonell19}, the Local Luminous AGN with Matched Analogs (LLAMA) survey \citep{Davies15,lin18,Caglar20} and the Keck OSIRIS Nearby AGN (KONA) survey \citep{ms18}. But so far, the studies based on these surveys were aimed to describe the sample, map its stellar kinematics and nuclear properties of the galaxies, while the origin, amount and distribution of the hot molecular hydrogen in the inner kiloparsec of nearby AGN are still not properly covered and mapped.

In the present work, we use K-band IFS of a sample of 36 AGN of the local Universe, observed with the Gemini Near-infrared Integral Field Spectrograph (NIFS), to map their molecular and ionised gas flux distributions at resolutions ranging from a few pc to $\approx$250\,pc. We investigate the origin of the H$_2$ emission, derive the H$_2$ excitation temperature and mass of hot molecular and ionised gas, available to feed the central AGN and star formation. This paper is organized as follows: Section \ref{sec:sample} presents the sample, observations and data reduction procedure, Sec.~\ref{sec:results} presents the results, which are discussed in Sec.~\ref{sec:disc}. We present our conclusions in Sec.~\ref{sec:conc}.  We use a $h=0.7, \Omega_m=0.3, \Omega_{\Lambda}=0.7$ cosmology throughout this paper.

\begin{table*}
	\centering
	\small
	\caption{\small The sample: (1) Galaxy name. The superscript letter in the name of some galaxies identifies the reference of previously published NIFS K-band, listed below. All galaxies were observed using adaptive optics, except NGC\,1125 and ESO578-G009 (marked with a $\dagger$ symbol).
 (2) Morphological classification from \citet{devaucouleurs91}, (3) Nuclear Activity Classification from  \citet{BAT105}.  (4) redshift, (5) adopted distance, (6) hard X-ray (14-195 keV) luminosity, (7) Gemini Program ID (prefix: GN-20), (8) Grating used during the observations. The central wavelengths are 2.2 and 2.3\,$\mu$m for the K and K$_{\rm long}$ (K$_{\rm l}$) gratings, respectively. The spectral coverage of both gratings is $\sim4000$\,\AA. (9) exposure time, (10) angular resolution and (11) spatial scale of 1$^{\prime\prime}$ projected at the galaxy distance. References to previous works on NIFS data: $a$) \citet{schonell19};  $b$) \citet{dahmer-hahn19}; $c$) \citet{rogemarN1068}, $d$) \citet{Diniz15}; $e$) \citet{Drehmer15}; $f$) \citet{Ilha16}; $g$) \citet{Gnilka20}; $h$) \citet{rogemarM1066_SP}; $i$) \citet{Husemann19}; $j$) \citet{rogemarN1275}; $k$) \citet{rogemarN4051}; $l$) \citet{sbN4151Exc}; $m$) \citet{Brum19}; $n$) \citet{SchonellN5548}; $o$) \citet{rogemarM79}; $p$) \citet{Diniz18} and $q$) \citet{Schonell14}. } 
	\label{tab:sample}
	\begin{tabular}{lcccccccccc} 
		\hline
Galaxy & Hubble type & Act. type & $z$ & $D$ & log $L_{\rm X}$ & Program ID & Grating & Exp. Time & FWHM$_{\rm PSF}$ & Scale \\ 
&   &  & & (Mpc)  & (erg\,s$^{-1}$) &  & & (sec) & ($^{\prime\prime}$) & (pc) \\
(1) & (2) & (3) & (4) & (5) & (6) & (7) & (8) & (9) & (10) & (11) \\
		\hline
	         \multicolumn{11}{c}{type 2}\\
	     	 \hline
NGC788$^a$       & SA0/a?(s)      &  Sy2	    &	0.0136  & 58.3 & 43.51 &  15B-Q-29  & K     & 11$\times$400 & 0.13  & 282	\\
NGC1052$^b$      &  E4            &  LINER      &	0.0050	& 21.4 & 42.24   & 10B-Q-25 & K$_{\rm l}$ &   4$\times$600     & 0.15 & 103	 \\
NGC1068$^c$      & (R)SA(rs)b     &  Sy1.9      &	0.0038  & 16.3 & 42.08   & 06B-C-9  & K     &   27$\times$90     & 0.11 & 78	 \\
NGC1125$^\dagger$          & (R')SB0/a?(r)  &  Sy2	    &	0.0109	& 47.1 & 42.64   & 18B-Q-140& K     &   8$\times$450     & 0.44 & 228	 \\
NGC1241          & SB(rs)b        &  Sy2	    &	0.0135  & 57.9 & 42.68   & 19A-Q-106& K     &   6$\times$600     & 0.14 & 280	 \\
NGC2110$^d$      & SAB0$^-$       &  Sy2	    &	0.0078  & 33.4 & 43.65   & 10B-Q-25 & K$_{\rm l}$ &   6$\times$600     & 0.15 & 162	 \\
NGC3393          & (R')SB(rs)a?   &  Sy2	    &	0.0125  & 53.6 & 42.98   & 12A-Q-2  & K     &   4$\times$600     & 0.13 & 260	 \\
NGC4258$^e$      & SAB(s)bc       &  Sy1.9      &	0.0015  & 6.4  & 41.06   & 07A-Q-25 & K     &   10$\times$600    & 0.20 & 31	 \\
NGC4388          & SA(s)b? edge-on&  Sy2	    &	0.0084  & 36.0 & 43.64   & 15A-Q-3  & K     &   2$\times$400     & 0.19 & 174	 \\
NGC5506$^a$      & Sa pec edge-on &  Sy1.9      &	0.0062  & 26.6 & 43.31   & 15A-Q-3  & K     &   10$\times$400    & 0.18 & 128	 \\
NGC5899$^a$      & SAB(rs)c       &  Sy2	    &	0.0086  & 36.9 & 42.53   & 13A-Q-48 & K$_{\rm l}$ &   10$\times$450    & 0.13 & 178	 \\
NGC6240$^f$      & I0: pec        &  Sy1.9      &	0.0245  & 105.0& 43.97   & 07A-Q-62 & K     &   16$\times$300    & 0.19 & 509	 \\
Mrk3$^g$         & S0?            &  Sy1.9      &	0.0135  & 57.9 & 43.79   & 10A-Q-56 & K$_{\rm l}$ &   6$\times$600     & 0.13 & 280	 \\
Mrk348           & SA(s)0/a:      &  Sy 2       &	0.0150	& 64.3 & 43.86   & 11B-Q-71 & K$_{\rm l}$ &   6$\times$400     & 0.18 & 311 \\
Mrk607$^a$       & Sa? edge-on    &  Sy2	    &	0.0089  & 38.1 & 42.36   & 12B-Q-45 & K$_{\rm l}$ &   12$\times$500    & 0.14 & 184	 \\
Mrk1066$^h$      & (R)SB0$^+$(s)  &  Sy2	    &	0.0120	& 51.4 & 42.51   & 08B-Q-30 & K$_{\rm l}$ &   8$\times$600     & 0.15 & 249	 \\
ESO578-G009$^i\dagger$  & Sc             &  Sy1.9      &	0.0350	& 150.0& 43.59   & 17A-FT-10& K     &   6$\times$600     & 0.35 & 727	 \\
CygnusA          & S?             &  Sy2	    &	0.0561  & 240.4& 45.04   & 06A-C-11 & K     &    7$\times$900     & 0.18 & 1165 \\ 
	     	 \hline
	     	 \multicolumn{11}{c}{type 1}\\
	     	 \hline
NGC1275$^j$      & E pec          &  Sy1.5     &    0.0176  & 75.5 & 43.76   & 19A-Q-106 & K    &   6$\times$600	  & 0.22   & 365\\
NGC3227$^a$      & SAB(s)a pec    &  Sy1.5     &	0.0039  & 16.7 & 42.58   & 16A-Q-6   & K    &   6$\times$400     & 0.12  & 81	 \\
NGC3516$^a$      & (R)SB0$^0$?(s) &  Sy1.2     &	0.0088  & 37.7 & 43.29   & 15A-Q-3   & K    &   10$\times$450    & 0.15  & 182	 \\
NGC4051$^k$      & SAB(rs)bc      &  Sy1.5     &	0.0023  & 9.9  & 41.69   & 06A-SV-123& K    &   6$\times$750     & 0.18  & 47	 \\
NGC4151$^l$      & (R')SAB(rs)ab? &  Sy1.5     &	0.0033  & 14.1 & 43.17   & 06B-C-9   & K    &   8$\times$90	   & 0.18  & 68	 \\
NGC4235          & SA(s)a edge-on &  Sy1.2     &	0.0080	& 34.3 & 42.74   & 16A-Q-6   & K    &   10$\times$400    & 0.13  & 166 \\
NGC4395$^m$      & Sd             &  Sy1	   &	0.0011  & 4.7  & 40.86   & 10A-Q-38  & K    &    9$\times$600    & 0.20  & 22 \\
NGC5548$^n$      & (R')SA0/a(s)   &  Sy1.5     &	0.0172  & 73.7 & 43.76   & 12A-Q-57  & K$_{\rm l}$&   12$\times$450    & 0.20  & 357	 \\
NGC6814          & SAB(rs)bc      &  Sy1.5     &	0.0052  & 22.3 & 42.59   & 13B-Q-5   & K    &    113$\times$120  & 0.18  & 108	 \\
Mrk79$^o$        & SBb            &  Sy1.5     &	0.0222  & 95.1 & 43.68   & 10A-Q-42  & K$_{\rm l}$&   6$\times$550     & 0.25  & 461	 \\
Mrk352$^p$       & 	SA0           &  Sy1.2     &	0.0149  & 63.8 & 43.19   & 12A-Q-42  & K$_{\rm l}$&   11$\times$400    & 0.37  & 309	 \\
Mrk509           & ?              &  Sy1.2     &	0.0344  & 147.4& 44.44   & 13A-Q-40  & K$_{\rm l}$&   6$\times$450     & 0.14  & 714	 \\
Mrk618           & SB(s)b pec     &  Sy1.2     &	0.0355  & 152.1& 43.73   & 18B-Q-138 & K    &   5$\times$600     & 0.18  & 737	 \\
Mrk766$^q$       & (R')SB(s)a?    &  Sy1.5     &	0.0129  & 55.3 & 42.99   & 10A-Q-42  & K$_{\rm l}$&   6$\times$550     &  0.19 & 268	 \\
Mrk926           & Sab            &  Sy1.5     &	0.0469  & 201.0& 44.76   & 18B-Q-138 & K    &   6$\times$600     & 0.17 & 974 \\
Mrk1044          & S?             &  Sy1	    &	0.0165  & 70.7 & 42.86   & 18B-Q-109 & K$_{\rm l}$&   6$\times$600     & 0.19 & 342	 \\
Mrk1048          & RING pec       &  Sy1.5     &	0.0431	& 184.3& 44.12   & 18B-Q-138 & K    &   6$\times$600     & 0.22 & 893	 \\
MCG+08-11-011    & SB?            &  Sy1.5     &	0.0205  & 87.9 & 44.13   & 19A-Q-106 & K    &   6$\times$470     & 0.20  &425 	 \\

\hline
	\end{tabular}
\end{table*}

\section{The sample and Data}\label{sec:sample}

\subsection{The sample}

We used the catalog of hard X-ray (14--195 keV) sources detected in the first 105 months of observations of the Swift Burst Alert Telescope (BAT) survey \citep{BAT105} to select our parent sample of AGN. The hard X-ray measures direct emission from the AGN  and is much less sensitive to obscuration along the line-of-sight than softer X-ray and optical observations \citep[see ][for a discussion about the advantages of using BAT catalogue to select AGN]{Davies15}.   This sample was cross-correlated with the Gemini Science archive looking for data obtained with the Near-Infrared Integral Field Spectrograph (NIFS) in the K band. We emphasize that just like any other AGN selection method (e.g. based on optical emission lines), our selection criteria may not include all AGN available in the Gemini database. As the aim of this work is to study the hot molecular hydrogen emission from AGN hosts, we limit our search to redshifts $z<0.12$, for which the H$_2\,2.1218\,\mu$m is still accessible in the K band. 

We have found NIFS data for 36 galaxies obtained using the K and K$_{\rm long}$ gratings, most of them observed as part of the Brazilian Large Gemini Program NIFS survey of feeding and feedback processes in nearby active galaxies \citep{rogemar_sample}. The central wavelengths of the  K and K$_{\rm long}$ gratings are 2.20 and 2.30\,$\mu$m, respectively.  For both gratings, the bandwidth is $\sim4000$\,\AA\ and the spectral bin is $\sim$2.13\,\AA. 
Table  \ref{tab:sample} lists the galaxies of our sample, as well as their morphologies, nuclear activity types, redshifts, hard X-ray luminosities ($L_{\rm X}$) obtained from the Swift BAT survey \citep{BAT105}, Gemini Program ID, exposure time and the full width at half maximum (FWHM) of the Point Spread Function (PSF) measured from the flux distribution of the telluric standard stars for the type 2 AGN and from the flux distribution of the broad Br$\gamma$ emission line for the observed type 1 AGN. Although our redshift limit is $z<0.12$, the most distant galaxy (Cygnus A) we have found in the archive obeying all our criteria has $z\sim0.056$.  Except NGC\,1052, all galaxies in our sample are classified as hosting Seyfert nuclei. The nature of LINERs is still not well understood, as similar line ratios can be produced by distinct mechanisms (e.g. AGN, hot low-mass evolved stars and shocks). However, NGC\,1052 undoubtedly presents an AGN as indicated by by its optical emission lines \citep{heckman80,ho97,dahmer19b}, by the detection of a hidden broad line region seen in polarized light that shows a broad H$\alpha$ component \citep{barth99} and by its high hard X-ray luminosity, comparable to the luminosities in Seyfert nuclei (Table~\ref{tab:sample}).
We divide the sample into type 1  (1 to 1.5) and type 2  (1.9 and 2), according to the classification in the 105 month BAT catalogue \citep{BAT105}. The type 1 and type 2 sub-samples have the same number (18) of objects.   

\begin{figure}
    \centering
\includegraphics[width=0.45\textwidth]{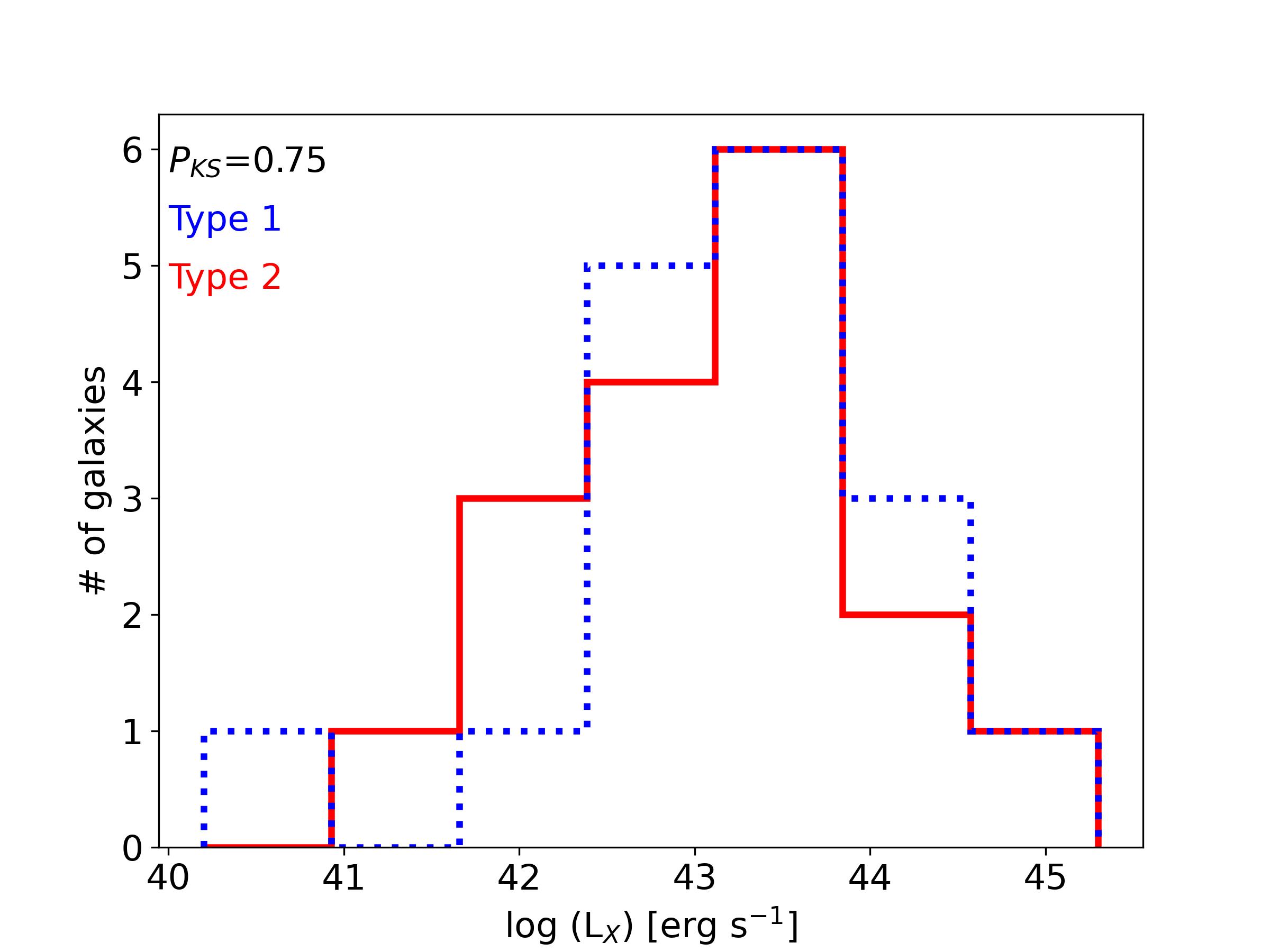} 
\includegraphics[width=0.45\textwidth]{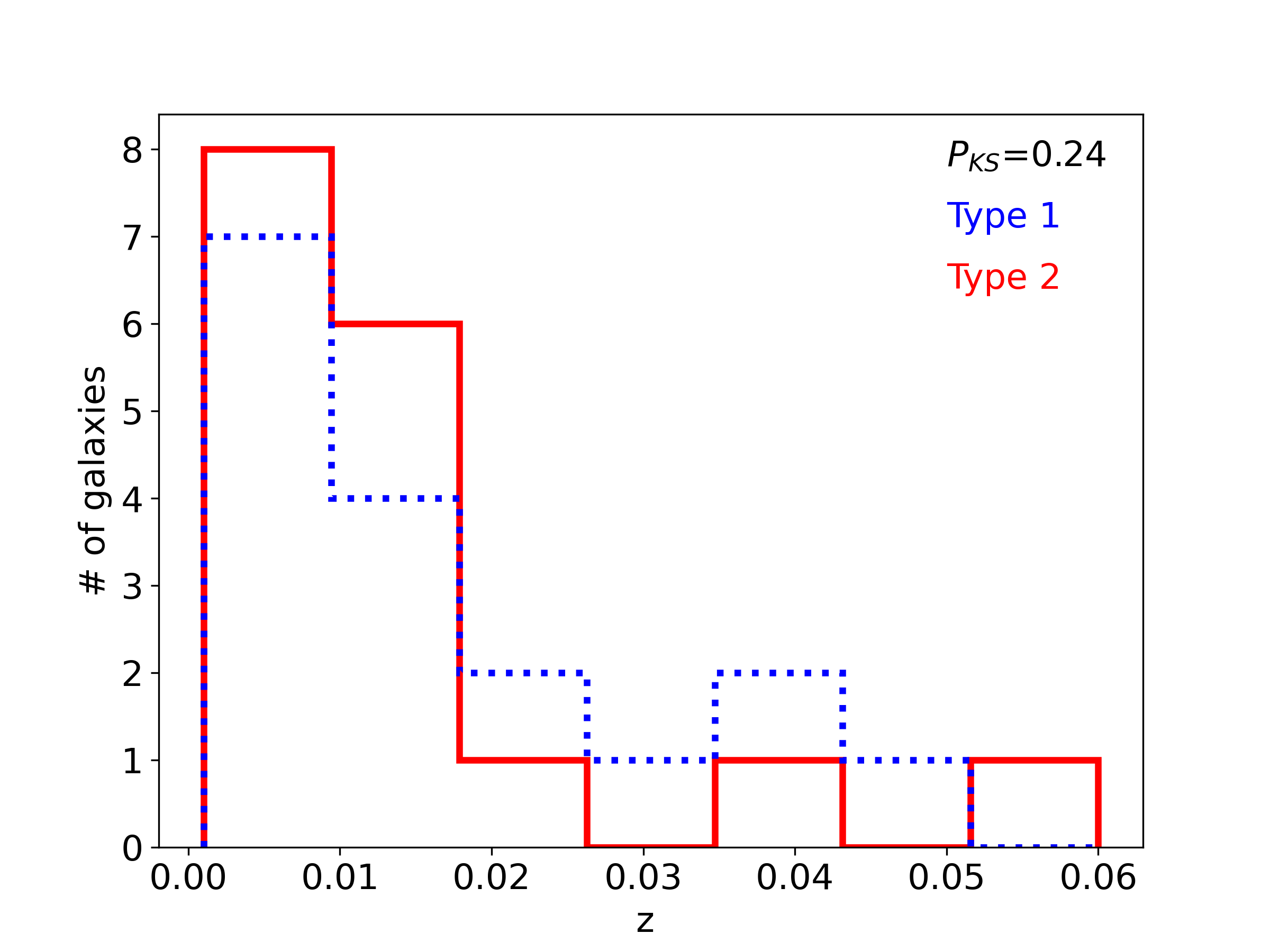} 
\caption{\small X-ray luminosity (top) and redshift (bottom) distributions of our sample, divided in type 1 (dashed-blue lines) and type 2 (continuous-red lines) AGN.  In each panel, we show the probability of the null hypothesis ($P_{\rm KS}$) that type 1 and type 2 AGN follow similar distributions computed using the Kolmogorov-Smirnov test. The $P_{\rm KS}$ values indicate that type 1 and type 2 AGN of our sample do not follow distinct distributions in terms of $L_{\rm X}$ and $z$.}
    \label{fig:histograms}
\end{figure}

In Figure~\ref{fig:histograms} we show the X-ray luminosity and redshift distribution of the galaxies of our sample. We divide the sample into type 1 (blue-dashed line) and type 2 (red-continuous line) as in the Table~\ref{tab:sample}. To test whether the distributions for the type 1 and type 2 are drawn from the same underlying distribution, 
we use the Kolmogorov-Smirnov (K-S) test and compute the probability of the null hypothesis ($P_{\rm KS}$).  $P_{\rm KS}<0.05$ implies that the null hypothesis, that the two distributions are drawn from the same underlying distribution, is rejected at a confidence level of 95 per cent. We find high values of $P_{\rm KS}$ for both $L_{\rm X}$ ($P_{\rm KS}=0.75$)  and $z$ ($P_{\rm KS}=0.24$) distributions, meaning that the type 1 and type 2 AGN in our sample most likely follow similar distributions in these two parameters. As type 1 and 2 AGN follow the same distributions in terms of X-ray luminosity and redshift, we compare these sub-samples in terms of other physical properties in the following sections.  

\subsection{Observations, Data Reduction and Measurements}

The observations were carried out with the NIFS instrument \citep{mcgregor03} on the Gemini North Telescope from 2006 to 2019. NIFS has a square field of view of $3\farcs0\times3\farcs0$, divided into 29 slices with an angular sampling of 0$\farcs$103$\times$0$\farcs$042. At the distances of the galaxies of our sample, the NIFS field of view varies from 60$\times$60 pc$^2$ to 3.5$\times$3.5 kpc$^2$.  Most of the observations (34/36) used Adaptive Optics (AO) by coupling NIFS with the ALTtitude conjugate Adaptive optics for the InfraRed (ALTAIR) system. Only NGC\,1125 and ESO578-G009 were observed without AO.  The resulting angular resolutions are shown in Table~\ref{tab:sample} and are in the range 0\farcs11--0\farcs44. Distinct observational strategies were used during the observations, which include the use of K or K$_{\rm long}$ gratings and spatial dithering, resulting in different spatial and spectral coverage among galaxies.  The spectral resolving power of NIFS in the K band is $R\approx5290$.  Telluric standard stars were observed just before and/or after the observations of each galaxy.

The data reduction followed the standard procedures \citep[e.g., ][]{rogemar_stellar} using the {\sc gemini iraf} package, including the trimming of the images, flat-fielding, cosmic ray rejection, sky subtraction, wavelength and s-distortion calibrations, removal of the telluric absorptions using the spectra of the telluric standard star and flux calibration  by interpolating a black body function 
to the spectrum of the telluric standard. Finally, the data cubes were created at an angular sampling of 0\farcs05$\times$0\farcs05 for each individual exposure and median combined using a sigma clipping algorithm to eliminate bad pixels and remaining cosmic rays and using the peak of the continuum emission of the galaxy as reference to perform the astrometry among the individual data cubes.

Results  on gas emission properties of individual sources have already been published for 22 galaxies of our sample based on the K-band data used here. The references to these studies are shown in Tab.~\ref{tab:sample}.

The K-band spectra of nearby active galaxies show plenty of H$_2$ emission lines and usually also present strong Br$\gamma$ emission \citep[e.g.][]{rogerio06}. 
To obtain the emission-line flux distributions, we integrated the fluxes within a spectral window of 1500 km\,s$^{-1}$ width centred at each emission line, after subtraction of the contribution of the underlying continuum fitted by a third order polynomial, similarly to the procedure adopted in \citet{sbN4151Exc}.  Our spectral window choice is very conservative, as the near-IR line widths in nearby galaxies are commonly much narrower than  1500 km\,s$^{-1}$ and so our choice warrants that we are computing the total line fluxes.
To minimize the effect of noise, we followed \citet{liu13} and first fitted each emission line by a combination of three Gaussian curves and the continuum  by a linear equation using the {\sc ifscube} code \citep{ifscube}. For type 1 AGN, we included an additional Gaussian to account for the broad Br$\gamma$ component. Before computing the fluxes of Br$\gamma$ in type 1 objects, we subtracted the contribution of this component and thus all measurements presented in this paper are for the narrow line components.

The fitting routine starts by modelling the spaxel corresponding to the peak of the continuum emission, using initial guesses for the centroid velocity and velocity dispersion of each component  provided by the user. Then the {\sc ifscube} code performs the fitting of the neighboring spaxles following a spiral loop and using the parameters from spaxels located at distances smaller than 0\farcs25 from the fitted spaxel as optimized guesses (as defined by the {\it refit} parameter). As mentioned above, we allow up to three Gaussian components to fit each line profile (4 for the Br$\gamma$ line in Sy\,1), but the code finds  the minimum number of Gaussians to reproduce the profiles by setting the initial guesses for the amplitudes of the unnecessary Gaussian functions to zero.  The fit of multi Gaussians has no physical motivation, it merely aims at reproducing the observed profiles. The emission-line flux distributions measured directly from the observed data cubes are consistent with those obtained from the fits of the line profiles, but the direct measurements produce noisier maps, as already discussed in \citet{liu13}.

\section{Results}\label{sec:results}

\subsection{Emission-line flux distributions and line-ratio maps}\label{resflux}

\begin{figure*}
    \centering
    \includegraphics[width=0.98\textwidth]{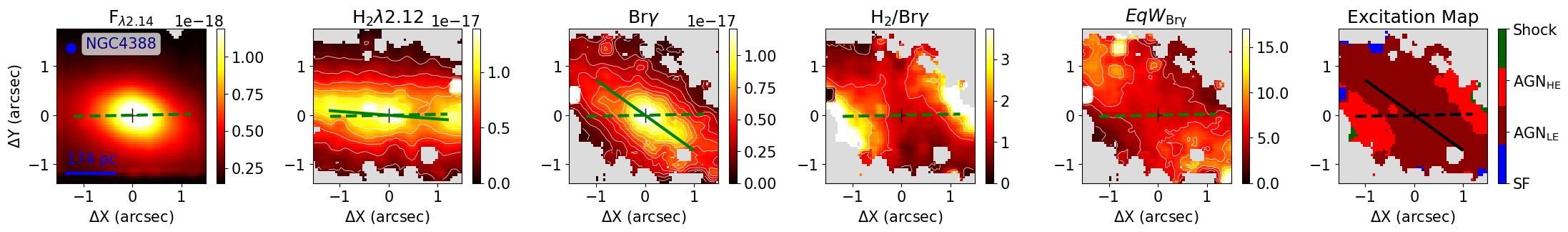}
    \includegraphics[width=0.98\textwidth]{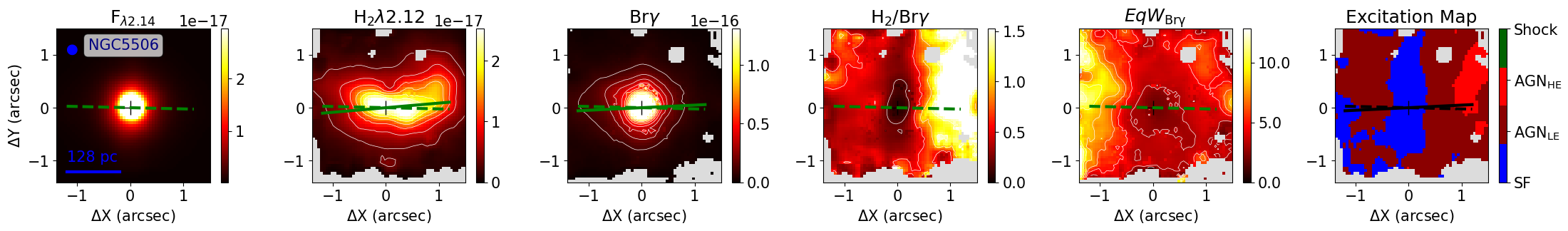}
    \includegraphics[width=0.98\textwidth]{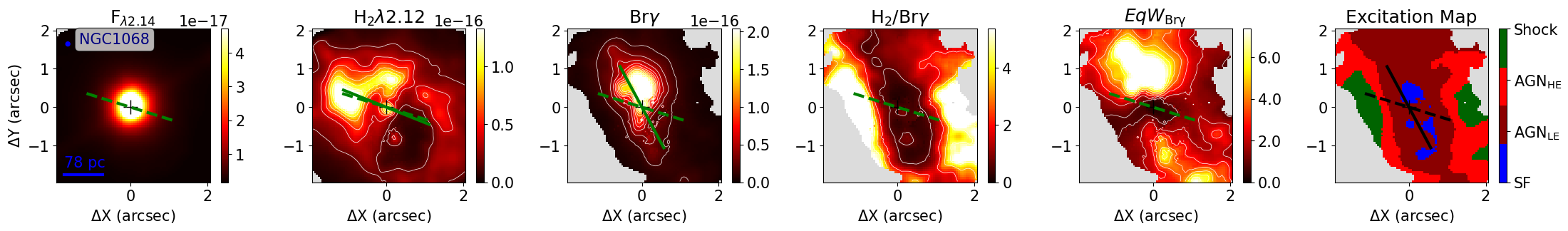}
\caption{\small {\bf Selected maps for type 2 AGN}. From left to right:  K-band continuum obtained within a spectral window of 100\,\AA\ centred at 2.14\,$\mu$m, H$_2\,2.1218\,\mu$m flux distribution, Br$\gamma$ flux distribution, H$_2$/Br$\gamma$ ratio map, Br$\gamma$ equivalent width map and excitation map. The color bars show the continuum in units of erg\,s$^{-1}$\,cm$^{-2}$\,\AA$^{-1}$\,spaxel$^{-1}$, the emission-line fluxes in erg\,s$^{-1}$\,cm$^{-2}$\,spaxel$^{-1}$ and the Br$\gamma$ equivalent width in \AA. The excitation map identifies with different colors the regions with typical $H_2$/Br$\gamma$ values for: star forming regions (SF: H$_2$/Br$\gamma<0.4$), low excitation AGN (AGN$_{\rm LE}$: $0.4\leq$ H$_2$/Br$\gamma<2$) and high excitation AGN (AGN$_{\rm HE}$: $2\leq$ H$_2$/Br$\gamma<6$) and shock-dominated objects (Shock: H$_2$/Br$\gamma>6$). In each row, the name of the galaxy is identified in the continuum image, the filled circle corresponds to the angular resolution, the spatial scale is shown in the bottom-left corner of the continuum image and the cross marks the position of the peak of the continuum emission. The dashed line indicates the orientation of the galaxy major axis obtained from the Hyperleda database. The continuous line on the flux maps shows the orientation of the emission-line flux distributions. The continuous line on the excitation map shows the orientation of the Br$\gamma$ emission. The gray regions indicate locations where emission lines are not detected above 3 times the continuum noise amplitude (3$\sigma$).} For all galaxies, north is up and east is to the left.
    \label{fig:mapsS2}
\end{figure*}

\begin{figure*}
    \centering
\includegraphics[width=0.98\textwidth]{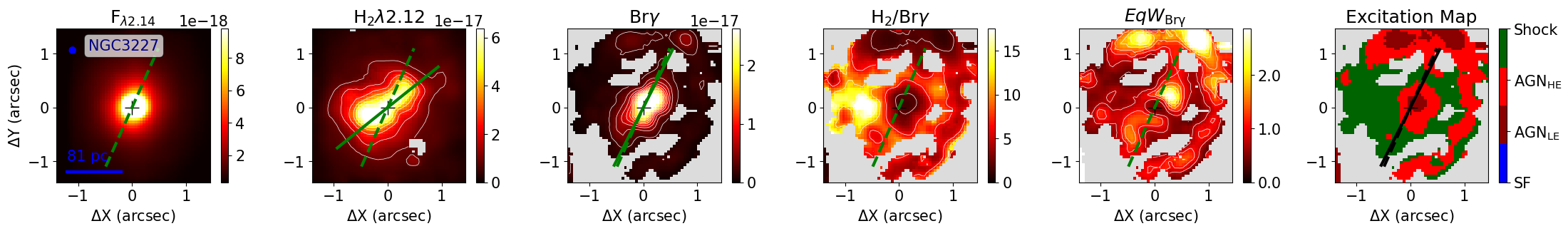}
\includegraphics[width=0.98\textwidth]{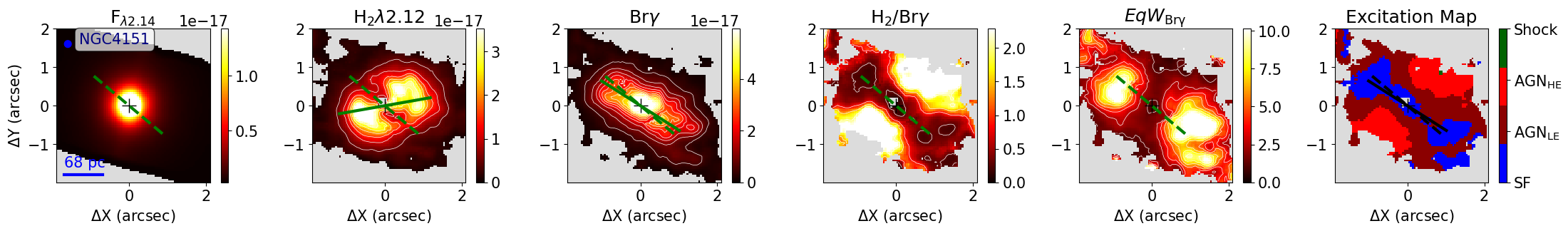}
\includegraphics[width=0.98\textwidth]{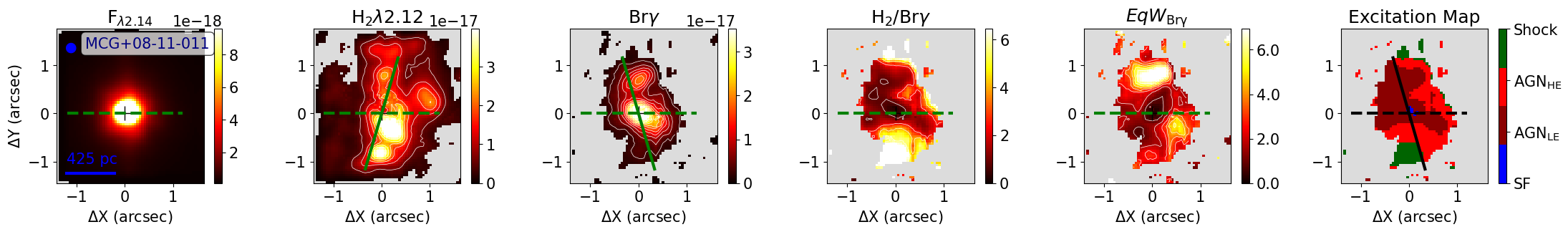}
\caption{\small {\bf Examples of maps for type 1 AGN}. Same as Fig.~\ref{fig:mapsS2}, but for type 1 AGN.}
    \label{fig:mapsS1}
\end{figure*}

In Figures~\ref{fig:mapsS2} and ~\ref{fig:mapsS1} we present examples of maps constructed from the Gemini NIFS data measurements, comprising: the continuum, H$_2\,2.1218\,\mu$m and Br$\gamma$ flux distributions, the H$_2\,2.1218\,\mu$m/Br$\gamma$ ratio map and Br$\gamma$ equivalent width ($EqW$) map for selected type 2 and type 1 AGN, respectively. The maps for the other galaxies are shown in Figures~\ref{fig:mapsS2ap} and ~\ref{fig:mapsS1ap} of the Appendix. We masked out regions where the amplitude of the line profile was smaller than three times the standard deviation of the continuum next to each line profile. The galaxies NGC\,3393 (Sy\,2) and Mrk\,352 (Sy\,1) do not present extended line emission and thus we do not show their corresponding maps.  All other galaxies show extended emission in both H$_2$ and Br$\gamma$ emission lines.  The only exception is NGC\,3516, for which the Br$\gamma$ emission is seen only from the unresolved nucleus.

\subsubsection{Flux distributions}
\label{sec:flux-dist}

The emission-line flux distributions present a wide variety of structures in both Br$\gamma$ and H$_2$ emission. In most galaxies, the H$_2$ emission is more extended than the Br$\gamma$ and the peak emission of both lines is observed at the galaxy nucleus. Other structures observed in the gas distribution comprise: nuclear spirals seen in molecular gas (e.g. Mrk\,79); galaxies in which both H$_2$ and Br$\gamma$ emission are seen mainly along the major axis of the large scale disk (e.g. NGC\,4235); galaxies in which the H$_2$ and Br$\gamma$ emission are distributed along distinct orientations (e.g. NGC\,4388); ring-like structures (e.g. Mrk\,1044); galaxies with elongated H$_2$ emission and round Br$\gamma$ flux distribution (e.g. Mrk\,766), among other emission structures. There is no clear difference between the emission-line flux distributions of type 1 and type 2 AGN. A qualitative inspection of the H$_2$ and Br$\gamma$ flux maps in each object (Figs.~\ref{fig:mapsS2}, \ref{fig:mapsS1}, \ref{fig:mapsS2ap} and \ref{fig:mapsS1ap}) shows that the  Br$\gamma$ usually traces a more collimated emission, while the H$_2$ emission spreads more over the whole field of view. This result is consistent with previous studies where the H$_2$ and ionised gas have been shown to present distinct flux distributions and kinematics \citep{rogemar_sample,schonell19}.

\begin{figure}
    \centering
\includegraphics[width=0.4\textwidth]{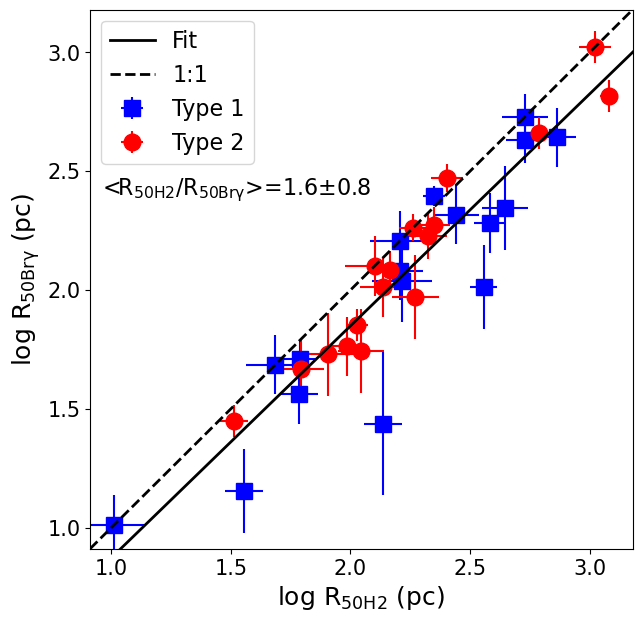}
\caption{\small Comparison of the radii that contain 50\,\% of the total flux of H$_2$2.1218\,$\mu$m and Br$\gamma$. The dashed line shows the 1:1 relation and the continuous line shows the best fit of the data by a first order polynomial, given by $\log{R_{\rm 50\,Br\gamma}}=(0.98\pm0.20)\times\log{R_{\rm 50\,H_2}} - (0.10\pm0.44)$.}
\label{fig:R50}
\end{figure}

We compute the radii that contain 50\,\% of the total flux ($R_{\rm 50}$) of H$_2$\,2.1218$\mu$m and Br$\gamma$ emission lines. Although this parameter can be affected by projection effects if the H$_2$ and Br$\gamma$ emission originate from distinct spatial locations in individual targets, the $R_{\rm 50}$ is useful to compare the H$_2$ and Br$\gamma$ emission in the whole sample.
The corresponding $R_{\rm 50}$ values for H$_2$ and Br$\gamma$ are shown in Table~\ref{tab:pas} and Figure~\ref{fig:R50} shows the comparison between $R_{\rm 50 H2}$ and $R_{\rm 50 Br\gamma}$.  The Br$\gamma$ flux distribution is more concentrated than that of H$_2$ and the  $R_{\rm 50}$ for the H$_2$ is on average 56\,\% larger than that of Br$\gamma$.

We use the {\sc cv2.moments} python package to compute the moments and orientation of the H$_2$ and Br$\gamma$ flux distributions. The orientations are shown as the continuous lines in the flux maps of Figs.~\ref{fig:mapsS2}, \ref{fig:mapsS1}, \ref{fig:mapsS2ap} and  \ref{fig:mapsS1ap}. The position angles (PAs) are listed in Table~\ref{tab:pas}, together with the orientation of the major axis of the galaxy obtained from the Hyperleda database \citep{paturel03}, measured from the 25\,mag\,arcsec$^2$ isophote in a B-band image. We performed Monte Carlo simulations with 100 iterations each to compute the uncertainties, by adding random noise with amplitude of the 20th percentile flux value of the corresponding map. The listed uncertainties in Tab.~\ref{tab:pas} correspond to the standard deviation of the mean of the simulations.

We have compared the PA offset ($\Delta$PA) between the major axis of the large scale disk and the orientation of the Br$\gamma$ and H$_2$ flux distributions. Although the orientations of both flux distributions vary, the K-S test ($P_{\rm KS}=0.97)$ indicates the distributions of the corresponding $\Delta$PA for  Br$\gamma$ and H$_2$ are not distinct. $\rm \Delta PA>30^\circ$ is usually used as a threshold to determine whether the stellar and gas disks are misaligned \citep{Jin16}. We find $\Delta$PA larger than this value in 15 galaxies (42\,\%) for the Br$\gamma$ and  in 16 galaxies (44\,\%) for H$_2$. These fractions are higher than those between the kinematic position angle of the stellar velocity field in the inner 3$^{\prime\prime}\times$3$^{\prime\prime}$ and the major axis of the large scale disk, of $\sim$18\,\% \citep{rogemar_stellar}, suggesting an origin in non-circular motions in the gas.
 The distributions of the $\Delta$PA between the orientation of the H$_2$ and Br$\gamma$ flux distributions (PA$_{\rm H2}$-PA$_{\rm Br\gamma}$) for type 1 and type 2 AGN are similar. We find $\rm\Delta PA>30^\circ$ for 15 galaxies (45\,\%). 3 galaxies (NGC\,1275, NGC\,5548 and Mrk\,766) of these show  $\rm \Delta PA>60^\circ$, for which the H$_2$ emission is observed mainly along the major axis of the galaxy.   

\begin{table*}
	\centering
	\scriptsize
	\caption{\small  (1) Galaxy name, (2) position angle of the large scale disk from Hyperleda database, (3) H$_2$ and (4) and Br$\gamma$ (4) position flux distribution position angles, (5) and (6) radii that contain half of the total flux of H$_2$\,2.1218$\mu$m and Br$\gamma$. All angles are measured east of north. 
	}
	\label{tab:pas}
	\begin{tabular}{lccccc} 
		\hline
Galaxy & $\Psi_0$ & PA$_{\rm H_2}$ &  PA$_{\rm Br\gamma}$ & $R_{\rm 50H_2}$ & $R_{\rm 50Br\gamma}$  \\
  & (deg) & (deg) & (deg)& (pc) & (pc)\\
  (1) & (2) & (3) & (4) & (5) & (6)  \\
		\hline
		\multicolumn{6}{c}{type 2}\\
		\hline
NGC788 & 108.1 & 39.8 $\pm$ 1.6 & 89.4 $\pm$ 3.2 & 254 $\pm$ 42 & 296 $\pm$ 42  \\		
NGC1052 & 112.7 & 44.0 $\pm$ 1.8 & 162.0 $\pm$ 7.7 & 62 $\pm$ 15 & 46 $\pm$ 15  \\
NGC1068 & 72.7 & 68.3 $\pm$ 1.2 & 28.3 $\pm$ 0.6 &  106 $\pm$ 11 & 71 $\pm$ 11    \\
NGC1125 & 53.5 & 46.5 $\pm$ 1.4 & 56.7 $\pm$ 1.2 &  137 $\pm$ 34 & 102 $\pm$ 34  \\
NGC1241 & 147.7 & 157.7 $\pm$ 4.6 & 46.9 $\pm$ 0.7 &  126 $\pm$ 42 & 126 $\pm$ 42  \\
NGC2110 & 175.1 & 143.4 $\pm$ 2.2 & 152.5 $\pm$ 0.9 & 145 $\pm$ 24 & 121 $\pm$ 24 \\  
NGC4258 & 150.0 & 20.3 $\pm$ 2.6 & 13.2 $\pm$ 7.5 & 32 $\pm$ 4 & 28 $\pm$ 4   \\
NGC4388 & 91.1 & 85.6 $\pm$ 1.0 & 54.3 $\pm$ 0.8 & 183 $\pm$ 26 & 183 $\pm$ 26  \\
NGC5506 & 88.7 & 94.9 $\pm$ 1.3 & 92.8 $\pm$ 4.3 & 96 $\pm$ 19 & 57 $\pm$ 19  \\
NGC5899 & 20.8 & 10.3 $\pm$ 1.0 & 170.7 $\pm$ 1.7 & 80 $\pm$ 26 & 53 $\pm$ 26  \\
NGC6240 & 12.2 & 0.9 $\pm$ 0.7 & 178.5 $\pm$ 0.8 & 610 $\pm$ 76 & 458 $\pm$ 768 \\ 
Mrk3 & 15.0 & 73.7 $\pm$ 1.3 & 80.9 $\pm$ 1.2 & 210 $\pm$ 42 & 168 $\pm$ 42  \\
Mrk348 & 87.0 & 160.8 $\pm$ 2.2 & 7.5 $\pm$ 5.9 & 186 $\pm$ 46 & 93 $\pm$ 46  \\
Mrk607 & 137.3 & 134.8 $\pm$ 1.2 & 140.6 $\pm$ 2.2 & 110 $\pm$ 27 & 55 $\pm$ 27  \\
Mrk1066 & 112.3 & 118.8 $\pm$ 1.5 & 126.0 $\pm$ 0.3 & 224 $\pm$ 37 & 186 $\pm$ 37   \\
ESO578-G009 & 27.6 & 29.9 $\pm$ 0.6 & 29.3 $\pm$ 0.7 & 1199 $\pm$ 109 & 654 $\pm$ 109  \\
CygnusA & 151.0 & 176.1 $\pm$ 1.8 & 140.6 $\pm$ 1.0 & 1049 $\pm$ 174 & 1049 $\pm$ 174 \\
		\hline
		\multicolumn{6}{c}{type 1}\\
		\hline
NGC1275 & 110.0 & 42.2 $\pm$ 6.1 & 137.9 $\pm$ 2.2 & 164 $\pm$ 54 & 109 $\pm$ 54 \\  
NGC3227 & 156.0 & 128.8 $\pm$ 3.1 & 153.1 $\pm$ 2.2 & 60 $\pm$ 12 & 36 $\pm$ 12\\  
NGC3516 & 55.0 & 138.0 $\pm$ 4.0 & --  &  $\pm$ 27 & 27 $\pm$ 27  \\
NGC4051 & 139.4 & 96.9 $\pm$ 1.5 & 117.4 $\pm$ 4.6 &  35 $\pm$ 7 & 14 $\pm$ 7  \\
NGC4151 & 50.0 & 100.0 $\pm$ 1.0 & 57.5 $\pm$ 0.4 & 61 $\pm$ 10 & 51 $\pm$ 10  \\
NGC4235 & 49.0 & 43.4 $\pm$ 0.5 & 42.6 $\pm$ 0.6 & 224 $\pm$ 24 & 249 $\pm$ 24 \\
NGC4395 & 127.8 & 106.4 $\pm$ 3.2 & 104.3 $\pm$ 5.2 & 10 $\pm$ 4 & 10 $\pm$ 4   \\
NGC5548 & 118.2 & 146.0 $\pm$ 3.8 & 46.2 $\pm$ 2.3 & 160 $\pm$ 53 & 160 $\pm$ 53  \\
NGC6814 & 107.6 & 121.2 $\pm$ 1.3 & 153.2 $\pm$ 1.3 & 48 $\pm$ 16 & 48 $\pm$ 16 \\
Mrk79 & 73.0 & 176.2 $\pm$ 3.9 & 171.9 $\pm$ 35.1 & 276 $\pm$ 69 & 207 $\pm$ 69  \\
Mrk509 & 80.0 & 139.4 $\pm$ 5.1 & 96.3 $\pm$ 4.3 & 536 $\pm$ 107 & 428 $\pm$ 107  \\
Mrk618 & 85.0 & 39.4 $\pm$ 2.2 & 24.8 $\pm$ 3.9 & 442 $\pm$ 110 & 221 $\pm$ 110    \\
Mrk766 & 73.1 & 62.0 $\pm$ 1.4 & 135.1 $\pm$ 1.1 & 160 $\pm$ 40 & 120 $\pm$ 40  \\
Mrk926 & 104.1 & 71.7 $\pm$ 4.2 & 104.0 $\pm$ 2.9 & 730 $\pm$ 146 & 438 $\pm$ 146  \\
Mrk1044 & 177.5 & 169.2 $\pm$ 2.2 & 6.5 $\pm$ 7.9 & 359 $\pm$ 51 & 102 $\pm$ 51  \\
Mrk1048 & 80.3 & 135.4 $\pm$ 1.5 & 167.4 $\pm$ 3.0 & 536 $\pm$ 134 & 536 $\pm$ 134  \\
MCG+08-11-01 & 90.0 & 163.6 $\pm$ 1.3 & 16.0 $\pm$ 2.4 & 383 $\pm$ 63 & 191 $\pm$ 63 \\
\hline
	\end{tabular}
\end{table*}

\subsubsection{Line ratios}

The H$_2\,2.1218\,\mu$m/Br$\gamma$ emission-line ratio is commonly used to investigate the main source of the H$_2$ excitation \citep{reunanen02,ardila04,ardila05,sbN4151Exc,rogemarMrk1066exc,rogemarN1068,rogerio13,rogemarHe,colina15,Schonell14,schonell19,dahmer19b,fazeli20}. Small values ($\rm H_2/Br\gamma\lesssim0.4$) are usually observed in H\,{\sc ii} regions and star forming galaxies, while AGN present $\rm0.4\lesssim H_2/Br\gamma\lesssim6.0$ and higher values are usually observed in LINERs and shock-dominated regions \citep[e.g.][]{rogerio13,colina15,rogemar21_exc}. In the near-IR the line ratio limits are empirical and their excitation mechanisms are less understood than those of the optical lines  \citep[e.g.][]{ardila04,ardila05}. However, it is worth mentioning that the H$_2$/Br$\gamma$ line ratio can be affected by the geometry of the H{\sc ii} region \citep{puxley90} and the velocity of the shock \citep{wilgenbus2000}. Both properties affect the dissociation of the H$_2$ molecule, making the fraction of H$_2$, and the H$_2$/Br$\gamma$ ratio, to change. The H$_2$/Br$\gamma$ maps for our sample (Figs.~\ref{fig:mapsS2}, \ref{fig:mapsS1}, \ref{fig:mapsS2ap} and \ref{fig:mapsS1ap}) show values ranging from nearly zero, as seen in the rings of star forming regions of Mrk\,1044 and for NGC\,4151 -- in which the H$_2$ emission decreases due to the dissociation of the molecule by the strong AGN radiation field \citep[e.g.][]{sbN4151Exc} -- to values of up to $\rm H_2/Br\gamma\sim30$ for NGC\,1275, where the H$_2$ emission originates in shocks produced by AGN winds \citep{rogemarN1275}. 

 We build excitation maps (fifth column of Figs.\,\ref{fig:mapsS2}, \ref{fig:mapsS1}, \ref{fig:mapsS2ap} and \ref{fig:mapsS1ap}) to spatially locate the regions where different excitation mechanisms may be occurring. 
All galaxies present spaxels dominated by AGN excitation ($0.4<\rm H_2/Br\gamma<6$), and in order to map the variation of this excitation, we have divided the AGN regions into low ($0.4\leq\rm H_2/Br\gamma<2$) and high excitation ($2\leq\rm H_2/Br\gamma<6$). This separation allow us to further investigate the origin of the H$_2$ emission in the AGN. A similar separation was done in \citet{rogemar_N1275} to split the high line ratio region in the H$_2$\,2.1218\,$\mu$m/Br$\gamma$ vs. [Fe\,{\sc ii}]1.2570\,$\mu$m/Pa$\beta$ diagnostic diagram.

Table\,\ref{tab:h2br} presents the median H$_2$/Br$\gamma$ ratio over the whole FoV (H$_2$/Br$\gamma_{\rm MED}$) for each galaxy of our sample, the median value within $r<125$\,pc (H$_2$/Br$\gamma_{r<125 {\rm pc}}$), the nuclear ratio, computed using an aperture with radius equal to the angular resolution of the data for each galaxy (H$_2$/Br$\gamma_{\rm nuc}$), and the extra-nuclear line ratio, measured as the median value of spaxels located at distances from the galaxy nucleus larger than the angular resolution (H$_2$/Br$\gamma_{\rm extra}$). The median values of H$_2$/Br$\gamma$ are within the AGN range for 31 (91\,\%) galaxies of our sample. The exceptions are NGC\,6240, NGC\,1275 and Mrk\,3 that show H$_2$/Br$\gamma$ median values of 11.76, 11.10 and 0.17, respectively. The high H$_2$/Br$\gamma_{\rm MED}$ values are consistent with shocks as the dominant H$_2$ excitation mechanism \citep{Ilha16,sanchez18,rogemarN1275}. The low ratios observed for Mrk\,3, NGC\,5506 and NGC\,4151  may be explained by the 
 dissociation of the H$_2$ molecule by the AGN radiation field in these galaxies as proposed by previous works \citep{Gnilka20,sbN4151Exc}.
As seen in Tab.~\ref{tab:h2br}, type 1 and type 2 AGN show similar values of H$_2$/Br$\gamma$ median values.

We use K-S statistics to test whether the distributions of H$_2$/Br$\gamma_{\rm MED}$ for type 1 and type 2 AGN are distinct.
We find $P_{\rm KS}=0.96$, meaning that likely the H$_2$/Br$\gamma_{\rm MED}$ distributions of type 1 and type 2 AGN are drawn from the same underlying distribution. However, our observations cover spatial scales from a few tens of pc to a few kpc, and so, for the most distant objects the H$_2$/Br$\gamma_{\rm MED}$ is dominated by the extra-nuclear regions, while in the closest galaxies, the contribution of the nuclear emission is higher. In order to avoid this problem, we compute the H$_2$/Br$\gamma$ within the inner 125\,pc radius for all objects.  This aperture corresponds to the lowest spatial resolution in our sample (for ESO578-G009).  For three galaxies (NGC\,4258, NGC\,4051 and NGC\,4395) the FoV is smaller than 125\,pc radius and thus, we use the whole FoV to compute the H$_2$/Br$\gamma$ in these cases. The K-S test indicates again that type 1 and type 2 AGN do not have distinct distributions of H$_2$/Br$\gamma$ within the inner 125\,pc radius. Similarly, we do not find a statistically significant difference in the nuclear H$_2$/Br$\gamma$ distributions (computed for an aperture corresponding to the angular resolution of the data for each galaxy) for type 1 and type 2 AGN. Finally, we test whether H$_2$/Br$\gamma$ (median, within 125\,pc radius and nuclear) and the hard X-ray luminosity are correlated using the Pearson test, resulting that these parameters do not present a statistically significant correlation.

\begin{table*}
	\centering
	\scriptsize
	\caption{\small H$_2$/Br$\gamma$ line ratio: (1) galaxy name, (2) median values of the line ratio for the whole FoV (H$_2$/Br$\gamma_{\rm MED}$), (3) in the inner $r<125$\,pc  (H$_2$/Br$\gamma_{r<125\rm pc}$), (4) wihin the central angular resolution element (H$_2$/Br$\gamma_{\rm nuc}$) and (5) for spaxels at distances larger than the angular resolution radius (H$_2$/Br$\gamma_{\rm extra}$). }
	\label{tab:h2br}
	\begin{tabular}{lcccc} 
		\hline
Galaxy & H$_2$/Br$\gamma_{\rm MED}$ & H$_2$/Br$\gamma_{r<125\rm pc}$ & H$_2$/Br$\gamma_{\rm nuc}$ & H$_2$/Br$\gamma_{\rm extra}$ \\
(1) & (2) & (3) & (4) & (5)  \\
		\hline
		\multicolumn{5}{c}{type 2}\\
		\hline
NGC788  & 0.96$\pm$0.91 & 0.82$\pm$0.35 & 0.53$\pm$0.05 & 0.96$\pm$0.91  \\
NGC1052 & 2.32$\pm$1.07 & 2.32$\pm$1.07 & 2.15$\pm$0.27 & 2.36$\pm$1.09  \\
NGC1068 & 1.78$\pm$3.18 & 1.30$\pm$3.72 & 0.54$\pm$0.10 & 1.79$\pm$3.18  \\
NGC1125 & 0.78$\pm$0.62 & 0.39$\pm$0.06 & 0.38$\pm$0.04 & 0.88$\pm$0.62  \\
NGC1241 & 3.11$\pm$2.21 & 3.21$\pm$1.58 & 2.67$\pm$0.09 & 3.18$\pm$2.24  \\
NGC2110 & 3.61$\pm$1.57 & 3.15$\pm$1.57 & 1.36$\pm$0.38 & 3.65$\pm$1.55  \\
NGC4258 & 1.84$\pm$0.90 & 1.84$\pm$0.90 & 2.59$\pm$1.47 & 1.81$\pm$0.85  \\
NGC4388 & 1.25$\pm$1.16 & 1.31$\pm$0.38 & 1.15$\pm$0.16 & 1.26$\pm$1.17  \\
NGC5506 & 0.51$\pm$0.68 & 0.43$\pm$0.45 & 0.09$\pm$0.02 & 0.51$\pm$0.68  \\
NGC5899 & 4.35$\pm$4.13 & 4.27$\pm$4.38 & 2.40$\pm$0.13 & 4.41$\pm$4.15  \\
NGC6240 & 11.78$\pm$7.49 & 8.55$\pm$4.70 & 8.65$\pm$3.96 & 11.95$\pm$7.86\\  
Mrk3 & 0.17$\pm$0.15 & 0.07$\pm$0.04 & 0.04$\pm$0.01 & 0.18$\pm$0.15  \\
Mrk348 & 1.30$\pm$0.84 & 0.89$\pm$0.28 & 0.56$\pm$0.09 & 1.34$\pm$0.83  \\
Mrk607 & 1.06$\pm$0.69 & 0.93$\pm$0.37 & 0.51$\pm$0.06 & 1.08$\pm$0.69  \\
Mrk1066 & 1.39$\pm$1.51 & 1.41$\pm$0.55 & 1.41$\pm$0.16 & 1.39$\pm$1.51  \\
ESO578-G009 & 1.60$\pm$1.77 & 2.97$\pm$0.77 & 2.17$\pm$1.17 & 1.59$\pm$1.79\\  
CygnusA & 2.06$\pm$1.75 & 1.92$\pm$0.14 & 1.93$\pm$0.27 & 2.08$\pm$1.76  \\
{\bf Mean} & 2.35$\pm$2.58 & 2.10$\pm$1.96 & 1.71$\pm$1.94 & 2.38$\pm$2.62 \\
		\hline
		\multicolumn{5}{c}{type 1}\\
		\hline
NGC1275 & 11.09$\pm$5.27 & 12.40$\pm$3.95 & 7.56$\pm$2.33 & 11.42$\pm$5.37 \\ 
NGC3227 & 5.72$\pm$3.84 & 5.89$\pm$3.81 & 1.84$\pm$0.41 & 5.75$\pm$3.83  \\
NGC3516 & 2.31$\pm$0.86 & 2.47$\pm$0.74 & 2.14$\pm$0.42 & 2.61$\pm$0.99  \\
NGC4051 & 3.16$\pm$2.03 & 3.16$\pm$2.03 & 0.53$\pm$0.08 & 3.19$\pm$2.02  \\
NGC4151 & 0.76$\pm$1.16 & 0.79$\pm$1.19 & 0.21$\pm$0.07 & 0.78$\pm$1.16  \\
NGC4235 & 3.71$\pm$1.88 & 3.10$\pm$1.29 & --            & 3.71$\pm$1.88  \\
NGC4395 & 1.01$\pm$0.88 & 1.01$\pm$0.88 & 0.79$\pm$0.06 & 1.02$\pm$0.89  \\
NGC5548 & 1.37$\pm$0.48 & 1.27$\pm$0.33 & 1.17$\pm$0.14 & 1.39$\pm$0.49  \\
NGC6814 & 5.55$\pm$6.00 & 5.73$\pm$41.48 & 18.57$\pm$149.12 & 5.34$\pm$6.45\\  
Mrk79 & 2.83$\pm$1.61 & 2.59$\pm$0.79 & 2.56$\pm$0.74 & 2.87$\pm$1.66  \\
Mrk509 & 0.91$\pm$0.64 & 0.41$\pm$0.60 & 0.23$\pm$0.19 & 0.91$\pm$0.63  \\
Mrk618 & 1.24$\pm$1.97 & 0.44$\pm$0.04 & 0.44$\pm$0.04 & 1.28$\pm$1.99  \\
Mrk766 & 0.80$\pm$1.02 & 0.34$\pm$0.12 & 0.13$\pm$0.04 & 0.82$\pm$1.03  \\
Mrk926 & 1.83$\pm$0.80 & 1.71$\pm$0.19 & 1.72$\pm$0.28 & 1.85$\pm$0.82  \\
Mrk1044 & 0.68$\pm$0.52 & 0.05$\pm$0.01 & --            & 0.68$\pm$0.52  \\
Mrk1048 & 1.75$\pm$1.42 & 3.25$\pm$1.12 & 3.07$\pm$1.90 & 1.69$\pm$1.32  \\
MCG+08-11-01 & 2.19$\pm$2.65 & 1.01$\pm$0.77 & 0.70$\pm$0.41 & 2.31$\pm$2.66  \\
{\bf Mean} & 2.76$\pm$2.57 & 2.68$\pm$2.96 & 2.78$\pm$4.59 & 2.80$\pm$2.62 \\
\hline
	\end{tabular}
\end{table*}

\begin{figure}
    \centering
\includegraphics[width=0.45\textwidth]{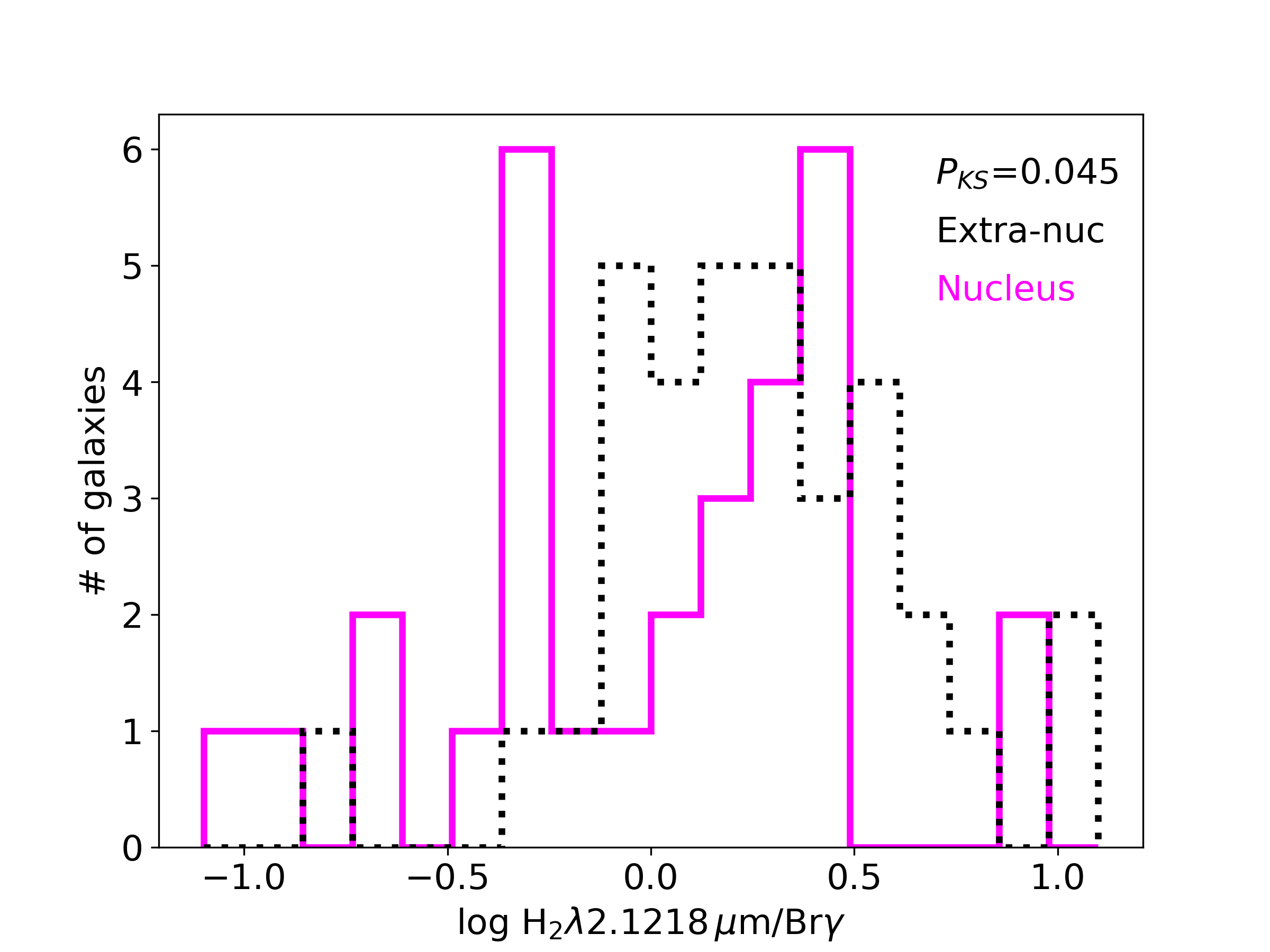}
\caption{\small Nuclear (magenta) and extra-nuclear (dotted black) H$_2$/Br$\gamma$ distributions of the galaxies of our sample. }
\label{fig:h2brdistr}
\end{figure}

Figure~\ref{fig:h2brdistr} shows the only distinct distributions we found: the H$_2$/Br$\gamma_{\rm nuc}$ and H$_2$/Br$\gamma_{\rm extra}$ ones. Using the K-S test we obtain  $P_{\rm KS}=0.045$ indicating that the distributions are distinct: on average, the nucleus presents lower H$_2$/Br$\gamma$ ratios than the extra-nuclear regions.

\subsection{Br$\gamma$ Equivalent width maps}

The fifth column of Figs.~\ref{fig:mapsS2}, \ref{fig:mapsS1}, \ref{fig:mapsS2ap} and \ref{fig:mapsS1ap} shows the Br$\gamma$ equivalent width ($EqW_{\rm Br\gamma}$) maps. The emission structures are better seen in the $EqW_{\rm Br\gamma}$ than in the flux distribution maps, as the former measure the emission relative to the continuum. For example, in Mrk\,3 (Fig.~\ref{fig:mapsS2ap}) the $EqW_{\rm Br\gamma}$ map shows knots of emission not easily observed in the Br$\gamma$ flux map. Young ($<$10 Myr) stellar populations present $EqW_{\rm Br\gamma}\gtrsim50$\,\AA, as predicted by evolutionary photoionisation models \citep{dors08,rogemarN7582}. All galaxies of our sample clearly show smaller values ($EqW_{\rm Br\gamma}\lesssim30$\,\AA) than those predicted for young stellar population models.  Even galaxies with known active star-formation, as NGC\,6240 \citep[e.g.][]{keel90,Lutz03,Pasquali04} and  NGC\,3227 \citep{Gonzalez97,Schinnerer00}, present low $EqW_{\rm Br\gamma}$ values, which suggest the near-infrared continuum  is dominated by the contribution of old stellar populations and the AGN featureless continuum in the nucleus.

Differently from the Br$\gamma$ flux distributions,  which usually present the emission peak at the nucleus, the highest values of $EqW_{\rm Br\gamma}$  are seen away from the nucleus in most galaxies of our sample. In addition, a visual inspection of the $EqW_{\rm Br\gamma}$ maps shows a drop in the $EqW_{\rm Br\gamma}$  values at the nucleus. This drop is more prominent in type 1 AGN than in type 2, possibly due to a stronger dilution of the Br$\gamma$ emission by the nuclear continuum. 

\begin{figure}
\centering
\includegraphics[width=0.45\textwidth]{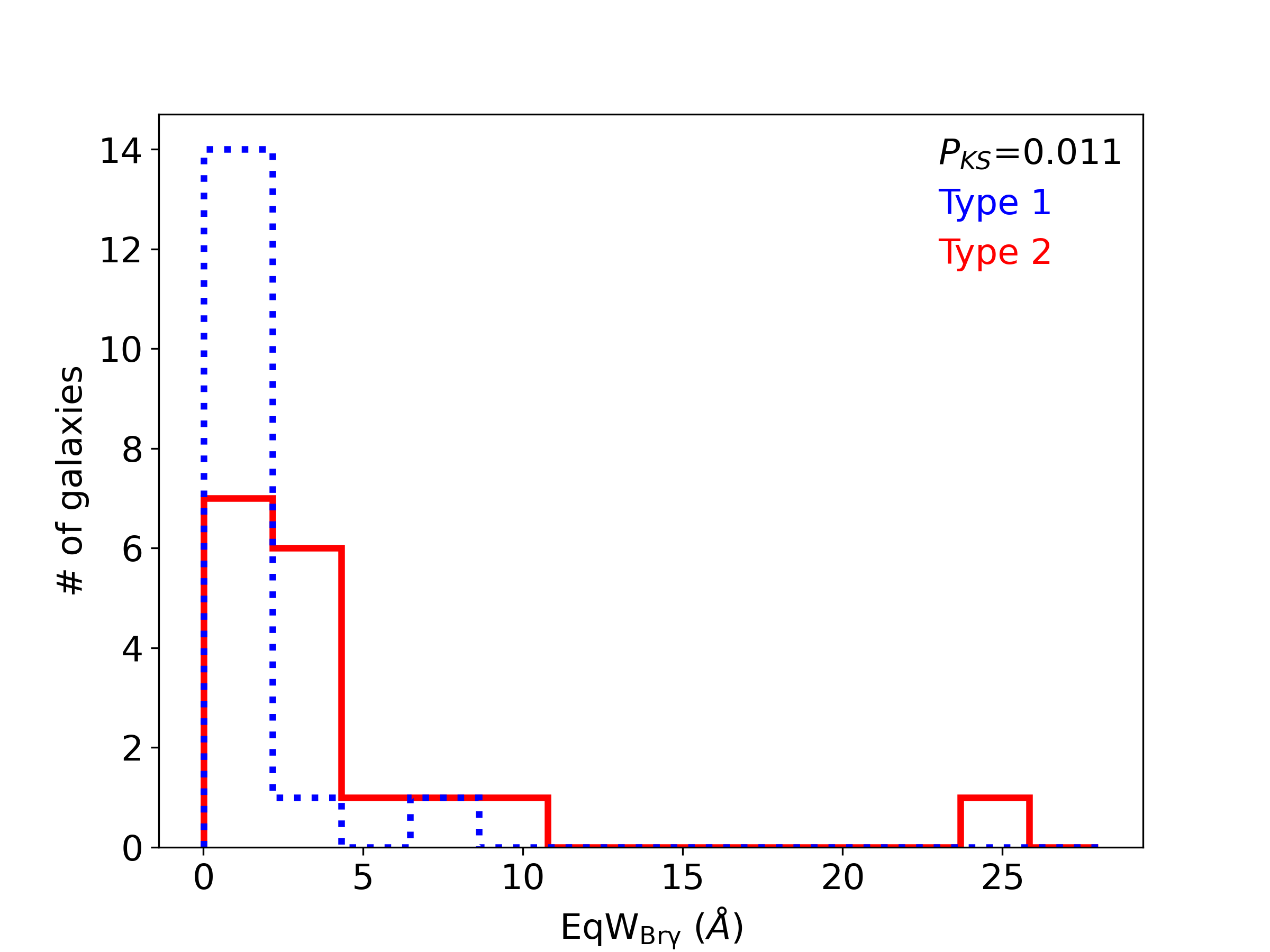}
\caption{\small Nuclear Br$\gamma$ equivalent width distributions for the type 1 (dashed-blue lines) and type 2 (continuous-red lines) AGN in our sample. The equivalent widths are measured within apertures corresponding to the angular resolution of the data. }
\label{fig:eqw}
\end{figure}

We measure the $EqW_{\rm Br\gamma}$ values for the nucleus of each galaxy within an aperture corresponding to the angular resolution of the data and compare the $EqW_{\rm Br\gamma}$  distributions of type 1 and type 2 AGN and find that they follow distinct distributions ($P_{\rm KS}=0.011$). Figure~\ref{fig:eqw} shows the $EqW_{\rm Br\gamma}$ distributions for type 1 AGN (in blue) and type 2 (in red) AGN.  A slightly higher $P_{\rm KS}$ (0.045) is obtained by comparing the Br$\gamma$ equivalent widths within the same spatial region of all galaxies (125 pc radius) implying the null hypothesis in which type 1 and type 2 AGN follow the same underlying distribution is rejected at the 95\,\% confidence level. On the other hand, the comparison of the median $EqW_{\rm Br\gamma}$ distributions over the whole FoV results in $P_{\rm KS}=0.465$, meaning that the null hypothesis cannot be rejected and likely type 1 and type 2 AGN originate from the same underlying $EqW_{\rm Br\gamma}$ distribution.  The Pearson test shows that $EqW_{\rm Br\gamma}$ does not correlate with $L_{\rm X}$.  This result suggests that there is no intrinsic difference between these type 1 and type 2 AGN, just circunstantial due orientation so that, at larger scales, where the central source is not seen, no clear distinction can be made between the two.

\begin{table*}
	\centering
	\small
	\caption{\small Ionised and hot H$_2$ gas masses and H$_2$ excitation temperatures. (1) Name of the galaxy; (2 \& 3) Hot Molecular and ionised hydrogen masses within the inner 125\,pc radius and (4 \& 5) in the whole field of view; (6 \& 7) H$_2$ vibrational and rotational temperatures. }
	\label{tab:masses}
	\begin{tabular}{lcccccc} 
		\hline
	Galaxy & $M_{{\rm H_2}\, r< \rm 125pc}$ & $M_{{\rm H\,II}\, r< 125\rm pc}$ & $M_{\rm H_2\,FoV}$ & $M_{\rm HII\, FoV}$& $T_{\rm vib}$ &  $T_{\rm rot}$  \\  
	       &(10$^1$ M$_{\odot}$)&(10$^4$ M$_{\odot}$)&(10$^1$ M$_{\odot}$)&(10$^4$ M$_{\odot}$) & (K) & (K) \\
	       (1) & (2) & (3) & (4) & (5) & (6) & (7) \\

	       \hline
		\hline
		\multicolumn{7}{c}{type 2}\\
		\hline
NGC788      & 12.8$\pm$0.5   & 19.5$\pm$0.6   & 59.0$\pm$5.6     & 83.4$\pm$9.2    & 2149$\pm$80 & 810$\pm$567  \\
NGC1052     & 11.0$\pm$1.1   & 5.7$\pm$1.6    & 11.9$\pm$1.4     & 5.8$\pm$1.8     & 4233$\pm$306 &      --      \\
NGC1068     & 202.5$\pm$2.2  & 193.1$\pm$3.8  & 288.6$\pm$4.2    & 222.2$\pm$6.2   & 2474$\pm$11 & 1551$\pm$332  \\ 
NGC1125     & 21.3$\pm$0.7   & 64.1$\pm$0.8   & 41.7$\pm$3.9     & 94.1$\pm$6.4    & 3857$\pm$161 & 1345$\pm$633   \\
NGC1241     & 34.9$\pm$0.7   & 13.0$\pm$0.9   & 51.0$\pm$5.4     & 19.4$\pm$4.3    & 3746$\pm$123 & 832$\pm$753   \\
NGC2110     & 41.7$\pm$0.8   & 16.8$\pm$1.0   & 100.5$\pm$3.1    & 27.4$\pm$2.1    & 2512$\pm$23  &      --      \\
NGC4258     & --               & --               & 0.7$\pm$0.1      & 0.3$\pm$0.1     & 5132$\pm$2490&      --      \\
NGC4388     & 40.0$\pm$1.1   & 35.3$\pm$1.5   & 119.6$\pm$5.8    & 101.5$\pm$6.4   & 2725$\pm$33 & 1065$\pm$593   \\
NGC5506     & 45.7$\pm$0.7   & 174.1$\pm$1.4  & 66.8$\pm$2.0     & 204.8$\pm$4.3   & 2746$\pm$24 & 1154$\pm$358  \\ 
NGC5899     & 17.8$\pm$0.8   & 6.6$\pm$1.1    & 21.4$\pm$2.4     & 7.1$\pm$3.5     & 2853$\pm$76  &      --      \\
NGC6240     & 533.0$\pm$1.6  & 99.0$\pm$2.5   & 10926.4$\pm$88.5 & 992.5$\pm$34.2  & 2339$\pm$3 & 1247$\pm$74    \\ 
Mrk3        & 23.6$\pm$2.7   & 498.09.5$\pm$4.9  & 99.6$\pm$8.9  & 1167.2$\pm$28.3 & 4953$\pm$139 &      --      \\
Mrk348      & 17.6$\pm$0.6   & 27.3$\pm$0.9   & 42.6$\pm$6.1     & 39.4$\pm$4.7    & 2932$\pm$73  &      --      \\
Mrk607      & 7.1$\pm$0.4    & 11.3$\pm$0.7   & 12.0$\pm$1.2     & 13.1$\pm$1.5    & 4079$\pm$301 &      --      \\
Mrk1066     & 57.4$\pm$0.4   & 68.6$\pm$0.7   & 228.4$\pm$5.7    & 270.0$\pm$8.7   & 2614$\pm$19  &      --      \\
ESO578-G009 & 0.4$\pm$0.3    & 1.2$\pm$1.3    & 251.6$\pm$37.5   & 83.0$\pm$19.3   & 4223$\pm$298 & 757$\pm$654  \\
CygnusA     & 51.7$\pm$0.3   & 32.5$\pm$0.3   & 2163.2$\pm$63.4  & 1109.3$\pm$59.2 & 2777$\pm$32 & 1352$\pm$410  \\
{\bf Mean}  & 67.0$\pm$121.1 & 74.5$\pm$120.3 & 852.2$\pm$2494.1 & 261.7$\pm$393.1 & 3409$\pm$955 & 1096$\pm$302 \\
		\hline
		\multicolumn{7}{c}{type 1}\\
		\hline
NGC1275     & 1170.3$\pm$5.3 & 150.6$\pm$8.5  & 2327.5$\pm$56.7  & 229.8$\pm$20.5  & 2309$\pm$12 & 1021$\pm$351   \\
NGC3227     & 52.1$\pm$1.3   & 11.5$\pm$1.2   & 53.7$\pm$1.5     & 11.8$\pm$1.3    & 2906$\pm$27 & 1909$\pm$586 \\ 
NGC3516     & 8.7$\pm$2.2    & 2.0$\pm$0.3    & 16.2$\pm$2.7     & 2.1$\pm$1.9     & 3583$\pm$374 &      --      \\
NGC4051     & --               & --               & 7.4$\pm$0.3      & 4.7$\pm$0.6     & 3596$\pm$88 & 1354$\pm$876  \\
NGC4151     & 36.9$\pm$1.3   & 54.7$\pm$1.6   & 37.8$\pm$1.8     & 56.8$\pm$2.3    & 2495$\pm$45 & 1365$\pm$778   \\ 
NGC4235     & 4.0$\pm$0.5    & 0.2$\pm$0.1    & 13.7$\pm$1.2     & 1.9$\pm$0.5     & 5055$\pm$1425& --          \\
NGC4395     & --               & --               & 0.2$\pm$0.1      & 0.2$\pm$0.1     &  2866$\pm$70 & 2075$\pm$656  \\
NGC5548     & 33.3$\pm$5.5   & 32.1$\pm$6.8   & 79.7$\pm$28.5    & 71.2$\pm$32.8   & --           &      --     \\
NGC6814     & 7.8$\pm$0.8    & 1.4$\pm$0.4    & 8.6$\pm$1.4      & 1.4$\pm$0.6     & 3093$\pm$115 & 1811$\pm$812   \\
Mrk79       & 52.9$\pm$2.1   & 23.9$\pm$2.5   & 195.6$\pm$12.1   & 61.4$\pm$6.7    & 3392$\pm$95  &      --      \\
Mrk509      & 8.2$\pm$10.1   & 21.0$\pm$9.3   & 213.8$\pm$26.0   & 226.9$\pm$22.3  & 3539$\pm$320 &      --      \\
Mrk618      & 62.5$\pm$8.3   & 162.8$\pm$10.4 & 507.1$\pm$58.0   & 538.0$\pm$82.7  & 4009$\pm$291 & 752$\pm$703  \\ 
Mrk766      & 23.3$\pm$2.0   & 112.2$\pm$3.7  & 60.5$\pm$7.3     & 166.5$\pm$17.7  & 3278$\pm$164 &      --       \\
Mrk926      & 79.3$\pm$12.5  & 54.7$\pm$15.2  & 1340.8$\pm$106.5 & 507.2$\pm$53.0  & 3388$\pm$173 & 1498$\pm$932 \\ 
Mrk1044     & 0.1$\pm$0.1    & 53.5$\pm$4.9   & 11.5$\pm$3.1     & 81.7$\pm$14.1   & --           &      --       \\
Mrk1048     & 23.4$\pm$5.8   & 10.3$\pm$8.2   & 620.6$\pm$71.7   & 313.0$\pm$67.3  & 3022$\pm$145 & 875$\pm$517  \\
MCG+08-11-01& 117.6$\pm$18.5 & 138.2$\pm$19.7 & 1080.3$\pm$56.8  & 424.8$\pm$44.4  & 2560$\pm$44 & 1212$\pm$675  \\
{\bf Mean} & 99.3$\pm$269.6 & 49.0$\pm$54.8   & 386.8$\pm$622.2  & 158.8$\pm$179.5 & 3298$\pm$791 & 1254$\pm$570  \\
\hline
	\end{tabular}

\end{table*}

\subsection{Masses of hot molecular and ionised hydrogen}

We use the fluxes of the  H$_2\,$2.12\,$\mu$m and Br$\gamma$ emission lines to compute the mass of hot molecular ($M_{\rm H_2}$) and ionised ($M_{\rm H\,II}$) hydrogen.  The mass of hot molecular  gas ($M_{\rm H_2}$) can be estimated, under the assumptions of local thermal equilibrium and
excitation temperature of 2000\,K,  by \citep[e.g.][]{scoville82,rogemarN1068}:
 
\begin{equation}
 \left(\frac{M_{\rm H_2}}{M_\odot}\right)=5.0776\times10^{13}\left(\frac{F_{\rm H_{2}\,2.1218}}{\rm erg\,s^{-1}\,cm^{-2}}\right)\left(\frac{D}{\rm Mpc}\right)^2,
\label{mh2}
\end{equation}
where $F_{\rm H_{2}\,2.1218}$ is the H$_2\,2.1218\mu$m emission-line flux and $D$ is the distance to the galaxy.

 Following \citet{Osterbrock06} and \citet{sbN4151Exc}, we estimate the mass of ionised gas $M_{\rm H\,II}$  by

\begin{equation}
 \left(\frac{M_{\rm H\,II}}{M_\odot}\right)=3\times10^{19}\left(\frac{F_{\rm Br\gamma}}{\rm erg\,cm^{-2}\,s^{-1}}\right)\left(\frac{D}{\rm Mpc}\right)^2\left(\frac{N_e}{\rm cm^{-3}}\right)^{-1},
\label{mhii}
\end{equation}
where $F_{\rm Br\gamma}$ is the Br$\gamma$ flux (summing the fluxes of all spaxels) and $N_{e}$ is the electron density. Recent studies find that the electron densities in ionised outflows are underestimated using the [S\,{\sc ii}] line ratio
\citep{baron19,davies20}. However, most of the Br$\gamma$ emission in the inner kpc of nearby Seyfert galaxies originate from gas rotating in the plane of the disk \citep[e.g.][]{rogemar_sample,schonell19}, and thus we adopt $N_e=500$\,cm$^{-3}$, which is a typical value measured in AGN from the [S\,{\sc ii}]$\lambda\lambda$6717,6730 lines \citep[e.g.][]{dors14,dors20,brum17,Freitas18,kakkad18}. 
The distances to the galaxies adopted here are based on the galaxy redshift (Table~\ref{tab:sample}) for all targets.

\begin{figure}
\centering
\includegraphics[width=0.45\textwidth]{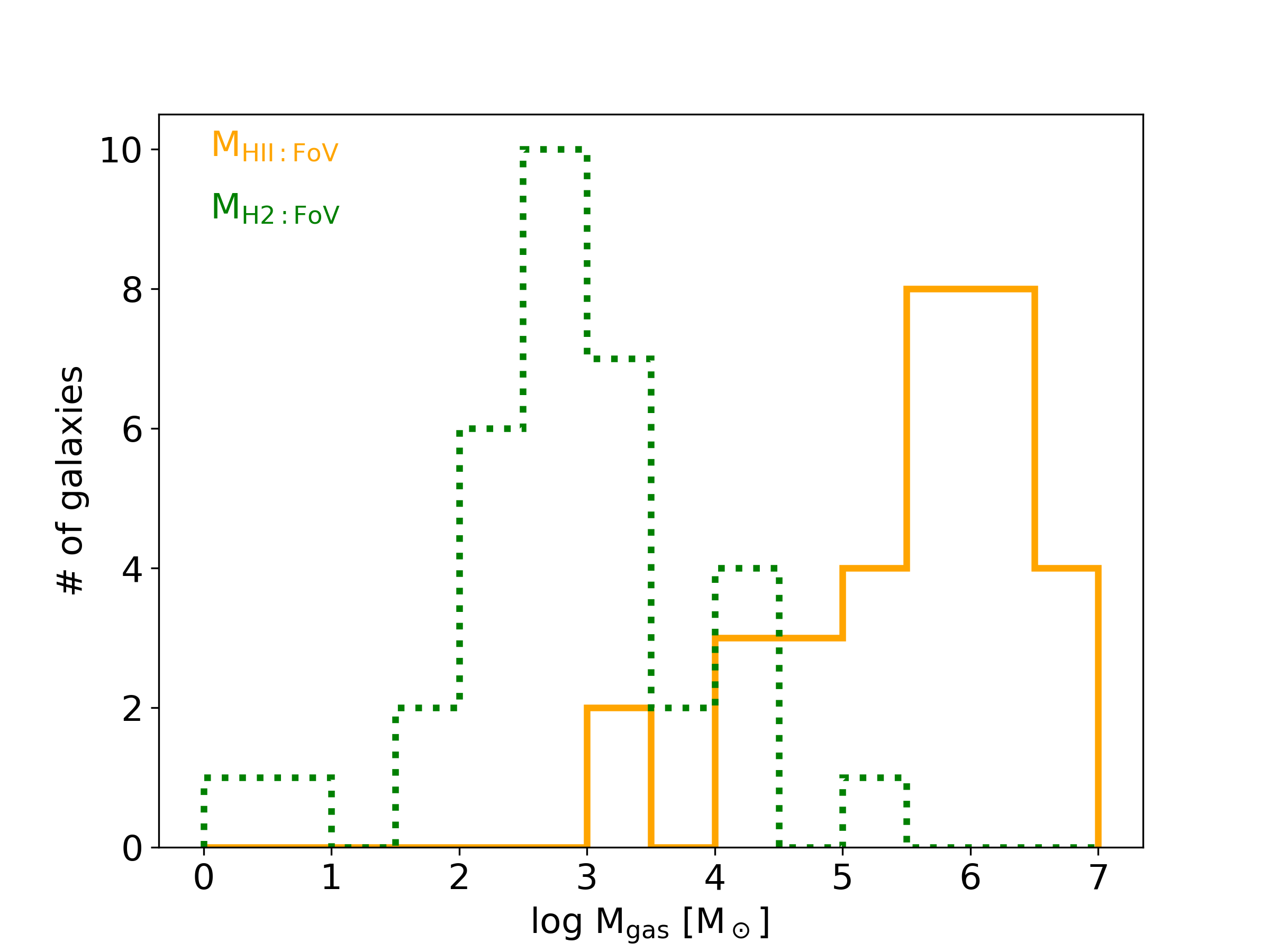}
\includegraphics[width=0.45\textwidth]{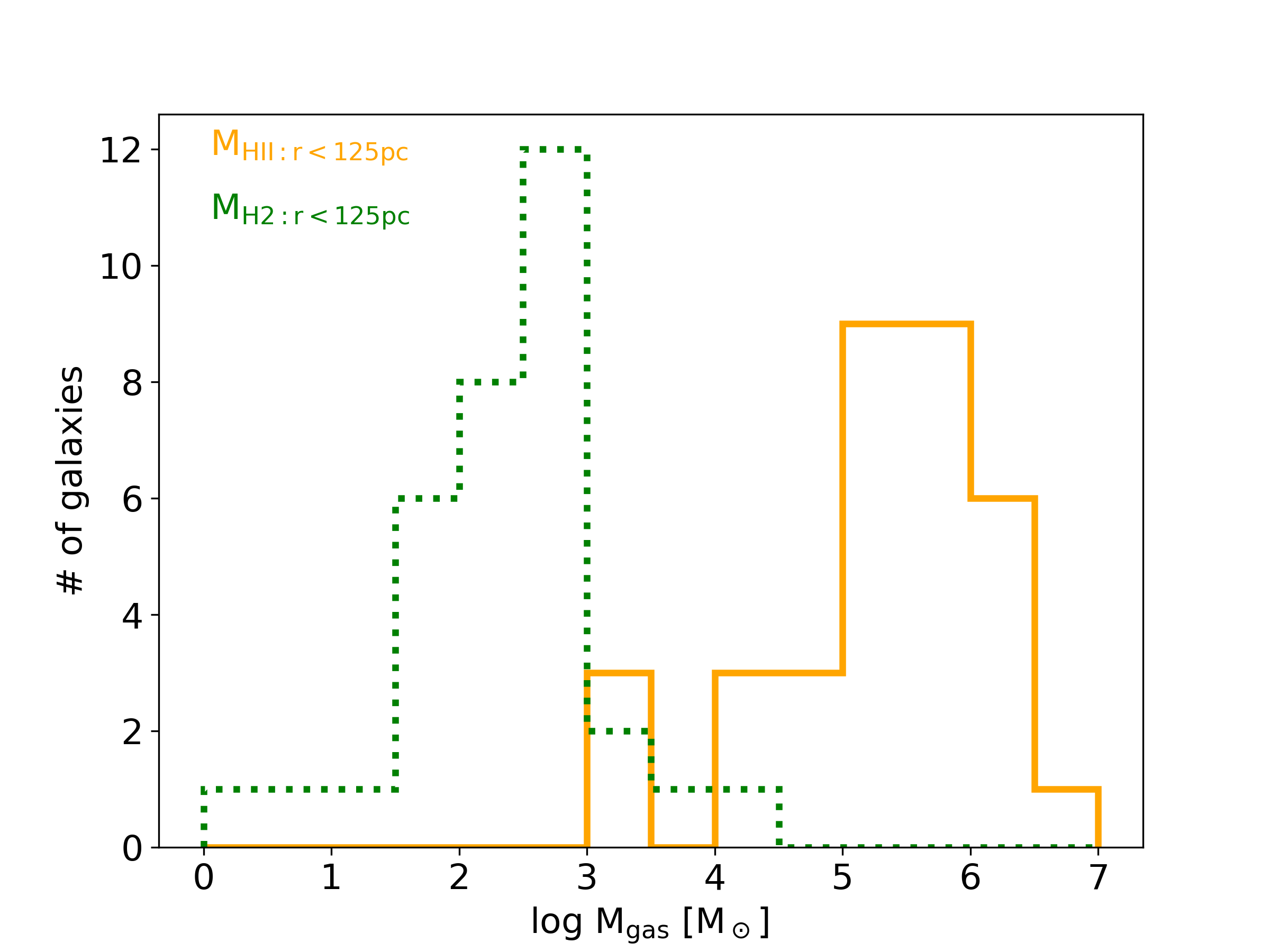}
\caption{\small Molecular (green) and ionised (orange) hydrogen mass distributions for the galaxies of our sample, computed using the whole FoV (top) and the inner 125 pc radius (bottom).}
\label{fig:mgas}
\end{figure}

Table~\ref{tab:masses} shows the mass of hot H$_2$ and ionised hydrogen for the galaxies of our sample computed in the whole NIFS FoV ($M_{\rm H_{2} FoV}$ and $M_{\rm H\,II FoV}$) and within the inner 125\,pc  ($M_{{\rm H_2}\, r<125 \rm pc}$ and $M_{{\rm H\,II}\, r< 125\rm pc}$). The corresponding distributions are shown in Fig.~\ref{fig:mgas}. For three galaxies (NGC\,4258, NGC\,4041 and NGC\,4395) the FoV is smaller than 250 pc and so, we do not calculate the mass within the inner 125\,pc radius. 
We do not find statistically significant differences between the masses of ionised and molecular gas of type 1 and type 2 AGN. 

\subsection{H$_2$ vibrational and rotational temperatures}

The H$_2$\,1--0\,S(1)$\,2.1218\,\mu$m/2-1\,S(1)$\,2.2477\,\mu$m and 1--0\,S(2)$\,2.0338\,\mu$m/1--0\,S(0)$\,2.2235\,\mu$m line ratios  can be used to estimate the vibrational and rotational temperatures of H$_2$. Using $N_i/g_i=F_i\lambda_i/(A_i g_i)={\rm exp}[-E_i/(k_B T_{\rm exc})]$, where $N_i$ are the column densities in the upper level, $g_i$ the statistical weights, $F_i$ and $\lambda_i$ are the line fluxes and wavelengths, $A_i$ are the transition probabilities, $E_i$ are the energies of the upper level, $T_{\rm exc}$ is the excitation temperature and $k_B$ is the Boltzmann constant. Using the transition probabilities from \citet{turner77}, the rotational temperature is given by:

\begin{equation}\label{eq:temp_rot}
T_{\rm rot} \cong -{\frac{ 1113}{\ln\left(0.323\frac{F_{\rm H_{2}\,2.0338}}{F_{\rm H_{2}\,2.2235}}\right)}}  
\end{equation}
and the vibrational temperature by 
\begin{equation}\label{eq:temp_vib}
T_{\rm vib}\cong {\frac{5594}{ \ln\left(1.355\frac{F_{\rm H_{2}\,2.1218}}{F_{\rm H_{2}\,2.2477}}\right)}}.
\end{equation}

We were able to estimate the H$_2$ vibrational temperature for 32 galaxies. For  NGC\,5548 and Mrk\,1044 at least one of the H$_2$ emission lines used to determine $T_{\rm vib}$ is not detected in our NIFS data with an amplitude larger than twice the standard deviation of the adjacent continuum. 
We estimate the H$_2$ rotational temperature for 21 galaxies. 11 galaxies of our sample were observed using the K$_{\rm long}$ grating (Tab.~\ref{tab:sample}), for which the spectral range does not include the H$_2$\,2.0338\,$\mu$m emission line, needed to estimate the H$_2$ rotational temperature. In addition, this line is not detected for NGC\,3516 and NGC\,4235. 
In  Figures~\ref{fig:temperarutesT2} and \ref{fig:temperarutesT1} we show the maps of $T_{\rm rot}$ and $T_{\rm vib}$ for the type 2 and type 1 AGN and in Figure~\ref{fig:temperarutesVib} we show the $T_{\rm vib}$ maps for the objects where the spectral range does not include the  1--0\,S(2) line.  Table~\ref{tab:masses} shows the median temperatures for each galaxy. The median values of the vibrational temperature are in the range from  $\sim$2100 to $\sim$4300 K, while the rotational temperatures are in the range $\sim$450--1900 K. There is no difference between the mean temperatures in type 1 and type 2 AGN.

\begin{figure*}
    \centering
\includegraphics[width=0.49\textwidth]{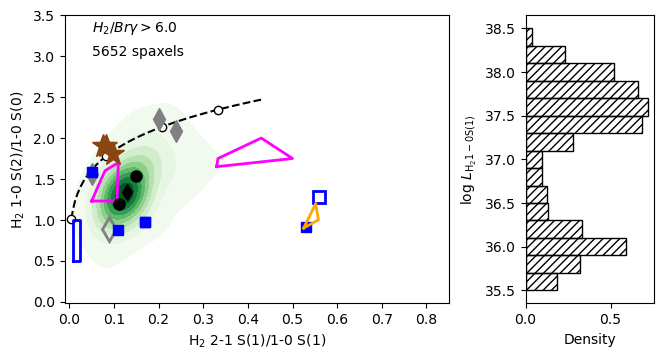} 
\includegraphics[width=0.49\textwidth]{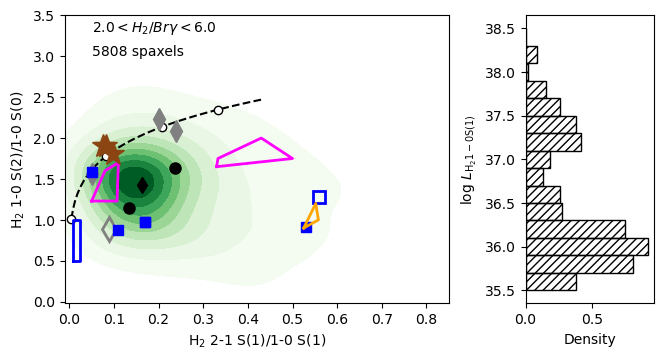} 
\includegraphics[width=0.49\textwidth]{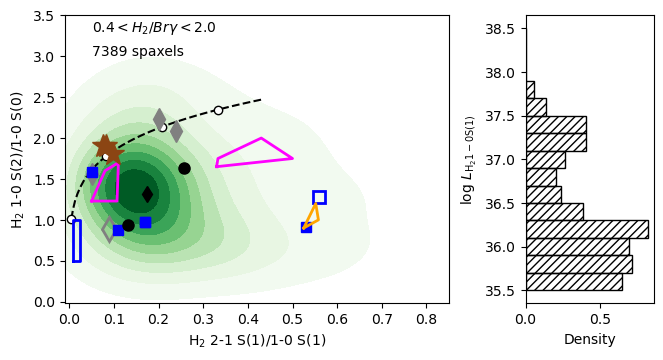} 
\includegraphics[width=0.49\textwidth]{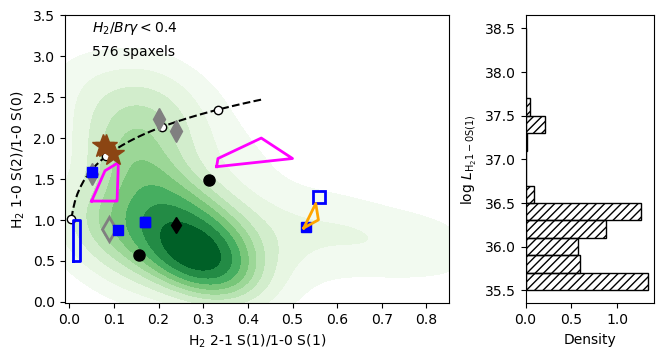}
\caption{\small Density plots for the H$_2$\,2--1 S(1)/1--0\,S(1) vs. 1--0\,S(2)/1--0\,S(0) diagnostic diagram for spaxels with  H$_2\, 2.1218\,\mu$m/Br$\gamma>6.0$ (top left), $2.0<$H$_2\, 2.1218\,\mu$m/Br$\gamma<6.0$ (top right),  $0.4<$H$_2\, 2.1218\,\mu$m/Br$\gamma<2.0$ (bottom left) and  H$_2\, 2.1218\,\mu$m/Br$\gamma<0.4$ (botom right). Black circles in all diagrams correspond to the 25th percentile and 75th percentile of the observed line ratio. Black diamonds represent the  median values.} The black dashed curve corresponds to the ratios for an isothermal and uniform density gas distribution for temperatures ranging from 1000 (left) to 4000 K (right) -- the open circles identify the ratios in steps of 1000 K.  The orange polygon represents the region occupied by the non-thermal UV excitation models of \citet{black87} and the pink polygons cover the region of the photoionisation models of \citet{Dors12}. The blue rectangle covers the locus of the thermal UV excitation models of \citet{Sternberg89}, as computed by \citet{mouri94} for gas densities ($n_t$) of 10$^5$ and 10$^6$ cm$^{-3}$ and UV scaling factors relative to the local interstellar radiation field $\chi$ from 10$^2$ to 10$^{4}$. The filled blue squares are the UV thermal models from \citet{davies03} for $10^3<n_t<10^6$ cm$^{-3}$ and $10^2<\chi<10^5$, while the open blue square is from \citet{Sternberg89} for $n_t=10^3$ cm$^{-3}$ and $\chi=10^2$. The brown stars are from the thermal X-ray models of \citet{draine90}, the gray open diamond is from the shock model of \citet{Kwan77} and the gray filled diamonds represent the shock models from \citet{Smith95}. The  AGN photoionisation from \citet{rogerio13} span a wide range in both axes ($0\lesssim$H$_2$ 2--1 S(1)/1--0 S(1)$\lesssim0.6$  and  $0.5\lesssim$1--0 S(2)/1--0 S(0)$\lesssim2.5$) and therefore is not shown in the figure. The right panels show the  H$_2\,2.1218\,\mu$m luminosity distributions for the spaxels used in the left panels, in logarithmic units of erg\,s$^{-1}$.
    \label{fig:diagrams}
\end{figure*}

\section{Discussion}\label{sec:disc}
\subsection{Type 1 vs. type 2 AGN}

According to the AGN Unification Model \citep{antonucci93,urry95}, type 1 and type 2 AGN represent the same class of objects seen
at distinct viewing angles. However, some recent results suggest these two classes may be intrinsically distinct objects \citep[e.g., ][]{elitzur12,elitzur14,villarroel14,prieto14,audibert17}, the tori of type 2 AGN having on average smaller opening angles and more clouds along the line of sight than for type 1 \citep{ramos-almeida11}.

Our sample presents the same number of type 1 and type 2 AGN and they follow similar luminosity and redshift distributions (Fig.~\ref{fig:histograms}), thus we can compare both AGN types in terms of the emission-line properties. We find the type 1 and type 2 AGN of our sample present similar H$_2$/Br$\gamma$ line ratios, ionised and hot molecular gas masses and H$_2$ excitation temperature distributions. They also present no difference in terms of the $\Delta$PA between the Br$\gamma$ and H$_2$ flux distributions and between the orientation of the major axis of the hosts and the emission-line flux distributions. These results support the unification scenario.

We find a statistically significant difference between type 1 and type 2 AGN only for the emission line equivalent widths at the nucleus (Fig.~\ref{fig:eqw}). Seyfert 1 nuclei show smaller equivalent widths than Seyfert 2 nuclei.  The K-band  continuum of AGN can present a strong contribution of hot dust emission from the inner border of the torus  \citep[e.g.][]{rogemar_N4151_dust,Dumont20}. This component is present in about 90 per cent of Seyfert 1 nuclei, while only about 25 per cent of Seyfert 2 nuclei present this component \citep{rogerio09}.  Thus, the smaller values of equivalent widths we observe in type 1 AGN may be due to stronger contributions of hot dust emission to the underlying continuum than those in type 2 AGN.  Similar results are found for CO absorption-band heads at $\sim$2.3\,$\mu$m, which can be diluted by the hot dust emission  \citep{rogerio09,burtscher15,ms18}. 
We find a decrease in the nuclear $EqW_{\rm Br\gamma}$ in 88\% (15 objects) of Type 1 AGN, and in 47\% (8 objects) of the type 2 AGN, as compared to the  extra-nuclear $EqW_{\rm Br\gamma}$ values. The mean values of the ratio between the mean value of the nuclear $EqW_{\rm Br\gamma}$ (computed using an aperture with radius equal to the angular resolution of the data for each galaxy) and the mean extra-nuclear $EqW_{\rm Br\gamma}$ (computed using spaxels at distances larger than the angular resolution) is 1.35$\pm$0.27 and 0.58$\pm$0.14 for the type 2 and type 1 AGN in our sample, respectively. Considering only the galaxies that show a nuclear decrease in $EqW_{\rm Br\gamma}$, the mean ratios between the nuclear and extra-nuclear $EqW_{\rm Br\gamma}$ are 0.52$\pm$0.11 and 0.39$\pm$0.07 for type 2 and type 1 AGN, respectively.

The difference in the equivalent widths in type 1 and type 2 AGN can be reconciled with the AGN unification model, as in type 1 AGN we observe directly the contribution of the inner and hotter region of the dusty torus, while in type 2 AGN this structure is not visible in most objects.
Smaller $EqW_{\rm Br\gamma}$ values at the nucleus compared to extra-nuclear regions could also be produced if the latter includes a larger contribution of young stellar populations than the former. Our results can be compared to those of  \citet{burtscher15}, who investigated the dilution of the K-band CO\,2.3\,$\mu$m absorption feature in a sample of nearby AGN hosts. They computed the intrinsic CO equivalent width (not affected by dilution) and found a wide range of values (from 6 to 14\, \AA), probably due to differences in the ages of the stellar populations in their sample. Our $EqW_{\rm Br\gamma}$ maps (Figs.\,\ref{fig:mapsS2},\ref{fig:mapsS1},\ref{fig:mapsS2ap} and \ref{fig:mapsS1ap}) also show a wide range of values, which may be due to different stellar populations in the host galaxies. However, \citet{burtscher15} found no difference in the  intrinsic CO equivalent widths of diluted and undiluted sources (see their Fig. 3), which provides a additional support that the nuclear decrease in $EqW_{\rm Br\gamma}$ in our sample is due to dilution of the line by the AGN continuum.

\subsection{The origin of the H$_2$ emission}

The origin of the H$_2$ near-IR emission lines in active galaxies has been investigated in several theoretical and observational studies \citep[e.g][]{black87,hollenbach89,draine90,maloney96,ardila04,ardila05,lamperti17}, but it is still not clear which is the main excitation mechanism -- if there is a dominant -- of the H$_2$ molecule.  In summary, three main processes can produce the H$_2$ emission: (i) fluorescent excitation by the absorption of soft-UV photons (912--1108 \AA) in the Lyman and Werner bands \citep{black87}, (ii) excitation by shocks \citep{hollenbach89} and (iii) excitation by X-ray illumination \citep{maloney96}. The first process is usually referred as non-thermal, while the latter two are commonly reported as thermal processes, where the H$_2$ emitting gas is in local thermodynamic equilibrium (LTE). In some cases, thermal and non-thermal processes are observed simultaneously with the H$_2$\,1--0 transitions in LTE, while the higher energy ones (H$_2$\,2--1 and H$_2$\,3--2) being due to fluorescent excitation of the dense gas \citep{davies03, davies05}.

In thermal processes, the rotational and vibrational temperatures are similar, as the gas is in LTE, while for fluorescent excitation the vibrational temperature is high ($T_{\rm vib}\sim$5000 K) and the rotational temperature is about a tenth of $T_{\rm vib}$, as the highest energy levels are overpopulated due to non-local UV photons compared to the prediction for a Maxwell-Boltzmann population \citep{Sternberg89,ardila04}. As shown in Table~\ref{tab:masses}, for all galaxies, but NGC\,788, the median values of the rotational temperature are larger than 10 percent of the vibrational temperatures. On average, we find $\langle T_{\rm rot}/T_{\rm vib}\rangle=0.48\pm0.31$. This value is consistent with measurements based on single aperture spectra of AGN  \citep[e.g.][]{ardila05,rogerio13}. 

The H$_2$/Br$\gamma$ line ratio is useful in the investigation of the origin of the H$_2$ emission. In star-forming regions and starburst galaxies, this ratio is usually smaller than 0.4, while AGN present  $0.4<\rm H_2/Br\gamma<6.0$ and larger values are usually associated to shocks \citep{rogerio13}. We find the median values of H$_2$/Br$\gamma$ in our sample are within the range observed in AGN, but there is a trend of higher ratios being observed at larger distances from the nucleus for most galaxies, as seen in Tab.~\ref{tab:h2br} and Figs.~\ref{fig:mapsS2}, \ref{fig:mapsS1}, \ref{fig:mapsS2ap} and \ref{fig:mapsS1ap}.   
Since the H$_2$ line intensity generally decreases with distance from the nucleus, an explanation for the higher values of the H$_2$/Br$\gamma$ away from the nucleus is that the Br$\gamma$ decreases faster with radius, i.e. it is enhanced very close around the AGN (excited primarily by the AGN) while the H$_2$ is excited by processes that operate on more extended spatial scales such as shocks.

In order to further investigate the H$_2$ emission origin, we construct the H$_2$ 2--1 S(1)/1--0 S(1) vs. 1--0 S(2)/1--0 S(0) diagnostic diagrams shown in Figure~\ref{fig:diagrams}.  The density plots show the observed line ratios for all spaxels where we were able to detect all H$_2$ emission lines. We separate the data into four diagrams, according to the H$_2$/Br$\gamma$ presented in the excitation maps (see Figs.~\ref{fig:mapsS2}, \ref{fig:mapsS1}, \ref{fig:mapsS2ap}, \ref{fig:mapsS1ap}): H$_2 > 6.0$ -- indicative of shocks (top left panel),
$2.0<\rm H_2/Br\gamma<6.0$ -- high excitation AGN (top right panel), $0.4<\rm H_2/Br\gamma<2.0$ -- low excitation AGN (bottom left panel) and H$_2$/Br$\gamma<0.4$ -- typical value of starbursts (bottom right panel). The number of points in each plot is shown in  the top-left corner of the corresponding panel. 

The observed H$_2$ 2--1 S(1)/1--0\,S(1) and 1--0 S(2)/1--0\,S(0) line ratio distributions for H$_2$/Br$\gamma>0.4$ lie close to the region predicted by photoionisation (pink polygons) and shock models (gray symbols) suggesting thermal processes dominate the H$_2$ excitation. 
This result is in good agreement with previous works using single aperture measurements of the H$_2$ line fluxes \citep[e.g.][]{ardila04,ardila05,rogerio13}.  Here, we show not only the nuclear emission originates from thermal processes but that also the emission from locations furthest from the nucleus, as the vast majority (96 per cent) of the spaxels in our sample present H$_2$/Br$\gamma>$0.4.

Although the peak of the distributions of H$_2$\,2--1\,S(1)/1--0\,S(1) and 1--0\,S(2)/1--0\,S(0) is observed nearly at the same location for both spaxels with $0.4<\rm H_2/Br\gamma<2.0$, $2.0<\rm H_2/Br\gamma<6.0$ and H$_2$/Br$\gamma>6.0$, the distributions are distinct. For the $0.4<\rm H_2/Br\gamma<2.0$ (low excitation AGN) and $2.0<\rm H_2/Br\gamma<6.0$ (high excitation AGN) the luminosity distribution of the H$_2$2.1218$\mu$m is very similar and both H$_2$\,2--1\,S(1)/1--0\,S(1) and 1--0\,S(2)/1--0\,S(0) spread over a larger region in the diagnostic diagram. The peak in the diagnostic diagrams lie close to both the photoionisation model (pink polygons) and to the UV thermal models (filled blue squares) indicating that heating due to the AGN may be the cause of the gas excitation. Also, the fact that the points on the diagram are not distributed along the isothermal line indicates that the gas is not in LTE. For the typical AGN line ratio, $0.4<\rm H_2/Br\gamma<6.0$, the H$_2$ emission can either be produced by shocks (gray symbols) or by the AGN radiation field. 

The H$_2$/Br$\gamma>6.0$ shows a more concentrated distribution in the diagnostic diagram and is particularly elongated towards higher values of 1--0\,S(2)/1--0\,S(0). The H$_2$2.1218$\mu$m distribution is concentrated towards higher values, indicating the higher ratios is due to the higher H$_2$ luminosity. 
This line ratio is more sensitive to shocks, as seen from the wider range of values predicted by distinct shock models for this ratio than for the 2--1\,S(1)/1--0\,S(1) \citep{Kwan77,Smith95}. Thus, the H$_2$ emission from locations with  H$_2$/Br$\gamma>6.0$ likely originates from heating of the gas by shocks, as those produced by AGN winds \citep[e.g.][]{rogemarN1275}.

For H$_2$/Br$\gamma<0.4$, typical of star-forming regions, the median values of 2--1\,S(1)/1--0\,S(1) and  1--0\,S(2)/1--0\,S(0) -- 0.39 and 1.09, respectively -- are close to the predicted values by the models of \citet{black87} (orange polygon) for fluorescent excitation. The peak of the observed distribution of ratios is shifted to slightly lower values of 2--1 S(1)/1--0 S(1) than predicted by the models, but a large scatter is seen in both axes. This shift can be understood as a contamination of thermal excitation plus the dissociation of part of the H$_2$ molecules by the AGN radiation field, as seen in some objects \citep{sbN4151Exc,rogemarMrk1066exc,Gnilka20}.

\subsection{Mass reservoir in the central region and AGN feeding}

\begin{figure}
\centering
\includegraphics[width=0.45\textwidth]{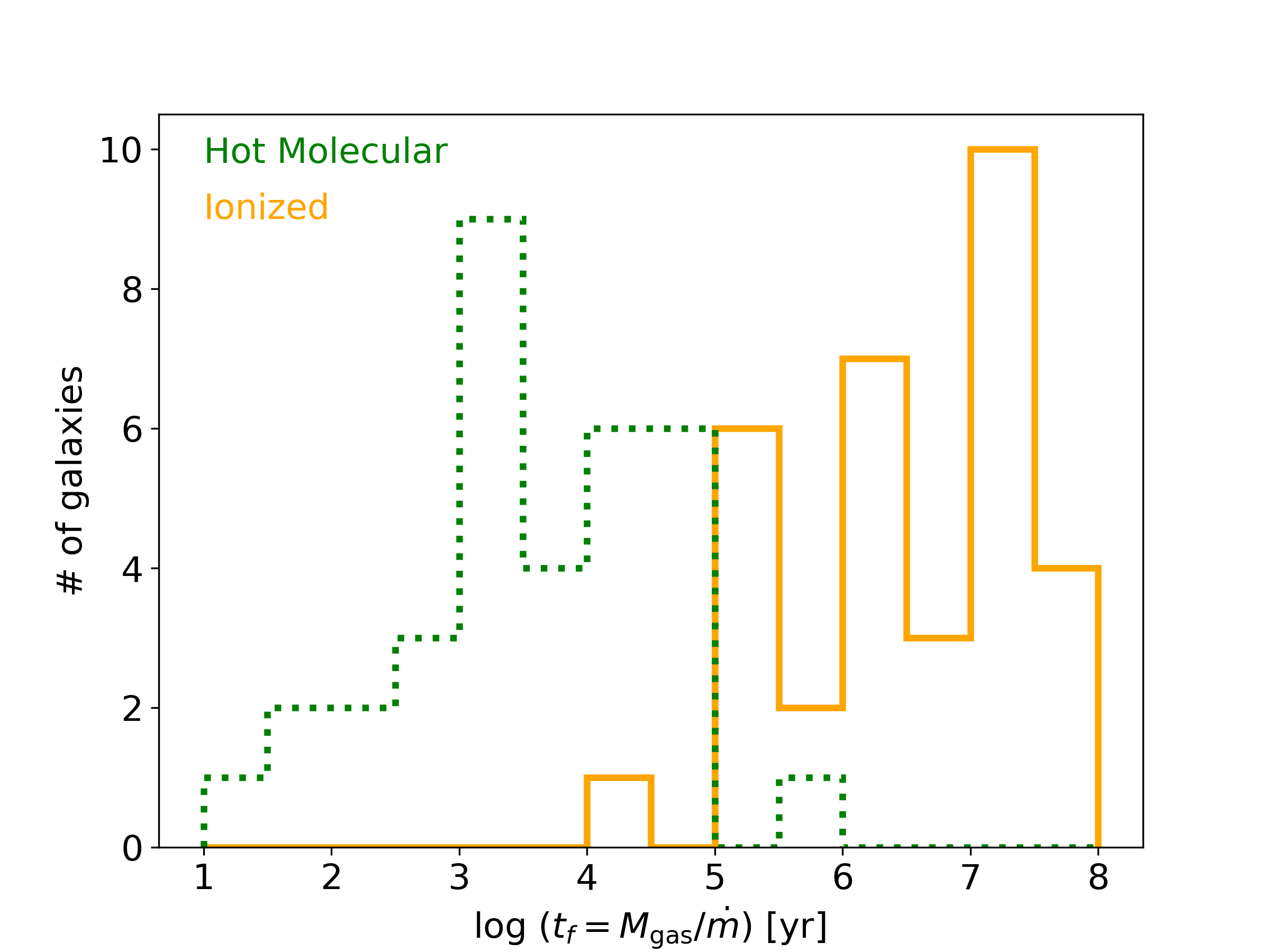}
\caption{\small Distribution of the AGN feeding time with the available masses of hot molecular (green) and ionised (orange) hydrogen mass in the inner 125\,pc radius.}
\label{fig:tf}
\end{figure}

The galaxies of our sample present masses of hot molecular gas ranging from a few tens of solar masses to 10$^4$ M$_\odot$ and of ionised gas in the range $\sim10^3-10^7$ M$_\odot$ within the inner 125\,pc radius (Tab.~\ref{tab:masses}). We can compare these masses with the mass accretion rate ($\dot{m}$) necessary to power the AGN, which is given by: 

\begin{equation}
 \dot{m}=\frac{L_{\rm bol}}{c^2\eta},
\end{equation}
where $c$ is the light speed, $\eta$ is the efficiency of conversion of the rest mass energy of the accreted material into radiation assumed to be 0.1 \citep{frank02} and  $L_{\rm bol}$ is the AGN bolometric luminosity, which can be estimated from the hard X-ray (14-195 keV) luminosities ($L_{\rm X}$) listed in Tab.~\ref{tab:sample}, by \citet{ichikawa17}: 
\begin{equation}
\log{L_{\rm bol}} = 0.0378(\log{L_{\rm X}})^2 - 2.03\log{L_{\rm X}}+61.6.
\end{equation}
 Figure~\ref{fig:tf} shows the feeding time ($t_f={M_{\rm gas}/\dot{m}}$) distributions for the galaxies computed using the hot molecular and ionised gas masses calculated in the previous sections within the inner 125\,pc. We find the mass of ionised gas alone can feed the central AGN for 10$^5$--10$^8$\,yr at the current accretion rates. The mass of hot molecular gas is on average 3 orders of magnitude smaller than that of ionised gas. We emphasize that the feeding times estimated above should be treated as an upper limit, as they are based on the total line fluxes and we do not separate the components associated to non-circular motions (e.g. due to inflows and outflows).

The hot molecular and ionised gas masses represent only the ``tip of the iceberg'' of the the total gas mass in the center of galaxies. \citet{dempsey18} find the mass of the NLR, measured from emission-line fluxes of the ionised gas, is underestimated because the gas behind the ionisation front is invisible in ionised transitions. This gas may be in the molecular phase, dominated by cold molecular gas, and we now know that the mass of cold molecular gas correlates with the H$_2$\,2.1218\,$\mu$m luminosity -- and thus with the hot H$_2$ mass \citep{dale05,ms06,mazzalay13}. These studies found that the amount of cold molecular gas is 10$^5$--10$^8$ times that of hot H$_2$. 
Thus, the cold molecular gas could provide fuel necessary to power the AGN for an activity cycle of 10$^5$ to 10$^6$ yr \citep[e.g.][]{novak11} and still remain available 
to form new stars in the nuclear region.

\section{Conclusions}\label{sec:conc}

We have used Gemini NIFS K-band observations to map the H$_2$ and Br$\gamma$ emission distribution within the inner  0.04--2\,kpc of a sample of 36 active galaxies with $0.001\lesssim z\lesssim0.056$ and hard X-ray luminosities in the range $41\lesssim \log L_{\rm X}/({\rm erg\, s^{-1}}) \lesssim45$.  The spatial resolutions at the galaxies range from 6 to 250\,pc and the field of view covers from the inner 75$\times$75~pc$^2$ to 3.6$\times$3.6~kpc$^2$. The main conclusions of this work are:

\begin{itemize}
    \item Extended H$_2\,2.1218\,\mu$m and Br$\gamma$ emission is observed in 34/36 galaxies of our sample.  There is no statistically significant difference between the orientation of the H$_2$ and Br$\gamma$ flux distributions relative to the orientation of the major axis of the host galaxy ($\Delta$PA).  We find $\Delta$PA larger than 30$^\circ$ in 15 galaxies (42\,\%) for the Br$\gamma$ and in 16 galaxies (44\,\%) for H$_2$. 
    
    \item The H$_2$ emission is usually more spread over the field of view, while the Br$\gamma$ shows a more collimated flux distribution in most cases and a steeper flux gradient, decreasing with the distance from the nucleus. We find offsets larger than 30$^\circ$ between the orientations of the H$_2\,$2.1218$\mu$m and Br$\gamma$ flux distributions in 45\,\% of our sample.  On average, the radius that contains 50\,per cent of the total H$_2$ emission is $\sim$60\,per cent larger than that for Br$\gamma$.

    \item  We derive the H$_2$ rotational and vibrational temperatures based on the observed  H$_2$ 1-0\,S(1)$\,2.1218\,\mu$m/2-1\,S(1)$\,2.2477\,\mu$m and 1-0\,S(2)$\,2.0338\,\mu$m/1-0S(0)$\,2.2235\,\mu$m  line ratios. The median values are in the ranges 2100--4300 K and 450--1900 K for the vibrational and rotational temperatures, respectively, with the mean ratio between the two of 0.43$\pm$0.15, indicating dominant thermal excitation for the gas.
    
     \item Type 1 and type 2 AGN show similar emission-line flux distributions, ratios, H$_2$ excitation temperatures and gas masses, supporting the AGN unification scenario.
    
     \item Type 1 and 2 AGN differ only in their nuclear Br$\gamma$ equivalent widths, which are smaller in type 1 AGN due to larger contributions of hot dust emission to the K-band continuum in type 1 nuclei.     
    
    \item The distribution of points in the H$_2$ 2--1 S(1)/1--0 S(1) vs. 1--0 S(2)/1--0 S(0) diagram in regions with H$_2$/Br$\gamma>0.4$  (96\,\% of all spaxels with flux measurements)  are consistent with predictions of photoionisation and shock models, confirming that the main excitation mechanism of the H$_2$ molecule are thermal processes, not only at the nucleus but also in the extranuclear regions.
    
    \item The gas thermal excitation usually increases outwards with H$_2$/Br$\gamma$ values increasing from $<2$ in the nucleus to values up to 6 outwards. 
    
    \item In locations with H$_2$/Br$\gamma>$6.0, the most likely H$_2$ excitation mechanism are shocks, as indicated by the H$_2$ 2--1 S(1)/1--0 S(1) and 1--0 S(2)/1--0 S(0) line ratios. This is observed mostly in locations away from the nucleus, for $\sim$40 per cent of the galaxies.
    
    \item Most of the regions with H$_2$/Br$\gamma<0.4$ (4\,\% of the spaxels with flux measurements) are consistent with fluorescent excitation of the H$_2$, but dissociation of the H$_2$ molecule by the AGN radiation cannot be ruled out in galaxies with small  H$_2$/Br$\gamma$ nuclear values. This is observed in $\sim$25 per cent of the sample. 
    
    \item  The mass of hot molecular and ionised gas in the inner 125\,pc radius are in the ranges $10^1-10^4$ M$_\odot$ and  $10^4-10^6$ M$_\odot$, respectively. The masses computed for the whole NIFS field of view are about one order of magnitude larger.
    
    \item The mass of ionised gas within the inner 125\,pc radius alone is more than enough to power the AGN in our sample for a duty cycle of $10^6$ yr at their current accretion rates. 
    
\end{itemize}

\section*{Acknowledgements}
RAR acknowledges financial support from Conselho Nacional de Desenvolvimento Cient\'ifico e Tecnol\'ogico (CNPq -- 202582/2018-3, 304927/2017-1,  400352/2016-8 and 312036/2019-1) and Funda\c c\~ao de Amparo \`a pesquisa do Estado do Rio Grande do Sul (FAPERGS -- 17/2551-0001144-9 and 16/2551-0000251-7). RR thanks CNPq, CAPES and FAPERGS for financial support.  MB thanks the financial support from Coordenação de Aperfeiçoamento de Pessoal de Nível Superior - Brasil (CAPES) - Finance Code 001. NZD acknowledges partial support from FONDECYT through project 3190769.  Based on observations obtained at the Gemini Observatory, which is operated by the Association of Universities for Research in Astronomy, Inc., under a cooperative agreement with the NSF on behalf of the Gemini partnership: the National Science Foundation (United States), National Research Council (Canada), CONICYT (Chile), Ministerio de Ciencia, Tecnolog\'{i}a e Innovaci\'{o}n Productiva (Argentina), Minist\'{e}rio da Ci\^{e}ncia, Tecnologia e Inova\c{c}\~{a}o (Brazil), and Korea Astronomy and Space Science Institute (Republic of Korea).  This research has made use of NASA's Astrophysics Data System Bibliographic Services. This research has made use of the NASA/IPAC Extragalactic Database (NED), which is operated by the Jet Propulsion Laboratory, California Institute of Technology, under contract with the National Aeronautics and Space Administration.

\section*{Data availability}
Most of the data used in this paper is available in the Gemini Science Archive at https://archive.gemini.edu/searchform with the project codes listed in Table~\ref{tab:sample}. Processed datacubes used will be shared on reasonable request to the corresponding author.




\bibliographystyle{mnras}
\bibliography{paper} 

\begin{thebibliography}{}
\makeatletter
\relax
\def\mn@urlcharsother{\let\do\@makeother \do\$\do\&\do\#\do\^\do\_\do\%\do\~}
\def\mn@doi{\begingroup\mn@urlcharsother \@ifnextchar [ {\mn@doi@}
  {\mn@doi@[]}}
\def\mn@doi@[#1]#2{\def\@tempa{#1}\ifx\@tempa\@empty \href
  {http://dx.doi.org/#2} {doi:#2}\else \href {http://dx.doi.org/#2} {#1}\fi
  \endgroup}
\def\mn@eprint#1#2{\mn@eprint@#1:#2::\@nil}
\def\mn@eprint@arXiv#1{\href {http://arxiv.org/abs/#1} {{\tt arXiv:#1}}}
\def\mn@eprint@dblp#1{\href {http://dblp.uni-trier.de/rec/bibtex/#1.xml}
  {dblp:#1}}
\def\mn@eprint@#1:#2:#3:#4\@nil{\def\@tempa {#1}\def\@tempb {#2}\def\@tempc
  {#3}\ifx \@tempc \@empty \let \@tempc \@tempb \let \@tempb \@tempa \fi \ifx
  \@tempb \@empty \def\@tempb {arXiv}\fi \@ifundefined
  {mn@eprint@\@tempb}{\@tempb:\@tempc}{\expandafter \expandafter \csname
  mn@eprint@\@tempb\endcsname \expandafter{\@tempc}}}

\bibitem[\protect\citeauthoryear{{Antonucci}}{{Antonucci}}{1993}]{antonucci93}
{Antonucci} R.,  1993, \mn@doi [\araa] {10.1146/annurev.aa.31.090193.002353},
  \href {https://ui.adsabs.harvard.edu/abs/1993ARA&A..31..473A} {31, 473}

\bibitem[\protect\citeauthoryear{{Audibert}, {Riffel}, {Sales}, {Pastoriza}  \&
  {Ruschel-Dutra}}{{Audibert} et~al.}{2017}]{audibert17}
{Audibert} A.,  {Riffel} R.,  {Sales} D.~A.,  {Pastoriza} M.~G.,
  {Ruschel-Dutra} D.,  2017, \mn@doi [\mnras] {10.1093/mnras/stw2477}, \href
  {https://ui.adsabs.harvard.edu/abs/2017MNRAS.464.2139A} {464, 2139}

\bibitem[\protect\citeauthoryear{{Barbosa}, {Storchi-Bergmann}, {McGregor},
  {Vale}  \& {Rogemar Riffel}}{{Barbosa} et~al.}{2014}]{barbosa14}
{Barbosa} F.~K.~B.,  {Storchi-Bergmann} T.,  {McGregor} P.,  {Vale} T.~B.,
  {Rogemar Riffel} A.,  2014, \mn@doi [\mnras] {10.1093/mnras/stu1637}, \href
  {https://ui.adsabs.harvard.edu/abs/2014MNRAS.445.2353B} {445, 2353}

\bibitem[\protect\citeauthoryear{{Baron} \& {Netzer}}{{Baron} \&
  {Netzer}}{2019}]{baron19}
{Baron} D.,  {Netzer} H.,  2019, \mn@doi [\mnras] {10.1093/mnras/stz1070},
  \href {https://ui.adsabs.harvard.edu/abs/2019MNRAS.486.4290B} {486, 4290}

\bibitem[\protect\citeauthoryear{{Barth}, {Filippenko}  \& {Moran}}{{Barth}
  et~al.}{1999}]{barth99}
{Barth} A.~J.,  {Filippenko} A.~V.,   {Moran} E.~C.,  1999, \mn@doi [\apjl]
  {10.1086/311976}, \href
  {https://ui.adsabs.harvard.edu/abs/1999ApJ...515L..61B} {515, L61}

\bibitem[\protect\citeauthoryear{{Black} \& {van Dishoeck}}{{Black} \& {van
  Dishoeck}}{1987}]{black87}
{Black} J.~H.,  {van Dishoeck} E.~F.,  1987, \mn@doi [\apj] {10.1086/165740},
  \href {https://ui.adsabs.harvard.edu/abs/1987ApJ...322..412B} {322, 412}

\bibitem[\protect\citeauthoryear{{Brum}, {Riffel}, {Storchi-Bergmann},
  {Robinson}, {Schnorr M{\"u}ller}  \& {Lena}}{{Brum} et~al.}{2017}]{brum17}
{Brum} C.,  {Riffel} R.~A.,  {Storchi-Bergmann} T.,  {Robinson} A.,  {Schnorr
  M{\"u}ller} A.,   {Lena} D.,  2017, \mn@doi [\mnras] {10.1093/mnras/stx964},
  \href {https://ui.adsabs.harvard.edu/abs/2017MNRAS.469.3405B} {469, 3405}

\bibitem[\protect\citeauthoryear{{Brum} et~al.,}{{Brum} et~al.}{2019}]{Brum19}
{Brum} C.,  et~al., 2019, \mn@doi [\mnras] {10.1093/mnras/stz893}, \href
  {https://ui.adsabs.harvard.edu/abs/2019MNRAS.486..691B} {486, 691}

\bibitem[\protect\citeauthoryear{{Burtscher} et~al.,}{{Burtscher}
  et~al.}{2015}]{burtscher15}
{Burtscher} L.,  et~al., 2015, \mn@doi [\aap] {10.1051/0004-6361/201525817},
  \href {https://ui.adsabs.harvard.edu/abs/2015A&A...578A..47B} {578, A47}

\bibitem[\protect\citeauthoryear{{Caglar} et~al.,}{{Caglar}
  et~al.}{2020}]{Caglar20}
{Caglar} T.,  et~al., 2020, \mn@doi [\aap] {10.1051/0004-6361/201936321}, \href
  {https://ui.adsabs.harvard.edu/abs/2020A&A...634A.114C} {634, A114}

\bibitem[\protect\citeauthoryear{{Colina} et~al.,}{{Colina}
  et~al.}{2015}]{colina15}
{Colina} L.,  et~al., 2015, \mn@doi [\aap] {10.1051/0004-6361/201425567}, \href
  {https://ui.adsabs.harvard.edu/abs/2015A&A...578A..48C} {578, A48}

\bibitem[\protect\citeauthoryear{{Daddi} et~al.,}{{Daddi}
  et~al.}{2010}]{Daddi+10}
{Daddi} E.,  et~al., 2010, \mn@doi [\apj] {10.1088/0004-637X/713/1/686}, \href
  {https://ui.adsabs.harvard.edu/abs/2010ApJ...713..686D} {713, 686}

\bibitem[\protect\citeauthoryear{{Dahmer-Hahn} et~al.,}{{Dahmer-Hahn}
  et~al.}{2019a}]{dahmer-hahn19}
{Dahmer-Hahn} L.~G.,  et~al., 2019a, \mn@doi [\mnras] {10.1093/mnras/sty3051},
  \href {https://ui.adsabs.harvard.edu/abs/2019MNRAS.482.5211D} {482, 5211}

\bibitem[\protect\citeauthoryear{{Dahmer-Hahn} et~al.,}{{Dahmer-Hahn}
  et~al.}{2019b}]{dahmer19b}
{Dahmer-Hahn} L.~G.,  et~al., 2019b, \mn@doi [\mnras] {10.1093/mnras/stz2453},
  \href {https://ui.adsabs.harvard.edu/abs/2019MNRAS.489.5653D} {489, 5653}

\bibitem[\protect\citeauthoryear{{Dale}, {Sheth}, {Helou}, {Regan}  \&
  {H{\"u}ttemeister}}{{Dale} et~al.}{2005}]{dale05}
{Dale} D.~A.,  {Sheth} K.,  {Helou} G.,  {Regan} M.~W.,   {H{\"u}ttemeister}
  S.,  2005, \mn@doi [\aj] {10.1086/429134}, \href
  {https://ui.adsabs.harvard.edu/abs/2005AJ....129.2197D} {129, 2197}

\bibitem[\protect\citeauthoryear{{Davies}, {Sternberg}, {Lehnert}  \&
  {Tacconi-Garman}}{{Davies} et~al.}{2003}]{davies03}
{Davies} R.~I.,  {Sternberg} A.,  {Lehnert} M.,   {Tacconi-Garman} L.~E.,
  2003, \mn@doi [\apj] {10.1086/378634}, \href
  {https://ui.adsabs.harvard.edu/abs/2003ApJ...597..907D} {597, 907}

\bibitem[\protect\citeauthoryear{{Davies}, {Sternberg}, {Lehnert}  \&
  {Tacconi-Garman}}{{Davies} et~al.}{2005}]{davies05}
{Davies} R.~I.,  {Sternberg} A.,  {Lehnert} M.~D.,   {Tacconi-Garman} L.~E.,
  2005, \mn@doi [\apj] {10.1086/444495}, \href
  {https://ui.adsabs.harvard.edu/abs/2005ApJ...633..105D} {633, 105}

\bibitem[\protect\citeauthoryear{{Davies}, {Maciejewski}, {Hicks}, {Tacconi},
  {Genzel}  \& {Engel}}{{Davies} et~al.}{2009}]{davies09}
{Davies} R.~I.,  {Maciejewski} W.,  {Hicks} E.~K.~S.,  {Tacconi} L.~J.,
  {Genzel} R.,   {Engel} H.,  2009, \mn@doi [\apj]
  {10.1088/0004-637X/702/1/114}, \href
  {https://ui.adsabs.harvard.edu/abs/2009ApJ...702..114D} {702, 114}

\bibitem[\protect\citeauthoryear{{Davies} et~al.,}{{Davies}
  et~al.}{2014}]{davies14}
{Davies} R.~I.,  et~al., 2014, \mn@doi [\apj] {10.1088/0004-637X/792/2/101},
  \href {https://ui.adsabs.harvard.edu/abs/2014ApJ...792..101D} {792, 101}

\bibitem[\protect\citeauthoryear{{Davies} et~al.,}{{Davies}
  et~al.}{2015}]{Davies15}
{Davies} R.~I.,  et~al., 2015, \mn@doi [\apj] {10.1088/0004-637X/806/1/127},
  \href {https://ui.adsabs.harvard.edu/abs/2015ApJ...806..127D} {806, 127}

\bibitem[\protect\citeauthoryear{{Davies} et~al.,}{{Davies}
  et~al.}{2020}]{davies20}
{Davies} R.,  et~al., 2020, \mn@doi [\mnras] {10.1093/mnras/staa2413}, \href
  {https://ui.adsabs.harvard.edu/abs/2020MNRAS.tmp.2061D} {}

\bibitem[\protect\citeauthoryear{{Dempsey} \& {Zakamska}}{{Dempsey} \&
  {Zakamska}}{2018}]{dempsey18}
{Dempsey} R.,  {Zakamska} N.~L.,  2018, \mn@doi [\mnras]
  {10.1093/mnras/sty941}, \href
  {https://ui.adsabs.harvard.edu/abs/2018MNRAS.477.4615D} {477, 4615}

\bibitem[\protect\citeauthoryear{{Diniz}, {Riffel}, {Storchi-Bergmann}  \&
  {Winge}}{{Diniz} et~al.}{2015}]{Diniz15}
{Diniz} M.~R.,  {Riffel} R.~A.,  {Storchi-Bergmann} T.,   {Winge} C.,  2015,
  \mn@doi [\mnras] {10.1093/mnras/stv1694}, \href
  {https://ui.adsabs.harvard.edu/abs/2015MNRAS.453.1727D} {453, 1727}

\bibitem[\protect\citeauthoryear{{Diniz}, {Riffel}  \& {Dors}}{{Diniz}
  et~al.}{2018}]{Diniz18}
{Diniz} M.~R.,  {Riffel} R.~A.,   {Dors} O.~L.,  2018, \mn@doi [Research Notes
  of the American Astronomical Society] {10.3847/2515-5172/aabbaf}, \href
  {https://ui.adsabs.harvard.edu/abs/2018RNAAS...2....3D} {2, 3}

\bibitem[\protect\citeauthoryear{{Dors}, {Storchi-Bergmann}, {Riffel}  \&
  {Schimdt}}{{Dors} et~al.}{2008}]{dors08}
{Dors} O.~L. J.,  {Storchi-Bergmann} T.,  {Riffel} R.~A.,   {Schimdt} A.~A.,
  2008, \mn@doi [\aap] {10.1051/0004-6361:20078960}, \href
  {https://ui.adsabs.harvard.edu/abs/2008A&A...482...59D} {482, 59}

\bibitem[\protect\citeauthoryear{{Dors}, {Riffel}, {Cardaci}, {H{\"a}gele},
  {Krabbe}, {P{\'e}rez-Montero}  \& {Rodrigues}}{{Dors} et~al.}{2012}]{Dors12}
{Dors} Oli~L. J.,  {Riffel} R.~A.,  {Cardaci} M.~V.,  {H{\"a}gele} G.~F.,
  {Krabbe} {\'A}.~C.,  {P{\'e}rez-Montero} E.,   {Rodrigues} I.,  2012, \mn@doi
  [\mnras] {10.1111/j.1365-2966.2012.20600.x}, \href
  {https://ui.adsabs.harvard.edu/abs/2012MNRAS.422..252D} {422, 252}

\bibitem[\protect\citeauthoryear{{Dors}, {Cardaci}, {H{\"a}gele}  \&
  {Krabbe}}{{Dors} et~al.}{2014}]{dors14}
{Dors} O.~L.,  {Cardaci} M.~V.,  {H{\"a}gele} G.~F.,   {Krabbe} {\^A}.~C.,
  2014, \mn@doi [\mnras] {10.1093/mnras/stu1218}, \href
  {https://ui.adsabs.harvard.edu/abs/2014MNRAS.443.1291D} {443, 1291}

\bibitem[\protect\citeauthoryear{{Dors}, {Maiolino}, {Cardaci}, {H{\"a}gele},
  {Krabbe}, {P{\'e}rez-Montero}  \& {Armah}}{{Dors} et~al.}{2020}]{dors20}
{Dors} O.~L.,  {Maiolino} R.,  {Cardaci} M.~V.,  {H{\"a}gele} G.~F.,  {Krabbe}
  A.~C.,  {P{\'e}rez-Montero} E.,   {Armah} M.,  2020, \mn@doi [\mnras]
  {10.1093/mnras/staa1781}, \href
  {https://ui.adsabs.harvard.edu/abs/2020MNRAS.tmp.1925D} {}

\bibitem[\protect\citeauthoryear{{Draine} \& {Woods}}{{Draine} \&
  {Woods}}{1990}]{draine90}
{Draine} B.~T.,  {Woods} D.~T.,  1990, \mn@doi [\apj] {10.1086/169358}, \href
  {https://ui.adsabs.harvard.edu/abs/1990ApJ...363..464D} {363, 464}

\bibitem[\protect\citeauthoryear{{Drehmer}, {Storchi-Bergmann}, {Ferrari},
  {Cappellari}  \& {Riffel}}{{Drehmer} et~al.}{2015}]{Drehmer15}
{Drehmer} D.~A.,  {Storchi-Bergmann} T.,  {Ferrari} F.,  {Cappellari} M.,
  {Riffel} R.~A.,  2015, \mn@doi [\mnras] {10.1093/mnras/stv536}, \href
  {https://ui.adsabs.harvard.edu/abs/2015MNRAS.450..128D} {450, 128}

\bibitem[\protect\citeauthoryear{{Dumont}, {Seth}, {Strader}, {Greene},
  {Burtscher}  \& {Neumayer}}{{Dumont} et~al.}{2020}]{Dumont20}
{Dumont} A.,  {Seth} A.~C.,  {Strader} J.,  {Greene} J.~E.,  {Burtscher} L.,
  {Neumayer} N.,  2020, \mn@doi [\apj] {10.3847/1538-4357/ab5798}, \href
  {https://ui.adsabs.harvard.edu/abs/2020ApJ...888...19D} {888, 19}

\bibitem[\protect\citeauthoryear{{Durr{\'e}} \& {Mould}}{{Durr{\'e}} \&
  {Mould}}{2014}]{durre14}
{Durr{\'e}} M.,  {Mould} J.,  2014, \mn@doi [\apj]
  {10.1088/0004-637X/784/1/79}, \href
  {https://ui.adsabs.harvard.edu/abs/2014ApJ...784...79D} {784, 79}

\bibitem[\protect\citeauthoryear{{Durr{\'e}} \& {Mould}}{{Durr{\'e}} \&
  {Mould}}{2018}]{durre18}
{Durr{\'e}} M.,  {Mould} J.,  2018, \mn@doi [\apj] {10.3847/1538-4357/aae68e},
  \href {https://ui.adsabs.harvard.edu/abs/2018ApJ...867..149D} {867, 149}

\bibitem[\protect\citeauthoryear{{Elitzur}}{{Elitzur}}{2012}]{elitzur12}
{Elitzur} M.,  2012, \mn@doi [\apjl] {10.1088/2041-8205/747/2/L33}, \href
  {https://ui.adsabs.harvard.edu/abs/2012ApJ...747L..33E} {747, L33}

\bibitem[\protect\citeauthoryear{{Elitzur}, {Ho}  \& {Trump}}{{Elitzur}
  et~al.}{2014}]{elitzur14}
{Elitzur} M.,  {Ho} L.~C.,   {Trump} J.~R.,  2014, \mn@doi [\mnras]
  {10.1093/mnras/stt2445}, \href
  {https://ui.adsabs.harvard.edu/abs/2014MNRAS.438.3340E} {438, 3340}

\bibitem[\protect\citeauthoryear{{Fazeli}, {Eckart}, {Busch}, {Yttergren},
  {Combes}, {Misquitta}  \& {Straubmeier}}{{Fazeli} et~al.}{2020}]{fazeli20}
{Fazeli} N.,  {Eckart} A.,  {Busch} G.,  {Yttergren} M.,  {Combes} F.,
  {Misquitta} P.,   {Straubmeier} C.,  2020, \mn@doi [\aap]
  {10.1051/0004-6361/201937092}, \href
  {https://ui.adsabs.harvard.edu/abs/2020A&A...638A..36F} {638, A36}

\bibitem[\protect\citeauthoryear{{Fischer} et~al.,}{{Fischer}
  et~al.}{2017}]{fischer17}
{Fischer} T.~C.,  et~al., 2017, \mn@doi [\apj] {10.3847/1538-4357/834/1/30},
  \href {https://ui.adsabs.harvard.edu/abs/2017ApJ...834...30F} {834, 30}

\bibitem[\protect\citeauthoryear{{Frank}, {King}  \& {Raine}}{{Frank}
  et~al.}{2002}]{frank02}
{Frank} J.,  {King} A.,   {Raine} D.~J.,  2002, {Accretion Power in
  Astrophysics: Third Edition}

\bibitem[\protect\citeauthoryear{{Freitas} et~al.,}{{Freitas}
  et~al.}{2018}]{Freitas18}
{Freitas} I.~C.,  et~al., 2018, \mn@doi [\mnras] {10.1093/mnras/sty303}, \href
  {https://ui.adsabs.harvard.edu/abs/2018MNRAS.476.2760F} {476, 2760}

\bibitem[\protect\citeauthoryear{{Genzel} et~al.,}{{Genzel}
  et~al.}{2010}]{Genzel+10}
{Genzel} R.,  et~al., 2010, \mn@doi [\mnras]
  {10.1111/j.1365-2966.2010.16969.x}, \href
  {https://ui.adsabs.harvard.edu/abs/2010MNRAS.407.2091G} {407, 2091}

\bibitem[\protect\citeauthoryear{{Gnilka} et~al.,}{{Gnilka}
  et~al.}{2020}]{Gnilka20}
{Gnilka} C.~L.,  et~al., 2020, \mn@doi [\apj] {10.3847/1538-4357/ab8000}, \href
  {https://ui.adsabs.harvard.edu/abs/2020ApJ...893...80G} {893, 80}

\bibitem[\protect\citeauthoryear{{Gonzalez Delgado} \& {Perez}}{{Gonzalez
  Delgado} \& {Perez}}{1997}]{Gonzalez97}
{Gonzalez Delgado} R.~M.,  {Perez} E.,  1997, \mn@doi [\mnras]
  {10.1093/mnras/284.4.931}, \href
  {https://ui.adsabs.harvard.edu/abs/1997MNRAS.284..931G} {284, 931}

\bibitem[\protect\citeauthoryear{{Harrison}}{{Harrison}}{2017}]{harrison17}
{Harrison} C.~M.,  2017, \mn@doi [Nature Astronomy] {10.1038/s41550-017-0165},
  \href {https://ui.adsabs.harvard.edu/abs/2017NatAs...1E.165H} {1, 0165}

\bibitem[\protect\citeauthoryear{{Harrison}, {Costa}, {Tadhunter},
  {Fl{\"u}tsch}, {Kakkad}, {Perna}  \& {Vietri}}{{Harrison}
  et~al.}{2018}]{harrison18}
{Harrison} C.~M.,  {Costa} T.,  {Tadhunter} C.~N.,  {Fl{\"u}tsch} A.,  {Kakkad}
  D.,  {Perna} M.,   {Vietri} G.,  2018, \mn@doi [Nature Astronomy]
  {10.1038/s41550-018-0403-6}, \href
  {https://ui.adsabs.harvard.edu/abs/2018NatAs...2..198H} {2, 198}

\bibitem[\protect\citeauthoryear{{Heckman}}{{Heckman}}{1980}]{heckman80}
{Heckman} T.~M.,  1980, \aap, \href
  {https://ui.adsabs.harvard.edu/abs/1980A&A....87..152H} {500, 187}

\bibitem[\protect\citeauthoryear{{Heckman} \& {Best}}{{Heckman} \&
  {Best}}{2014}]{heckman14}
{Heckman} T.~M.,  {Best} P.~N.,  2014, \mn@doi [\araa]
  {10.1146/annurev-astro-081913-035722}, \href
  {https://ui.adsabs.harvard.edu/abs/2014ARA&A..52..589H} {52, 589}

\bibitem[\protect\citeauthoryear{{Hicks}, {Davies}, {Malkan}, {Genzel},
  {Tacconi}, {M{\"u}ller S{\'a}nchez}  \& {Sternberg}}{{Hicks}
  et~al.}{2009}]{hicks09}
{Hicks} E.~K.~S.,  {Davies} R.~I.,  {Malkan} M.~A.,  {Genzel} R.,  {Tacconi}
  L.~J.,  {M{\"u}ller S{\'a}nchez} F.,   {Sternberg} A.,  2009, \mn@doi [\apj]
  {10.1088/0004-637X/696/1/448}, \href
  {https://ui.adsabs.harvard.edu/abs/2009ApJ...696..448H} {696, 448}

\bibitem[\protect\citeauthoryear{{Ho}, {Filippenko}, {Sargent}  \& {Peng}}{{Ho}
  et~al.}{1997}]{ho97}
{Ho} L.~C.,  {Filippenko} A.~V.,  {Sargent} W. L.~W.,   {Peng} C.~Y.,  1997,
  \mn@doi [\apjs] {10.1086/313042}, \href
  {https://ui.adsabs.harvard.edu/abs/1997ApJS..112..391H} {112, 391}

\bibitem[\protect\citeauthoryear{{Hollenbach} \& {McKee}}{{Hollenbach} \&
  {McKee}}{1989}]{hollenbach89}
{Hollenbach} D.,  {McKee} C.~F.,  1989, \mn@doi [\apj] {10.1086/167595}, \href
  {https://ui.adsabs.harvard.edu/abs/1989ApJ...342..306H} {342, 306}

\bibitem[\protect\citeauthoryear{{Hopkins} \& {Quataert}}{{Hopkins} \&
  {Quataert}}{2010}]{hopkins10}
{Hopkins} P.~F.,  {Quataert} E.,  2010, \mn@doi [\mnras]
  {10.1111/j.1365-2966.2010.17064.x}, \href
  {https://ui.adsabs.harvard.edu/abs/2010MNRAS.407.1529H} {407, 1529}

\bibitem[\protect\citeauthoryear{{Husemann} et~al.,}{{Husemann}
  et~al.}{2019}]{Husemann19}
{Husemann} B.,  et~al., 2019, \mn@doi [\aap] {10.1051/0004-6361/201935283},
  \href {https://ui.adsabs.harvard.edu/abs/2019A&A...627A..53H} {627, A53}

\bibitem[\protect\citeauthoryear{{Ichikawa}, {Ricci}, {Ueda}, {Matsuoka},
  {Toba}, {Kawamuro}, {Trakhtenbrot}  \& {Koss}}{{Ichikawa}
  et~al.}{2017}]{ichikawa17}
{Ichikawa} K.,  {Ricci} C.,  {Ueda} Y.,  {Matsuoka} K.,  {Toba} Y.,  {Kawamuro}
  T.,  {Trakhtenbrot} B.,   {Koss} M.~J.,  2017, \mn@doi [\apj]
  {10.3847/1538-4357/835/1/74}, \href
  {https://ui.adsabs.harvard.edu/abs/2017ApJ...835...74I} {835, 74}

\bibitem[\protect\citeauthoryear{{Ilha}, {Bianchin}  \& {Riffel}}{{Ilha}
  et~al.}{2016}]{Ilha16}
{Ilha} G. d.~S.,  {Bianchin} M.,   {Riffel} R.~A.,  2016, \mn@doi [\apss]
  {10.1007/s10509-016-2760-x}, \href
  {https://ui.adsabs.harvard.edu/abs/2016Ap&SS.361..178I} {361, 178}

\bibitem[\protect\citeauthoryear{{Jin} et~al.,}{{Jin} et~al.}{2016}]{Jin16}
{Jin} Y.,  et~al., 2016, \mn@doi [\mnras] {10.1093/mnras/stw2055}, \href
  {https://ui.adsabs.harvard.edu/abs/2016MNRAS.463..913J} {463, 913}

\bibitem[\protect\citeauthoryear{{Kakkad} et~al.,}{{Kakkad}
  et~al.}{2018}]{kakkad18}
{Kakkad} D.,  et~al., 2018, \mn@doi [\aap] {10.1051/0004-6361/201832790}, \href
  {https://ui.adsabs.harvard.edu/abs/2018A&A...618A...6K} {618, A6}

\bibitem[\protect\citeauthoryear{{Keel}}{{Keel}}{1990}]{keel90}
{Keel} W.~C.,  1990, \mn@doi [\aj] {10.1086/115519}, \href
  {https://ui.adsabs.harvard.edu/abs/1990AJ....100..356K} {100, 356}

\bibitem[\protect\citeauthoryear{{Kormendy} \& {Ho}}{{Kormendy} \&
  {Ho}}{2013}]{kormendy13}
{Kormendy} J.,  {Ho} L.~C.,  2013, \mn@doi [\araa]
  {10.1146/annurev-astro-082708-101811}, \href
  {https://ui.adsabs.harvard.edu/abs/2013ARA&A..51..511K} {51, 511}

\bibitem[\protect\citeauthoryear{{Kwan}, {Gatley}, {Merrill}, {Probst}  \&
  {Weintraub}}{{Kwan} et~al.}{1977}]{Kwan77}
{Kwan} J.~H.,  {Gatley} I.,  {Merrill} K.~M.,  {Probst} R.,   {Weintraub}
  D.~A.,  1977, \mn@doi [\apj] {10.1086/155514}, \href
  {https://ui.adsabs.harvard.edu/abs/1977ApJ...216..713K} {216, 713}

\bibitem[\protect\citeauthoryear{{Lamperti} et~al.,}{{Lamperti}
  et~al.}{2017}]{lamperti17}
{Lamperti} I.,  et~al., 2017, \mn@doi [\mnras] {10.1093/mnras/stx055}, \href
  {https://ui.adsabs.harvard.edu/abs/2017MNRAS.467..540L} {467, 540}

\bibitem[\protect\citeauthoryear{{Lin} et~al.,}{{Lin} et~al.}{2018}]{lin18}
{Lin} M.-Y.,  et~al., 2018, \mn@doi [\mnras] {10.1093/mnras/stx2618}, \href
  {https://ui.adsabs.harvard.edu/abs/2018MNRAS.473.4582L} {473, 4582}

\bibitem[\protect\citeauthoryear{{Liu}, {Zakamska}, {Greene}, {Nesvadba}  \&
  {Liu}}{{Liu} et~al.}{2013}]{liu13}
{Liu} G.,  {Zakamska} N.~L.,  {Greene} J.~E.,  {Nesvadba} N. P.~H.,   {Liu} X.,
   2013, \mn@doi [\mnras] {10.1093/mnras/stt1755}, \href
  {https://ui.adsabs.harvard.edu/abs/2013MNRAS.436.2576L} {436, 2576}

\bibitem[\protect\citeauthoryear{{Lutz}, {Sturm}, {Genzel}, {Spoon},
  {Moorwood}, {Netzer}  \& {Sternberg}}{{Lutz} et~al.}{2003}]{Lutz03}
{Lutz} D.,  {Sturm} E.,  {Genzel} R.,  {Spoon} H.~W.~W.,  {Moorwood} A.~F.~M.,
  {Netzer} H.,   {Sternberg} A.,  2003, \mn@doi [\aap]
  {10.1051/0004-6361:20031165}, \href
  {https://ui.adsabs.harvard.edu/abs/2003A&A...409..867L} {409, 867}

\bibitem[\protect\citeauthoryear{{Maloney}, {Hollenbach}  \&
  {Tielens}}{{Maloney} et~al.}{1996}]{maloney96}
{Maloney} P.~R.,  {Hollenbach} D.~J.,   {Tielens} A.~G.~G.~M.,  1996, \mn@doi
  [\apj] {10.1086/177532}, \href
  {https://ui.adsabs.harvard.edu/abs/1996ApJ...466..561M} {466, 561}

\bibitem[\protect\citeauthoryear{{May} \& {Steiner}}{{May} \&
  {Steiner}}{2017}]{may17}
{May} D.,  {Steiner} J.~E.,  2017, \mn@doi [\mnras] {10.1093/mnras/stx886},
  \href {https://ui.adsabs.harvard.edu/abs/2017MNRAS.469..994M} {469, 994}

\bibitem[\protect\citeauthoryear{{May}, {Steiner}, {Menezes}, {Williams}  \&
  {Wang}}{{May} et~al.}{2020}]{may20}
{May} D.,  {Steiner} J.~E.,  {Menezes} R.~B.,  {Williams} D.~R.~A.,   {Wang}
  J.,  2020, \mn@doi [\mnras] {10.1093/mnras/staa1545}, \href
  {https://ui.adsabs.harvard.edu/abs/2020MNRAS.496.1488M} {496, 1488}

\bibitem[\protect\citeauthoryear{{Mazzalay} et~al.,}{{Mazzalay}
  et~al.}{2013}]{mazzalay13}
{Mazzalay} X.,  et~al., 2013, \mn@doi [\mnras] {10.1093/mnras/sts204}, \href
  {https://ui.adsabs.harvard.edu/abs/2013MNRAS.428.2389M} {428, 2389}

\bibitem[\protect\citeauthoryear{{McGregor} et~al.,}{{McGregor}
  et~al.}{2003}]{mcgregor03}
{McGregor} P.~J.,  et~al., 2003, {Gemini near-infrared integral field
  spectrograph (NIFS)}.
Proceedings of the SPIE, pp 1581--1591, \mn@doi{10.1117/12.459448}

\bibitem[\protect\citeauthoryear{{Mouri}}{{Mouri}}{1994}]{mouri94}
{Mouri} H.,  1994, \mn@doi [\apj] {10.1086/174184}, \href
  {https://ui.adsabs.harvard.edu/abs/1994ApJ...427..777M} {427, 777}

\bibitem[\protect\citeauthoryear{{M{\"u}ller S{\'a}nchez}, {Davies},
  {Eisenhauer}, {Tacconi}, {Genzel}  \& {Sternberg}}{{M{\"u}ller S{\'a}nchez}
  et~al.}{2006}]{ms06}
{M{\"u}ller S{\'a}nchez} F.,  {Davies} R.~I.,  {Eisenhauer} F.,  {Tacconi}
  L.~J.,  {Genzel} R.,   {Sternberg} A.,  2006, \mn@doi [\aap]
  {10.1051/0004-6361:20054387}, \href
  {https://ui.adsabs.harvard.edu/abs/2006A&A...454..481M} {454, 481}

\bibitem[\protect\citeauthoryear{{M{\"u}ller-S{\'a}nchez}, {Nevin},
  {Comerford}, {Davies}, {Privon}  \& {Treister}}{{M{\"u}ller-S{\'a}nchez}
  et~al.}{2018a}]{sanchez18}
{M{\"u}ller-S{\'a}nchez} F.,  {Nevin} R.,  {Comerford} J.~M.,  {Davies} R.~I.,
  {Privon} G.~C.,   {Treister} E.,  2018a, \mn@doi [\nat]
  {10.1038/s41586-018-0033-2}, \href
  {https://ui.adsabs.harvard.edu/abs/2018Natur.556..345M} {556, 345}

\bibitem[\protect\citeauthoryear{{M{\"u}ller-S{\'a}nchez}, {Hicks}, {Malkan},
  {Davies}, {Yu}, {Shaver}  \& {Davis}}{{M{\"u}ller-S{\'a}nchez}
  et~al.}{2018b}]{ms18}
{M{\"u}ller-S{\'a}nchez} F.,  {Hicks} E.~K.~S.,  {Malkan} M.,  {Davies} R.,
  {Yu} P.~C.,  {Shaver} S.,   {Davis} B.,  2018b, \mn@doi [\apj]
  {10.3847/1538-4357/aab9ad}, \href
  {https://ui.adsabs.harvard.edu/abs/2018ApJ...858...48M} {858, 48}

\bibitem[\protect\citeauthoryear{{Novak}, {Ostriker}  \& {Ciotti}}{{Novak}
  et~al.}{2011}]{novak11}
{Novak} G.~S.,  {Ostriker} J.~P.,   {Ciotti} L.,  2011, \mn@doi [\apj]
  {10.1088/0004-637X/737/1/26}, \href
  {https://ui.adsabs.harvard.edu/abs/2011ApJ...737...26N} {737, 26}

\bibitem[\protect\citeauthoryear{{Oh} et~al.,}{{Oh} et~al.}{2018}]{BAT105}
{Oh} K.,  et~al., 2018, \mn@doi [\apjs] {10.3847/1538-4365/aaa7fd}, \href
  {https://ui.adsabs.harvard.edu/abs/2018ApJS..235....4O} {235, 4}

\bibitem[\protect\citeauthoryear{{Osterbrock} \& {Ferland}}{{Osterbrock} \&
  {Ferland}}{2006}]{Osterbrock06}
{Osterbrock} D.~E.,  {Ferland} G.~J.,  2006, {Astrophysics of gaseous nebulae
  and active galactic nuclei}.
University Science Books

\bibitem[\protect\citeauthoryear{{Pasquali}, {Gallagher}  \& {de
  Grijs}}{{Pasquali} et~al.}{2004}]{Pasquali04}
{Pasquali} A.,  {Gallagher} J.~S.,   {de Grijs} R.,  2004, \mn@doi [\aap]
  {10.1051/0004-6361:20034183}, \href
  {https://ui.adsabs.harvard.edu/abs/2004A&A...415..103P} {415, 103}

\bibitem[\protect\citeauthoryear{{Paturel}, {Petit}, {Prugniel}, {Theureau},
  {Rousseau}, {Brouty}, {Dubois}  \& {Cambr{\'e}sy}}{{Paturel}
  et~al.}{2003}]{paturel03}
{Paturel} G.,  {Petit} C.,  {Prugniel} P.,  {Theureau} G.,  {Rousseau} J.,
  {Brouty} M.,  {Dubois} P.,   {Cambr{\'e}sy} L.,  2003, \mn@doi [\aap]
  {10.1051/0004-6361:20031411}, \href
  {https://ui.adsabs.harvard.edu/abs/2003A&A...412...45P} {412, 45}

\bibitem[\protect\citeauthoryear{{Prieto}, {Mezcua}, {Fern{\'a}ndez-Ontiveros}
  \& {Schartmann}}{{Prieto} et~al.}{2014}]{prieto14}
{Prieto} M.~A.,  {Mezcua} M.,  {Fern{\'a}ndez-Ontiveros} J.~A.,   {Schartmann}
  M.,  2014, \mn@doi [\mnras] {10.1093/mnras/stu1006}, \href
  {https://ui.adsabs.harvard.edu/abs/2014MNRAS.442.2145P} {442, 2145}

\bibitem[\protect\citeauthoryear{{Puxley}, {Hawarden}  \& {Mountain}}{{Puxley}
  et~al.}{1990}]{puxley90}
{Puxley} P.~J.,  {Hawarden} T.~G.,   {Mountain} C.~M.,  1990, \mn@doi [\apj]
  {10.1086/169386}, \href
  {https://ui.adsabs.harvard.edu/abs/1990ApJ...364...77P} {364, 77}

\bibitem[\protect\citeauthoryear{{Ramos Almeida} et~al.,}{{Ramos Almeida}
  et~al.}{2011}]{ramos-almeida11}
{Ramos Almeida} C.,  et~al., 2011, \mn@doi [\apj] {10.1088/0004-637X/731/2/92},
  \href {https://ui.adsabs.harvard.edu/abs/2011ApJ...731...92R} {731, 92}

\bibitem[\protect\citeauthoryear{{Reunanen}, {Kotilainen}  \&
  {Prieto}}{{Reunanen} et~al.}{2002}]{reunanen02}
{Reunanen} J.,  {Kotilainen} J.~K.,   {Prieto} M.~A.,  2002, \mn@doi [\mnras]
  {10.1046/j.1365-8711.2002.05181.x}, \href
  {https://ui.adsabs.harvard.edu/abs/2002MNRAS.331..154R} {331, 154}

\bibitem[\protect\citeauthoryear{{Ricci}, {Steiner}  \& {Menezes}}{{Ricci}
  et~al.}{2014}]{Ricci+14}
{Ricci} T.~V.,  {Steiner} J.~E.,   {Menezes} R.~B.,  2014, \mn@doi [\mnras]
  {10.1093/mnras/stu441}, \href
  {https://ui.adsabs.harvard.edu/abs/2014MNRAS.440.2419R} {440, 2419}

\bibitem[\protect\citeauthoryear{{Riffel}}{{Riffel}}{2020a}]{rogemarN1275}
{Riffel} R.~A.,  2020a, \mn@doi [\mnras] {10.1093/mnras/staa903}, \href
  {https://ui.adsabs.harvard.edu/abs/2020MNRAS.494.2004R} {494, 2004}

\bibitem[\protect\citeauthoryear{{Riffel}}{{Riffel}}{2020b}]{rogemarHe}
{Riffel} R.~A.,  2020b, \mn@doi [\mnras] {10.1093/mnras/staa903}, \href
  {https://ui.adsabs.harvard.edu/abs/2020MNRAS.494.2004R} {494, 2004}

\bibitem[\protect\citeauthoryear{{Riffel}, {Rodr{\'\i}guez-Ardila}  \&
  {Pastoriza}}{{Riffel} et~al.}{2006}]{rogerio06}
{Riffel} R.,  {Rodr{\'\i}guez-Ardila} A.,   {Pastoriza} M.~G.,  2006, \mn@doi
  [\aap] {10.1051/0004-6361:20065291}, \href
  {https://ui.adsabs.harvard.edu/abs/2006A&A...457...61R} {457, 61}

\bibitem[\protect\citeauthoryear{{Riffel}, {Storchi-Bergmann}, {Winge},
  {McGregor}, {Beck}  \& {Schmitt}}{{Riffel} et~al.}{2008}]{rogemarN4051}
{Riffel} R.~A.,  {Storchi-Bergmann} T.,  {Winge} C.,  {McGregor} P.~J.,  {Beck}
  T.,   {Schmitt} H.,  2008, \mn@doi [\mnras]
  {10.1111/j.1365-2966.2008.12936.x}, \href
  {https://ui.adsabs.harvard.edu/abs/2008MNRAS.385.1129R} {385, 1129}

\bibitem[\protect\citeauthoryear{{Riffel}, {Storchi-Bergmann}, {Dors}  \&
  {Winge}}{{Riffel} et~al.}{2009a}]{rogemarN7582}
{Riffel} R.~A.,  {Storchi-Bergmann} T.,  {Dors} O.~L.,   {Winge} C.,  2009a,
  \mn@doi [\mnras] {10.1111/j.1365-2966.2008.14250.x}, \href
  {https://ui.adsabs.harvard.edu/abs/2009MNRAS.393..783R} {393, 783}

\bibitem[\protect\citeauthoryear{{Riffel}, {Pastoriza}, {Rodr{\'\i}guez-Ardila}
   \& {Bonatto}}{{Riffel} et~al.}{2009b}]{rogerio09}
{Riffel} R.,  {Pastoriza} M.~G.,  {Rodr{\'\i}guez-Ardila} A.,   {Bonatto} C.,
  2009b, \mn@doi [\mnras] {10.1111/j.1365-2966.2009.15448.x}, \href
  {https://ui.adsabs.harvard.edu/abs/2009MNRAS.400..273R} {400, 273}

\bibitem[\protect\citeauthoryear{{Riffel}, {Storchi-Bergmann}  \&
  {McGregor}}{{Riffel} et~al.}{2009c}]{rogemar_N4151_dust}
{Riffel} R.~A.,  {Storchi-Bergmann} T.,   {McGregor} P.~J.,  2009c, \mn@doi
  [\apj] {10.1088/0004-637X/698/2/1767}, \href
  {https://ui.adsabs.harvard.edu/abs/2009ApJ...698.1767R} {698, 1767}

\bibitem[\protect\citeauthoryear{{Riffel}, {Storchi-Bergmann}  \&
  {Nagar}}{{Riffel} et~al.}{2010a}]{rogemarMrk1066exc}
{Riffel} R.~A.,  {Storchi-Bergmann} T.,   {Nagar} N.~M.,  2010a, \mn@doi
  [\mnras] {10.1111/j.1365-2966.2010.16308.x}, \href
  {https://ui.adsabs.harvard.edu/abs/2010MNRAS.404..166R} {404, 166}

\bibitem[\protect\citeauthoryear{{Riffel}, {Storchi-Bergmann}, {Riffel}  \&
  {Pastoriza}}{{Riffel} et~al.}{2010b}]{rogemarM1066_SP}
{Riffel} R.~A.,  {Storchi-Bergmann} T.,  {Riffel} R.,   {Pastoriza} M.~G.,
  2010b, \mn@doi [\apj] {10.1088/0004-637X/713/1/469}, \href
  {https://ui.adsabs.harvard.edu/abs/2010ApJ...713..469R} {713, 469}

\bibitem[\protect\citeauthoryear{{Riffel}, {Rodr{\'\i}guez-Ardila}, {Aleman},
  {Brotherton}, {Pastoriza}, {Bonatto}  \& {Dors}}{{Riffel}
  et~al.}{2013a}]{rogerio13}
{Riffel} R.,  {Rodr{\'\i}guez-Ardila} A.,  {Aleman} I.,  {Brotherton} M.~S.,
  {Pastoriza} M.~G.,  {Bonatto} C.,   {Dors} O.~L.,  2013a, \mn@doi [\mnras]
  {10.1093/mnras/stt026}, \href
  {https://ui.adsabs.harvard.edu/abs/2013MNRAS.430.2002R} {430, 2002}

\bibitem[\protect\citeauthoryear{{Riffel}, {Storchi-Bergmann}  \&
  {Winge}}{{Riffel} et~al.}{2013b}]{rogemarM79}
{Riffel} R.~A.,  {Storchi-Bergmann} T.,   {Winge} C.,  2013b, \mn@doi [\mnras]
  {10.1093/mnras/stt045}, \href
  {https://ui.adsabs.harvard.edu/abs/2013MNRAS.430.2249R} {430, 2249}

\bibitem[\protect\citeauthoryear{{Riffel}, {Vale}, {Storchi-Bergmann}  \&
  {McGregor}}{{Riffel} et~al.}{2014}]{rogemarN1068}
{Riffel} R.~A.,  {Vale} T.~B.,  {Storchi-Bergmann} T.,   {McGregor} P.~J.,
  2014, \mn@doi [\mnras] {10.1093/mnras/stu843}, \href
  {https://ui.adsabs.harvard.edu/abs/2014MNRAS.442..656R} {442, 656}

\bibitem[\protect\citeauthoryear{{Riffel}, {Storchi-Bergmann}, {Riffel},
  {Dahmer-Hahn}, {Diniz}, {Sch{\"o}nell}  \& {Dametto}}{{Riffel}
  et~al.}{2017}]{rogemar_stellar}
{Riffel} R.~A.,  {Storchi-Bergmann} T.,  {Riffel} R.,  {Dahmer-Hahn} L.~G.,
  {Diniz} M.~R.,  {Sch{\"o}nell} A.~J.,   {Dametto} N.~Z.,  2017, \mn@doi
  [\mnras] {10.1093/mnras/stx1308}, \href
  {https://ui.adsabs.harvard.edu/abs/2017MNRAS.470..992R} {470, 992}

\bibitem[\protect\citeauthoryear{{Riffel} et~al.,}{{Riffel}
  et~al.}{2018}]{rogemar_sample}
{Riffel} R.~A.,  et~al., 2018, \mn@doi [\mnras] {10.1093/mnras/stx2857}, \href
  {https://ui.adsabs.harvard.edu/abs/2018MNRAS.474.1373R} {474, 1373}

\bibitem[\protect\citeauthoryear{{Riffel}, {Storchi-Bergmann}, {Zakamska}  \&
  {Riffel}}{{Riffel} et~al.}{2020}]{rogemar_N1275}
{Riffel} R.~A.,  {Storchi-Bergmann} T.,  {Zakamska} N.~L.,   {Riffel} R.,
  2020, \mn@doi [\mnras] {10.1093/mnras/staa1922}, \href
  {https://ui.adsabs.harvard.edu/abs/2020MNRAS.tmp.2053R} {496, 4857}

\bibitem[\protect\citeauthoryear{{Riffel}, {Bianchin}, {Riffel},
  {Storchi-Bergmann}, {Sch{\"o}nell}, {Dahmer-Hahn}, {Dametto}  \&
  {Diniz}}{{Riffel} et~al.}{2021}]{rogemar21_exc}
{Riffel} R.~A.,  {Bianchin} M.,  {Riffel} R.,  {Storchi-Bergmann} T.,
  {Sch{\"o}nell} A.~J.,  {Dahmer-Hahn} L.~G.,  {Dametto} N.~Z.,   {Diniz}
  M.~R.,  2021, \mn@doi [\mnras] {10.1093/mnras/stab788}, \href
  {https://ui.adsabs.harvard.edu/abs/2021MNRAS.tmp..778R} {}

\bibitem[\protect\citeauthoryear{{Rodr{\'\i}guez-Ardila}, {Pastoriza},
  {Viegas}, {Sigut}  \& {Pradhan}}{{Rodr{\'\i}guez-Ardila}
  et~al.}{2004}]{ardila04}
{Rodr{\'\i}guez-Ardila} A.,  {Pastoriza} M.~G.,  {Viegas} S.,  {Sigut}
  T.~A.~A.,   {Pradhan} A.~K.,  2004, \mn@doi [\aap]
  {10.1051/0004-6361:20034285}, \href
  {https://ui.adsabs.harvard.edu/abs/2004A&A...425..457R} {425, 457}

\bibitem[\protect\citeauthoryear{{Rodr{\'\i}guez-Ardila}, {Riffel}  \&
  {Pastoriza}}{{Rodr{\'\i}guez-Ardila} et~al.}{2005}]{ardila05}
{Rodr{\'\i}guez-Ardila} A.,  {Riffel} R.,   {Pastoriza} M.~G.,  2005, \mn@doi
  [\mnras] {10.1111/j.1365-2966.2005.09638.x}, \href
  {https://ui.adsabs.harvard.edu/abs/2005MNRAS.364.1041R} {364, 1041}

\bibitem[\protect\citeauthoryear{{Rosario}, {Togi}, {Burtscher}, {Davies},
  {Shimizu}  \& {Lutz}}{{Rosario} et~al.}{2019}]{rosario19}
{Rosario} D.~J.,  {Togi} A.,  {Burtscher} L.,  {Davies} R.~I.,  {Shimizu}
  T.~T.,   {Lutz} D.,  2019, \mn@doi [\apjl] {10.3847/2041-8213/ab1262}, \href
  {https://ui.adsabs.harvard.edu/abs/2019ApJ...875L...8R} {875, L8}

\bibitem[\protect\citeauthoryear{Ruschel-Dutra}{Ruschel-Dutra}{2020}]{ifscube}
Ruschel-Dutra D.,  2020, danielrd6/ifscube v1.0,
  \mn@doi{10.5281/zenodo.3945237}, \url
  {https://doi.org/10.5281/zenodo.3945237}

\bibitem[\protect\citeauthoryear{{Sargent} et~al.,}{{Sargent}
  et~al.}{2014}]{Sargent+14}
{Sargent} M.~T.,  et~al., 2014, \mn@doi [\apj] {10.1088/0004-637X/793/1/19},
  \href {https://ui.adsabs.harvard.edu/abs/2014ApJ...793...19S} {793, 19}

\bibitem[\protect\citeauthoryear{{Scharw{\"a}chter}, {McGregor}, {Dopita}  \&
  {Beck}}{{Scharw{\"a}chter} et~al.}{2013}]{schawachter13}
{Scharw{\"a}chter} J.,  {McGregor} P.~J.,  {Dopita} M.~A.,   {Beck} T.~L.,
  2013, \mn@doi [\mnras] {10.1093/mnras/sts502}, \href
  {https://ui.adsabs.harvard.edu/abs/2013MNRAS.429.2315S} {429, 2315}

\bibitem[\protect\citeauthoryear{{Schinnerer}, {Eckart}  \&
  {Tacconi}}{{Schinnerer} et~al.}{2000}]{Schinnerer00}
{Schinnerer} E.,  {Eckart} A.,   {Tacconi} L.~J.,  2000, \mn@doi [\apj]
  {10.1086/308703}, \href
  {https://ui.adsabs.harvard.edu/abs/2000ApJ...533..826S} {533, 826}

\bibitem[\protect\citeauthoryear{{Sch{\"o}nell}, {Riffel}, {Storchi-Bergmann}
  \& {Winge}}{{Sch{\"o}nell} et~al.}{2014}]{Schonell14}
{Sch{\"o}nell} A.~J.,  {Riffel} R.~A.,  {Storchi-Bergmann} T.,   {Winge} C.,
  2014, \mn@doi [\mnras] {10.1093/mnras/stu1685}, \href
  {https://ui.adsabs.harvard.edu/abs/2014MNRAS.445..414S} {445, 414}

\bibitem[\protect\citeauthoryear{{Sch{\"o}nell}, {Storchi-Bergmann}, {Riffel}
  \& {Riffel}}{{Sch{\"o}nell} et~al.}{2017}]{SchonellN5548}
{Sch{\"o}nell} Astor~J. J.,  {Storchi-Bergmann} T.,  {Riffel} R.~A.,   {Riffel}
  R.,  2017, \mn@doi [\mnras] {10.1093/mnras/stw2263}, \href
  {https://ui.adsabs.harvard.edu/abs/2017MNRAS.464.1771S} {464, 1771}

\bibitem[\protect\citeauthoryear{{Sch{\"o}nell}, {Storchi-Bergmann}, {Riffel},
  {Riffel}, {Bianchin}, {Dahmer-Hahn}, {Diniz}  \& {Dametto}}{{Sch{\"o}nell}
  et~al.}{2019}]{schonell19}
{Sch{\"o}nell} A.~J.,  {Storchi-Bergmann} T.,  {Riffel} R.~A.,  {Riffel} R.,
  {Bianchin} M.,  {Dahmer-Hahn} L.~G.,  {Diniz} M.~R.,   {Dametto} N.~Z.,
  2019, \mn@doi [\mnras] {10.1093/mnras/stz523}, \href
  {https://ui.adsabs.harvard.edu/abs/2019MNRAS.485.2054S} {485, 2054}

\bibitem[\protect\citeauthoryear{{Scoville}, {Hall}, {Ridgway}  \&
  {Kleinmann}}{{Scoville} et~al.}{1982}]{scoville82}
{Scoville} N.~Z.,  {Hall} D.~N.~B.,  {Ridgway} S.~T.,   {Kleinmann} S.~G.,
  1982, \mn@doi [\apj] {10.1086/159618}, \href
  {https://ui.adsabs.harvard.edu/abs/1982ApJ...253..136S} {253, 136}

\bibitem[\protect\citeauthoryear{{Shimizu} et~al.,}{{Shimizu}
  et~al.}{2019}]{Shimizu19}
{Shimizu} T.~T.,  et~al., 2019, \mn@doi [\mnras] {10.1093/mnras/stz2802}, \href
  {https://ui.adsabs.harvard.edu/abs/2019MNRAS.490.5860S} {490, 5860}

\bibitem[\protect\citeauthoryear{{Silverman} et~al.,}{{Silverman}
  et~al.}{2015}]{Silverman+15}
{Silverman} J.~D.,  et~al., 2015, \mn@doi [\apjl]
  {10.1088/2041-8205/812/2/L23}, \href
  {https://ui.adsabs.harvard.edu/abs/2015ApJ...812L..23S} {812, L23}

\bibitem[\protect\citeauthoryear{{Smith}}{{Smith}}{1995}]{Smith95}
{Smith} M.~D.,  1995, \aap, \href
  {https://ui.adsabs.harvard.edu/abs/1995A&A...296..789S} {296, 789}

\bibitem[\protect\citeauthoryear{{Sternberg} \& {Dalgarno}}{{Sternberg} \&
  {Dalgarno}}{1989}]{Sternberg89}
{Sternberg} A.,  {Dalgarno} A.,  1989, \mn@doi [\apj] {10.1086/167193}, \href
  {https://ui.adsabs.harvard.edu/abs/1989ApJ...338..197S} {338, 197}

\bibitem[\protect\citeauthoryear{{Storchi-Bergmann} \&
  {Schnorr-M{\"u}ller}}{{Storchi-Bergmann} \&
  {Schnorr-M{\"u}ller}}{2019}]{sb19}
{Storchi-Bergmann} T.,  {Schnorr-M{\"u}ller} A.,  2019, \mn@doi [Nature
  Astronomy] {10.1038/s41550-018-0611-0}, \href
  {https://ui.adsabs.harvard.edu/abs/2019NatAs...3...48S} {3, 48}

\bibitem[\protect\citeauthoryear{{Storchi-Bergmann}, {McGregor}, {Riffel},
  {Sim{\~o}es Lopes}, {Beck}  \& {Dopita}}{{Storchi-Bergmann}
  et~al.}{2009}]{sbN4151Exc}
{Storchi-Bergmann} T.,  {McGregor} P.~J.,  {Riffel} R.~A.,  {Sim{\~o}es Lopes}
  R.,  {Beck} T.,   {Dopita} M.,  2009, \mn@doi [\mnras]
  {10.1111/j.1365-2966.2009.14388.x}, \href
  {https://ui.adsabs.harvard.edu/abs/2009MNRAS.394.1148S} {394, 1148}

\bibitem[\protect\citeauthoryear{{Turner}, {Kirby-Docken}  \&
  {Dalgarno}}{{Turner} et~al.}{1977}]{turner77}
{Turner} J.,  {Kirby-Docken} K.,   {Dalgarno} A.,  1977, \mn@doi [\apjs]
  {10.1086/190481}, \href
  {https://ui.adsabs.harvard.edu/abs/1977ApJS...35..281T} {35, 281}

\bibitem[\protect\citeauthoryear{{Urry} \& {Padovani}}{{Urry} \&
  {Padovani}}{1995}]{urry95}
{Urry} C.~M.,  {Padovani} P.,  1995, \mn@doi [\pasp] {10.1086/133630}, \href
  {https://ui.adsabs.harvard.edu/abs/1995PASP..107..803U} {107, 803}

\bibitem[\protect\citeauthoryear{{Villarroel} \& {Korn}}{{Villarroel} \&
  {Korn}}{2014}]{villarroel14}
{Villarroel} B.,  {Korn} A.~J.,  2014, \mn@doi [Nature Physics]
  {10.1038/nphys2951}, \href
  {https://ui.adsabs.harvard.edu/abs/2014NatPh..10..417V} {10, 417}

\bibitem[\protect\citeauthoryear{{Wilgenbus}, {Cabrit}, {Pineau des For{\^e}ts}
   \& {Flower}}{{Wilgenbus} et~al.}{2000}]{wilgenbus2000}
{Wilgenbus} D.,  {Cabrit} S.,  {Pineau des For{\^e}ts} G.,   {Flower} D.~R.,
  2000, \aap, \href {https://ui.adsabs.harvard.edu/abs/2000A&A...356.1010W}
  {356, 1010}

\bibitem[\protect\citeauthoryear{{de Vaucouleurs}, {de Vaucouleurs}, {Corwin},
  {Buta}, {Paturel}  \& {Fouque}}{{de Vaucouleurs}
  et~al.}{1991}]{devaucouleurs91}
{de Vaucouleurs} G.,  {de Vaucouleurs} A.,  {Corwin} Herold~G. J.,  {Buta}
  R.~J.,  {Paturel} G.,   {Fouque} P.,  1991, {Third Reference Catalogue of
  Bright Galaxies}

\makeatother
\end{thebibliography}




\appendix


\section{Gemini NIFS measurements}

\begin{figure*}
    \centering
\includegraphics[width=0.98\textwidth]{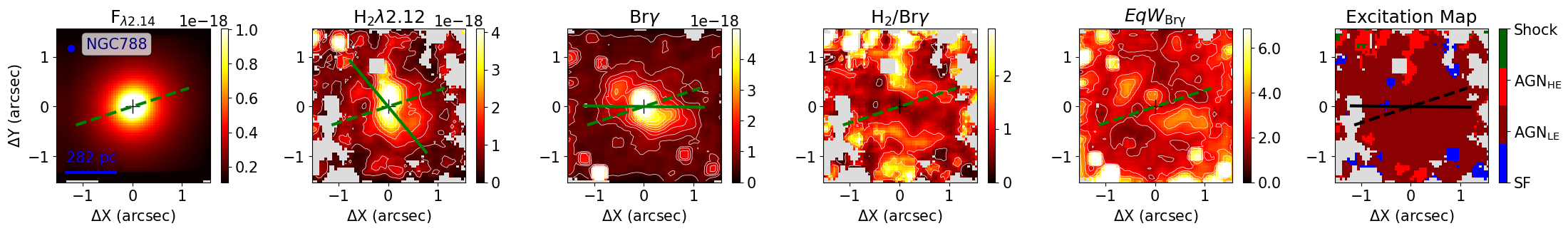}
\includegraphics[width=0.98\textwidth]{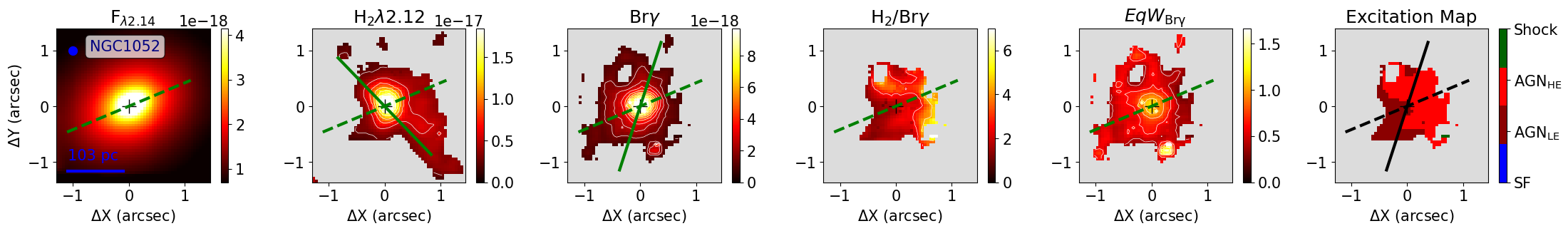}
\includegraphics[width=0.98\textwidth]{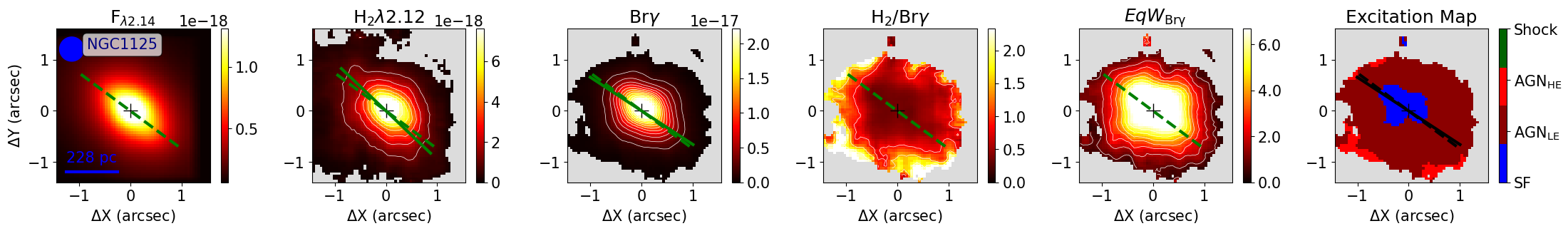}
\includegraphics[width=0.98\textwidth]{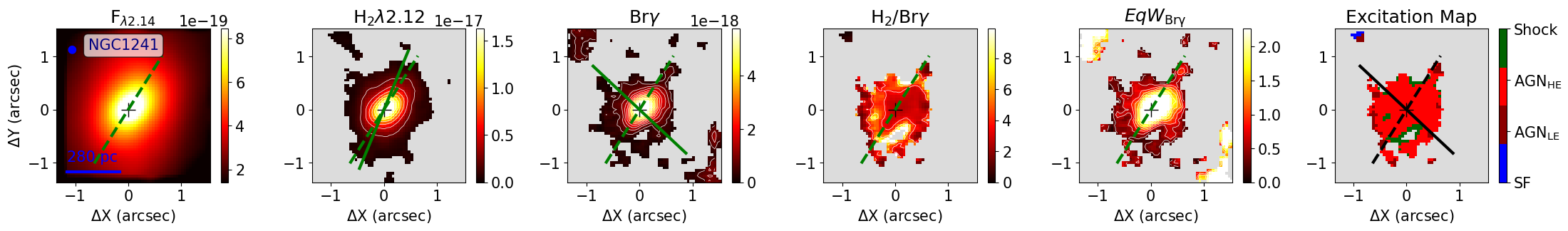}
\includegraphics[width=0.98\textwidth]{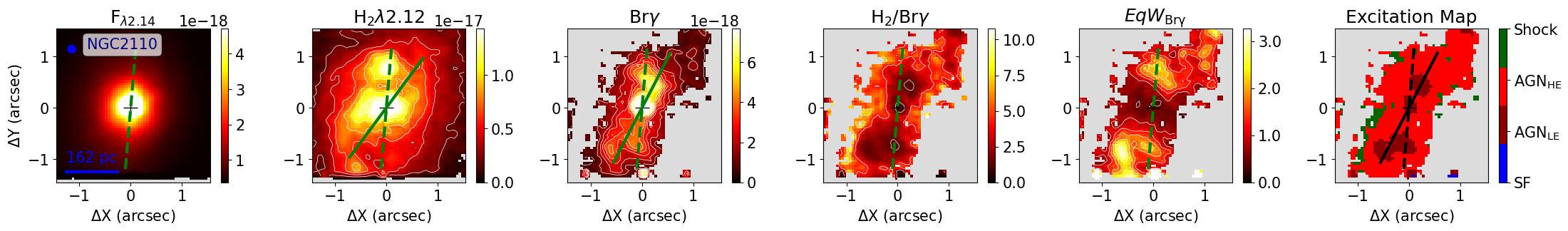}
    \includegraphics[width=0.98\textwidth]{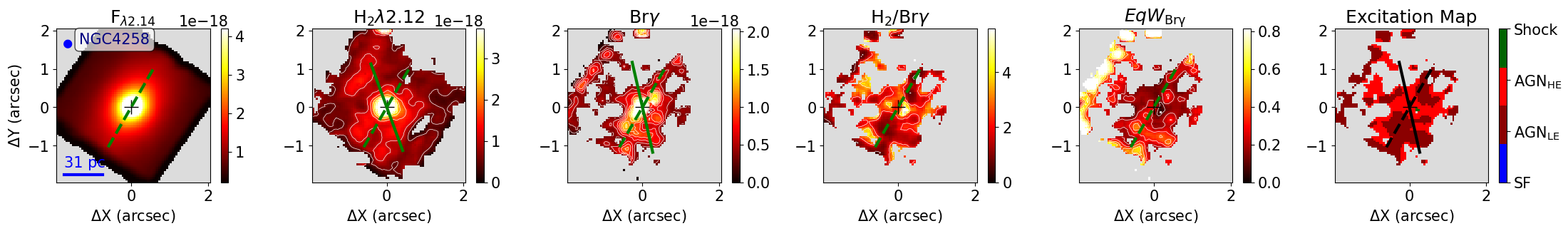}
    \includegraphics[width=0.98\textwidth]{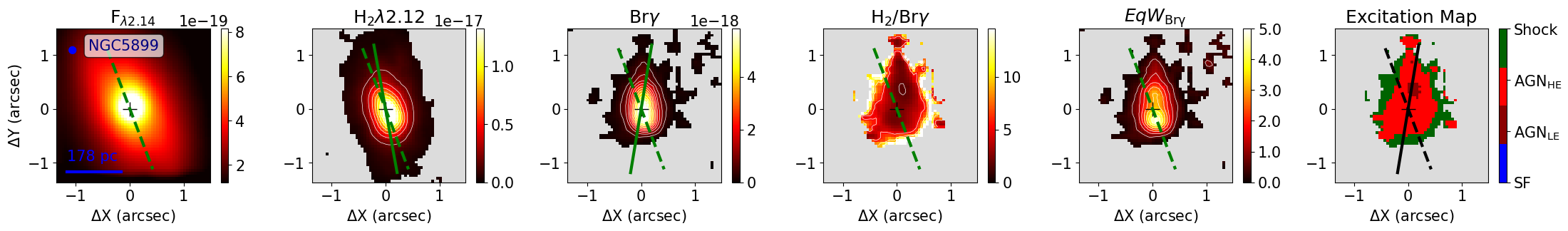}
    \includegraphics[width=0.98\textwidth]{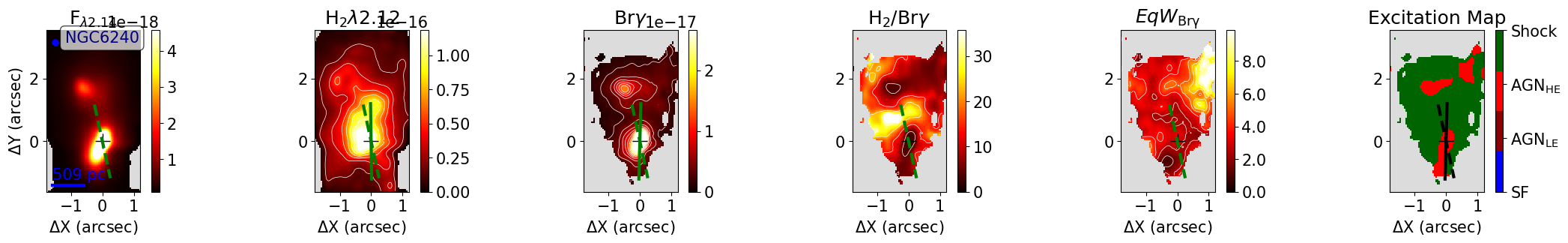}
\caption{\small {\bf Maps for type 2 AGN}. From left to right:  K-band continuum obtained within a spectral window of 100\,\AA\ centred at 2.14\,$\mu$m, H$_2\,2.1218\,\mu$m flux distribution, Br$\gamma$ flux distribution, H$_2$/Br$\gamma$ ratio map, Br$\gamma$ equivalent width map and excitation map. The color bars show the continuum in units of erg\,s$^{-1}$\,cm$^{-2}$\,\AA$^{-1}$\,spaxel$^{-1}$, the emission-line fluxes in erg\,s$^{-1}$\,cm$^{-2}$\,spaxel$^{-1}$ and the Br$\gamma$ equivalent width in \AA. The excitation map identifies the regions with typical $H_2$/Br$\gamma$ values for star forming galaxies (SF: H$_2$/Br$\gamma<0.4$), AGN with low excitation (AGN$_{\rm LE}$: $0.4\leq$ H$_2$/Br$\gamma<2$) and high excitation (AGN$_{\rm HE}$: $2\leq$  H$_2$/Br$\gamma<6$) and higher line ratios, usually observed in shock dominated objects (Shock: H$_2$/Br$\gamma>6$). In each row, the name of the galaxy is identified in the continuum image, the filled circle corresponds to the angular resolution of the data, the spatial scale is shown in the bottom-left corner of the continuum image and the cross marks the position of the peak of the continuum emission. The dashed line indicates the orientation of the galaxy major axis obtained from the Hyperleda database. The continuous line on the flux maps shows the orientation of the emission-line flux distributions. The continuous line on the excitation map shows the orientation of the Br$\gamma$ emission. The gray regions indicate locations where the emission lines are not detected within 2$\sigma$ above the continuum noise.} For all galaxies, north is up and east is to the left.
    \label{fig:mapsS2ap}
\end{figure*}

\begin{figure*}
      \setcounter{figure}{0}
    \centering
\includegraphics[width=0.98\textwidth]{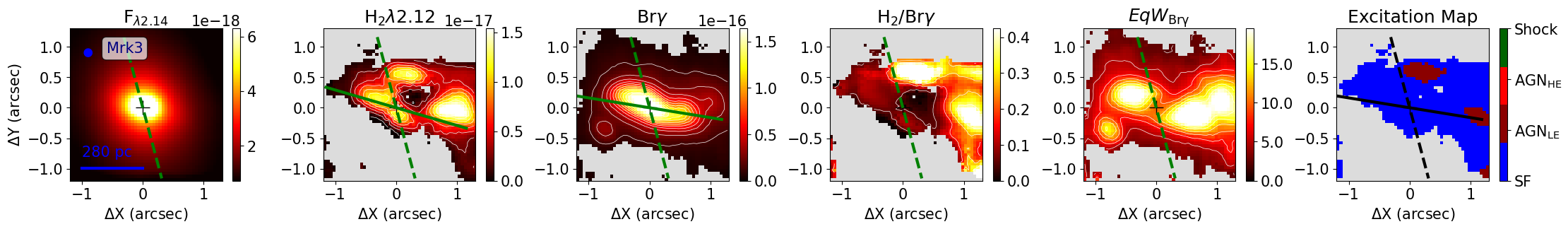}
\includegraphics[width=0.98\textwidth]{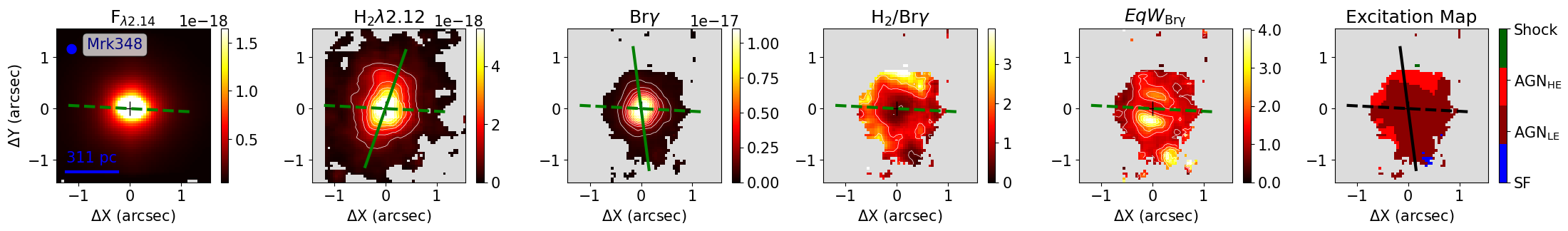}
    \includegraphics[width=0.98\textwidth]{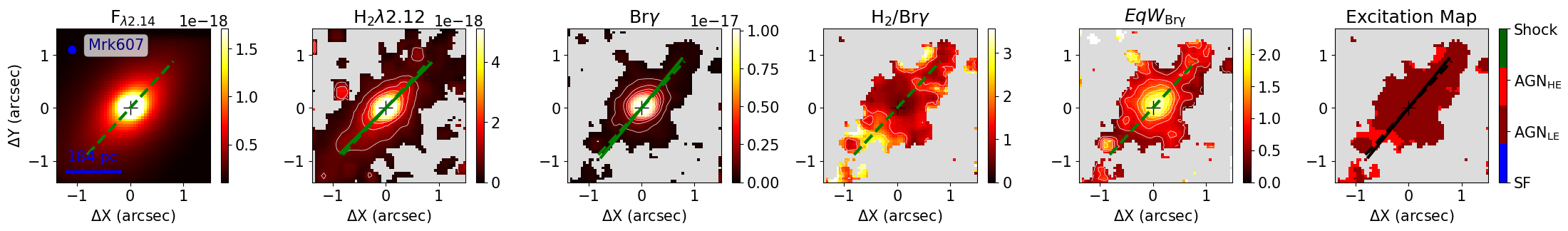}
      \includegraphics[width=0.98\textwidth]{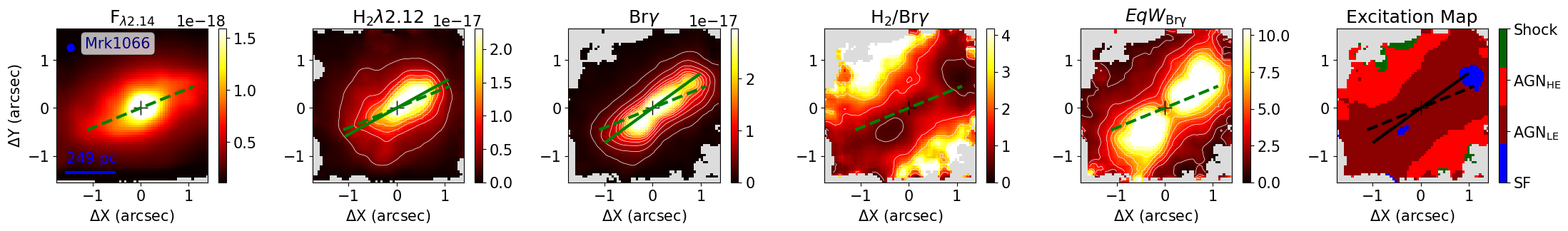}
\includegraphics[width=0.98\textwidth]{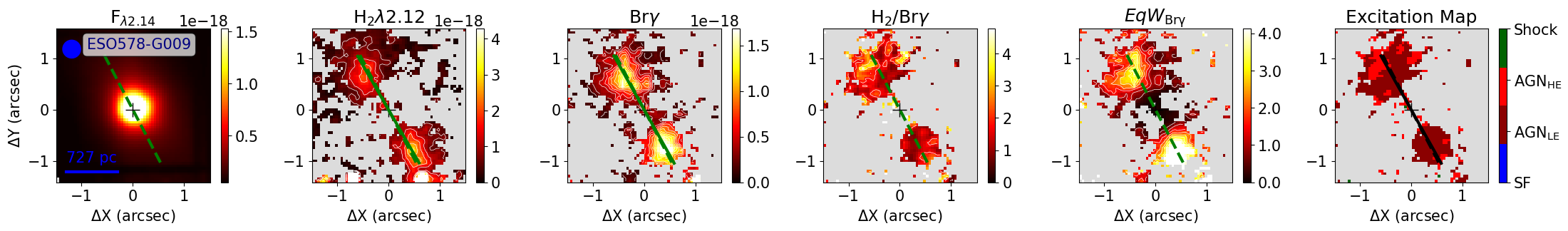}
\includegraphics[width=0.98\textwidth]{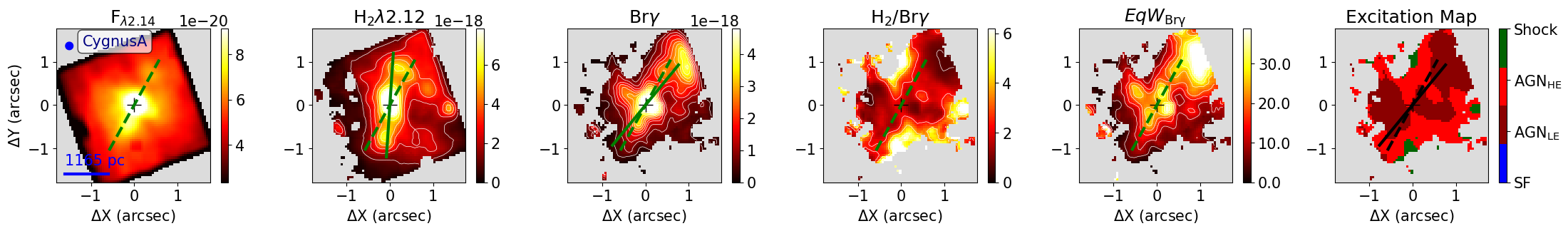}
\caption{\small continued. }
\end{figure*}

\begin{figure*}
    \centering
\includegraphics[width=0.98\textwidth]{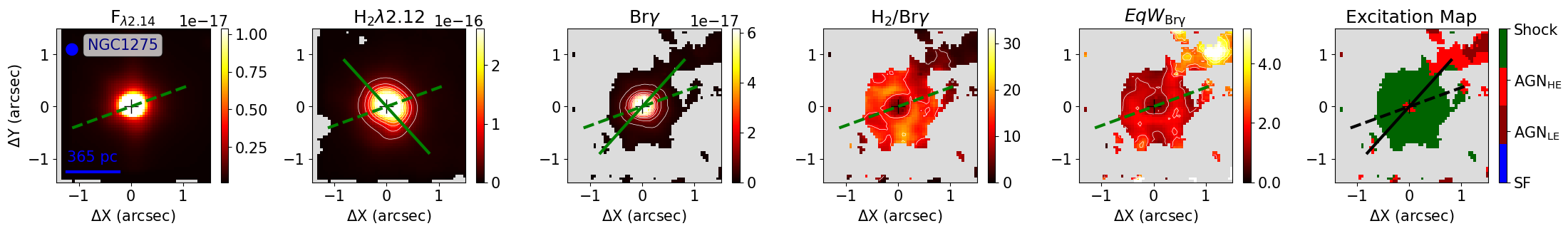}
\includegraphics[width=0.98\textwidth]{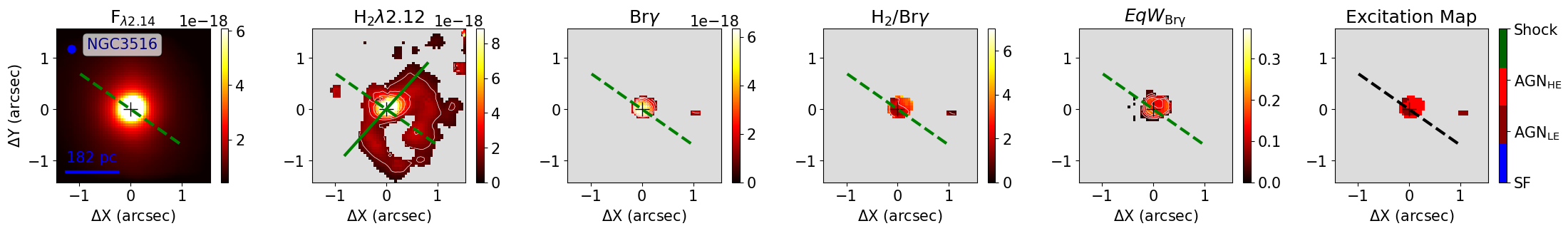}
\caption{\small {\bf Maps for type 1 AGN}. Same as Fig.~\ref{fig:mapsS2ap}, but for type 1 AGN.}
    \label{fig:mapsS1ap}
\end{figure*}

\begin{figure*}
      \setcounter{figure}{1}
    \centering
\includegraphics[width=0.98\textwidth]{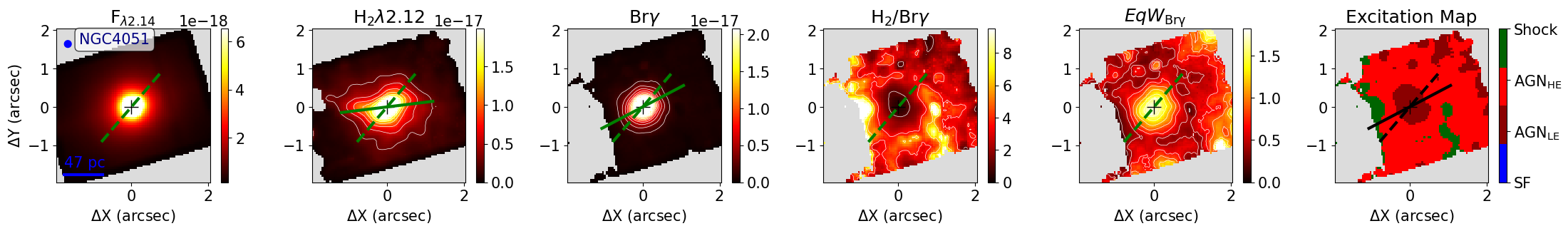}
\includegraphics[width=0.98\textwidth]{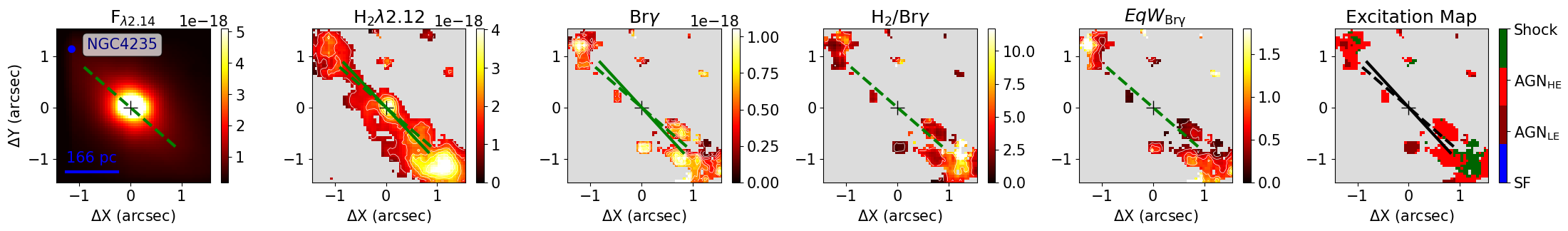}
\includegraphics[width=0.98\textwidth]{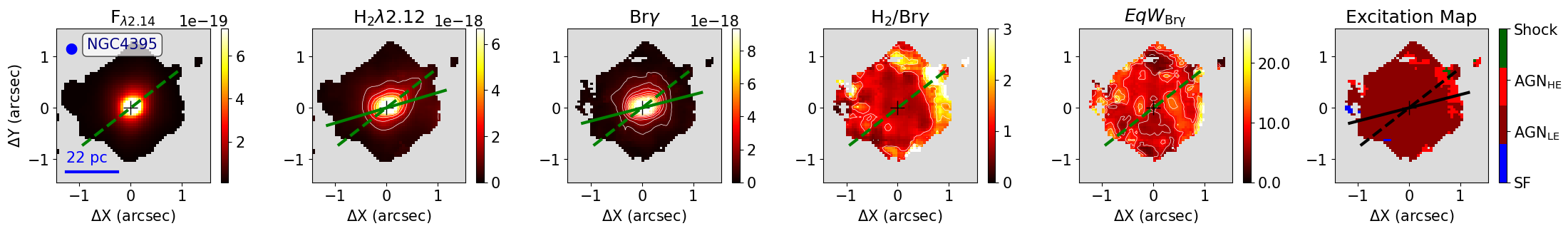}
\includegraphics[width=0.98\textwidth]{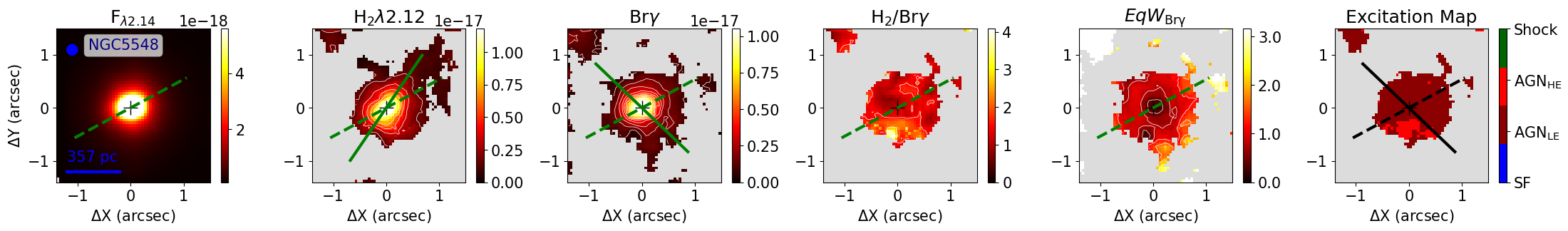}
\includegraphics[width=0.98\textwidth]{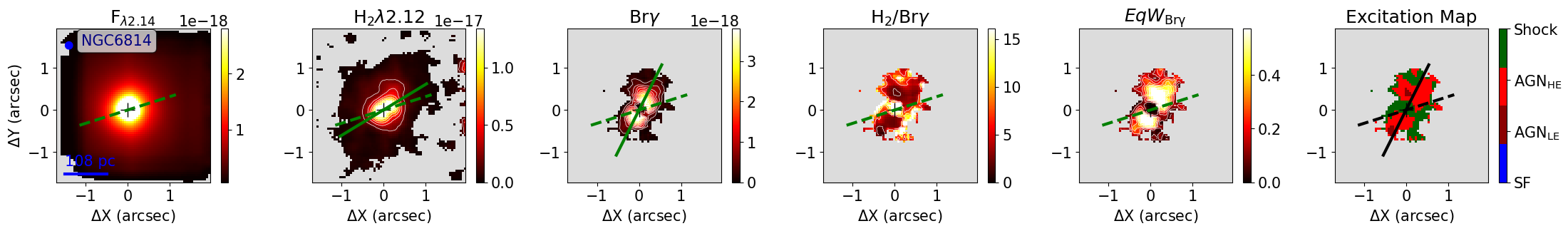}
\includegraphics[width=0.98\textwidth]{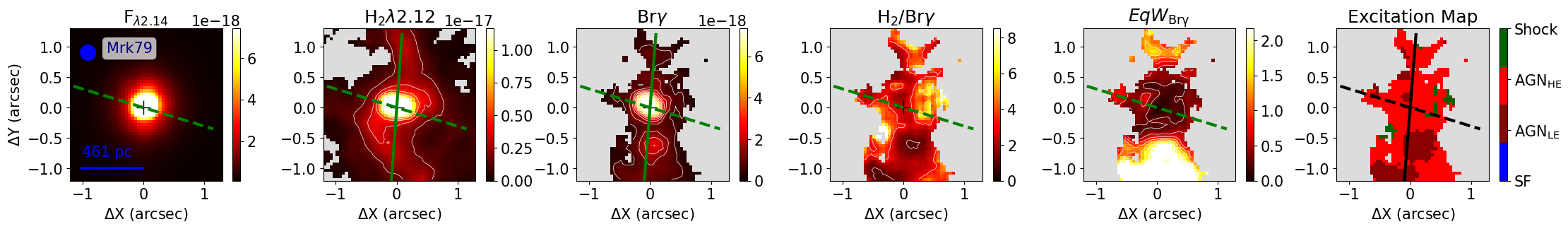}   
\includegraphics[width=0.98\textwidth]{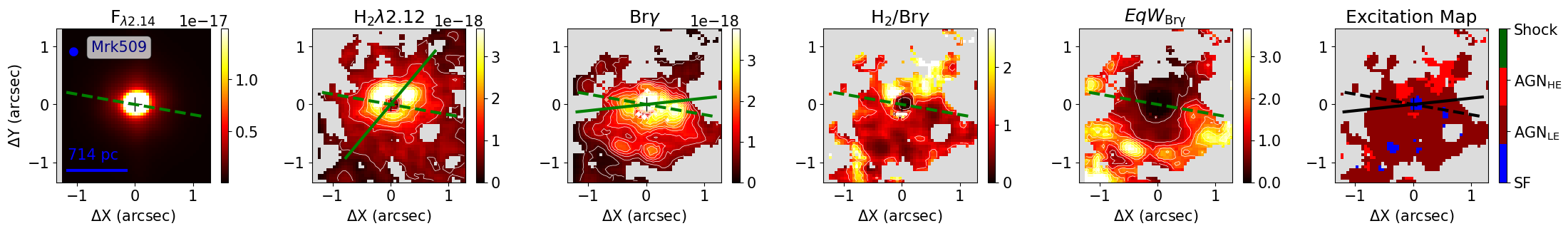}
\includegraphics[width=0.98\textwidth]{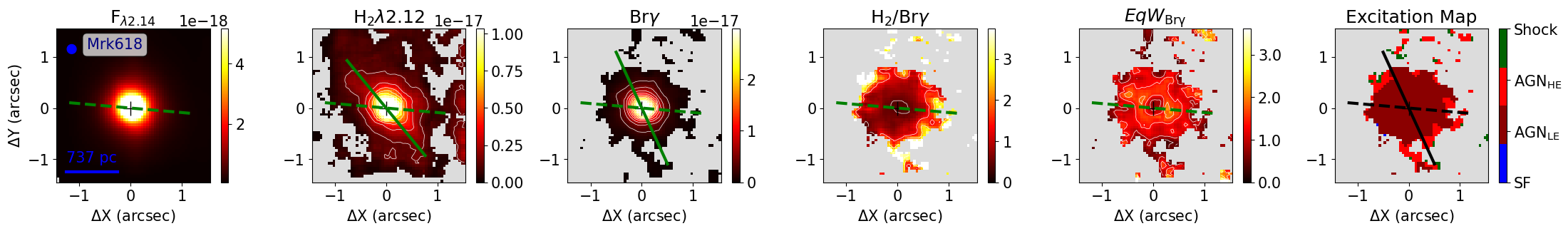}  
\includegraphics[width=0.98\textwidth]{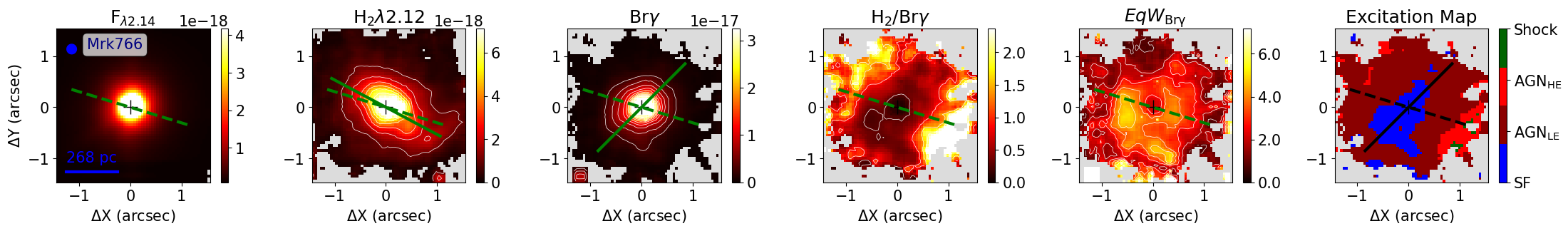}
\caption{\small continued. }
\end{figure*}

\begin{figure*}
      \setcounter{figure}{1}
    \centering
\includegraphics[width=0.98\textwidth]{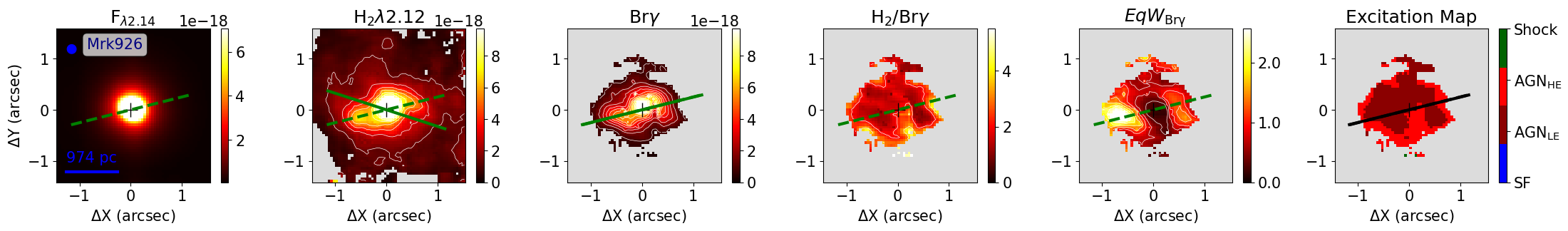}
\includegraphics[width=0.98\textwidth]{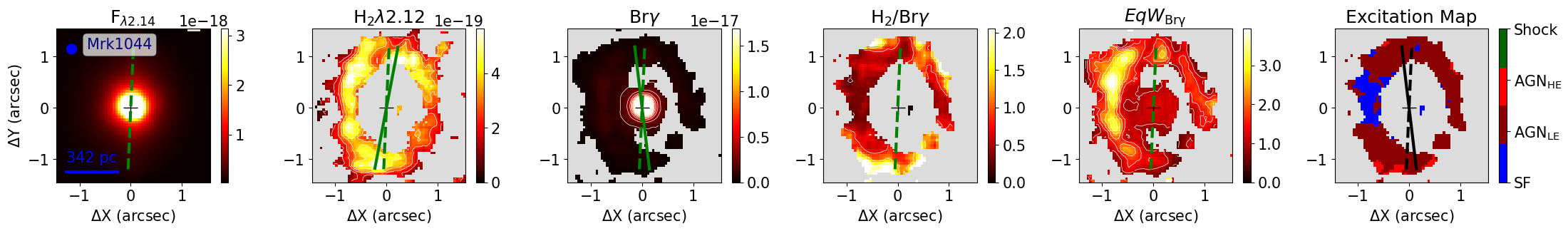}
\includegraphics[width=0.98\textwidth]{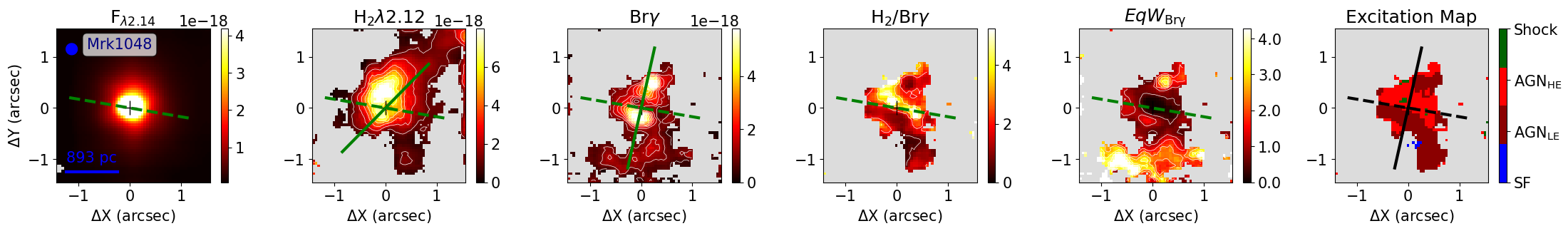} 
\caption{\small continued. }
\end{figure*}

\begin{figure*}
    \centering
	\begin{tabular}{c c c} 
\includegraphics[width=0.3\textwidth]{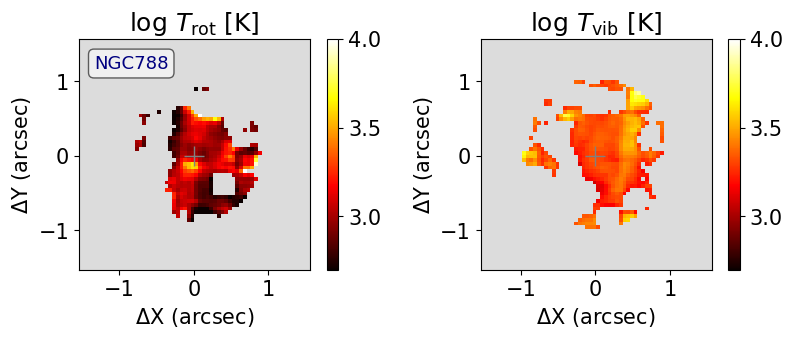} & 
\includegraphics[width=0.3\textwidth]{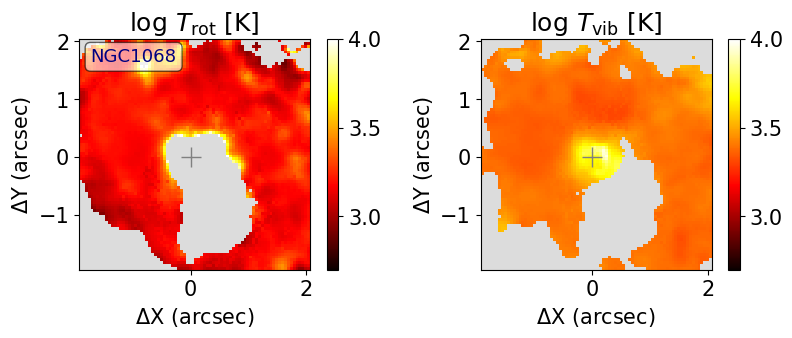}&
\includegraphics[width=0.3\textwidth]{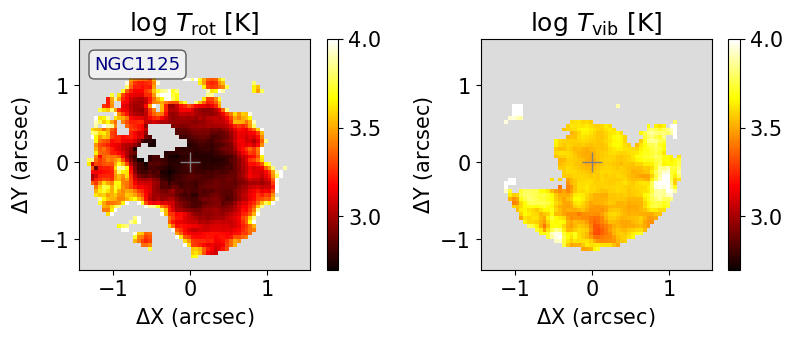} \\ 
\includegraphics[width=0.3\textwidth]{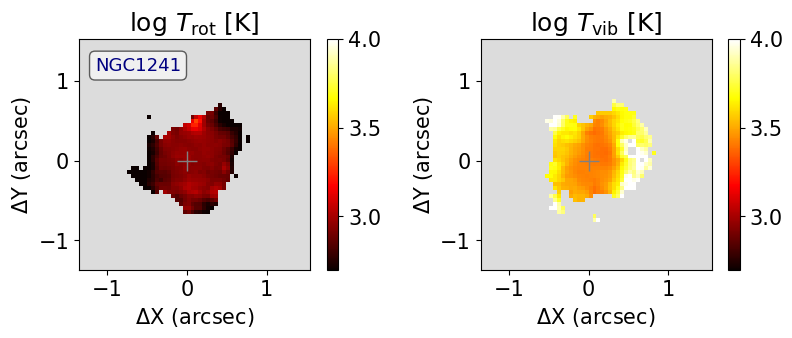}&
  \includegraphics[width=0.3\textwidth]{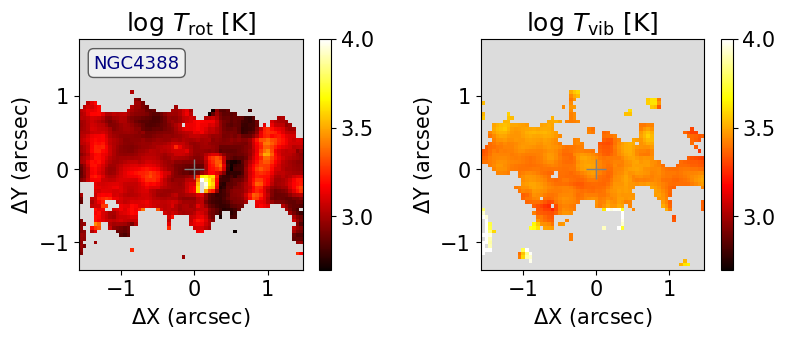} & 
\includegraphics[width=0.3\textwidth]{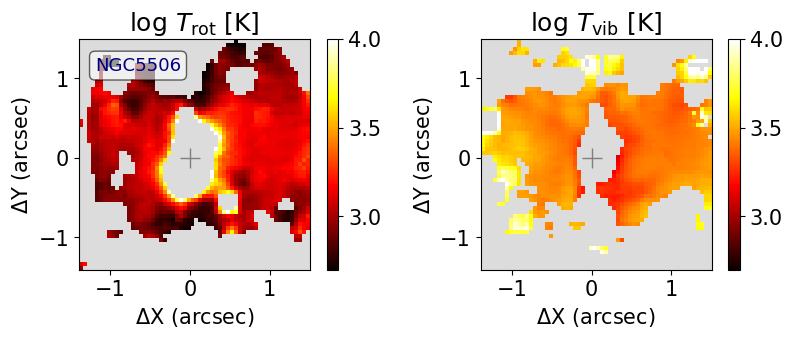}\\
\includegraphics[width=0.3\textwidth]{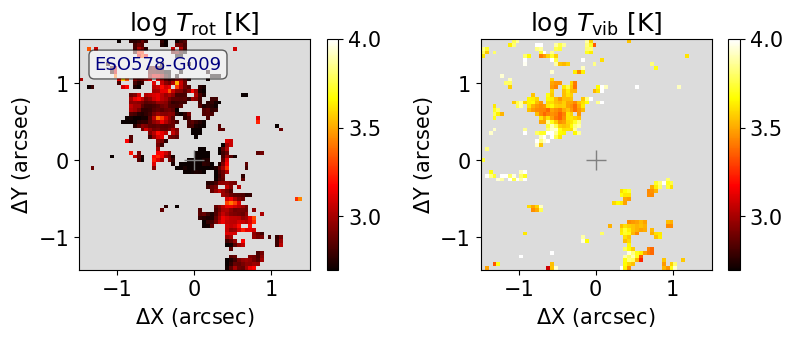} &
\includegraphics[width=0.3\textwidth]{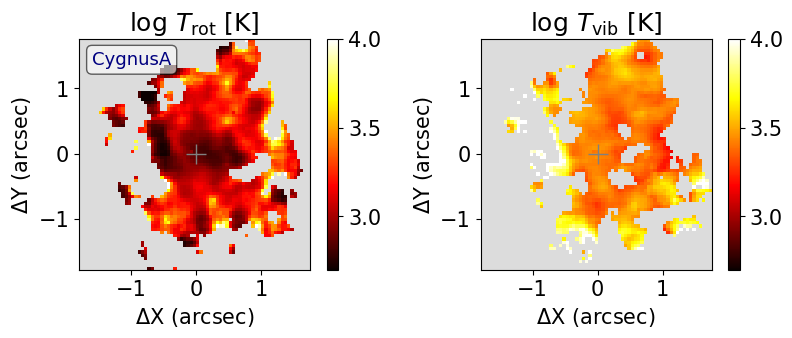} &
\includegraphics[width=0.3\textwidth]{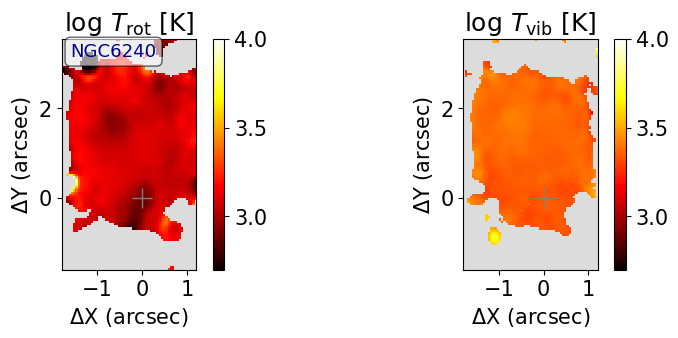} \\
   \end{tabular}
\caption{\small H$_2$  rotational and vibrational temperatures for type 2 in our sample calculated according to Eqs.~\ref{eq:temp_rot} and \ref{eq:temp_vib}. The gray regions indicate locations where at least one of the H$_2$ emission lines is not detected within 2$\sigma$ above the continuum noise. The colorbar shows the temperatures in logarithmic scale in units of K.}
    \label{fig:temperarutesT2}
\end{figure*}

\begin{figure*}
    \centering
	\begin{tabular}{c c c} 
\includegraphics[width=0.3\textwidth]{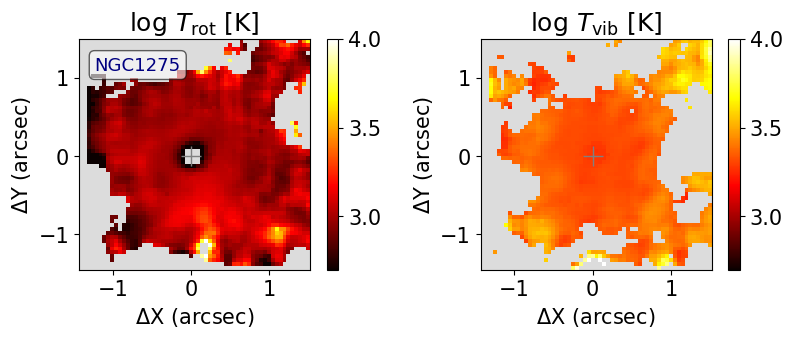}  &
\includegraphics[width=0.3\textwidth]{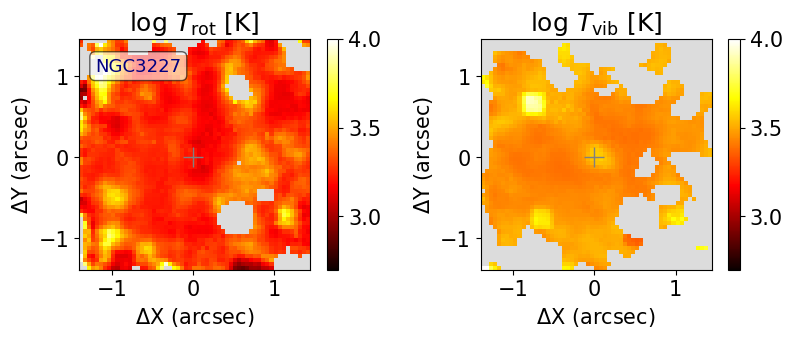} &
\includegraphics[width=0.3\textwidth]{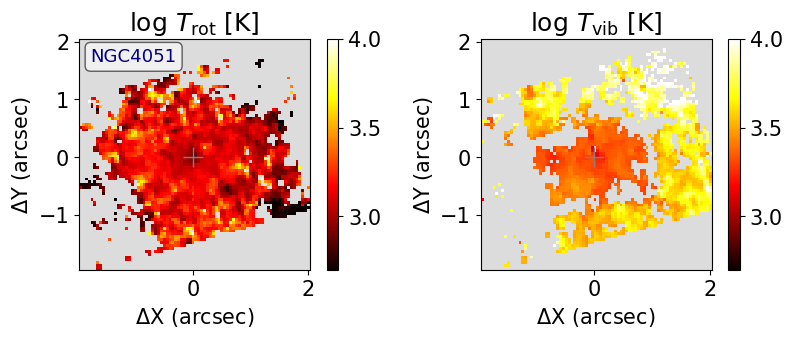}  \\
\includegraphics[width=0.3\textwidth]{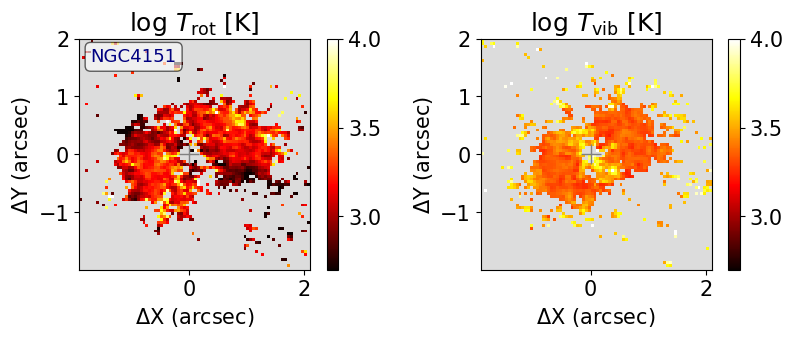} &
 \includegraphics[width=0.3\textwidth]{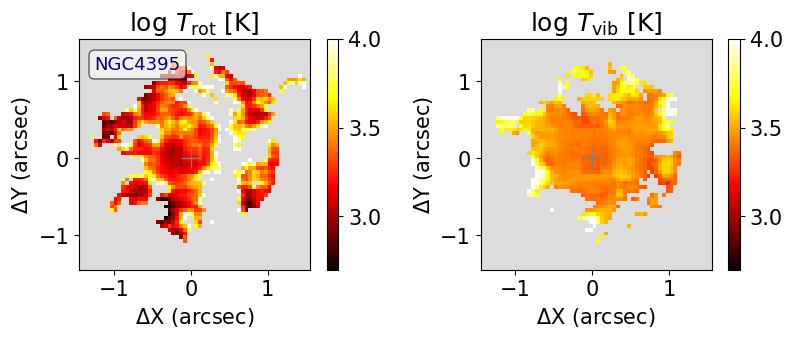} &
\includegraphics[width=0.3\textwidth]{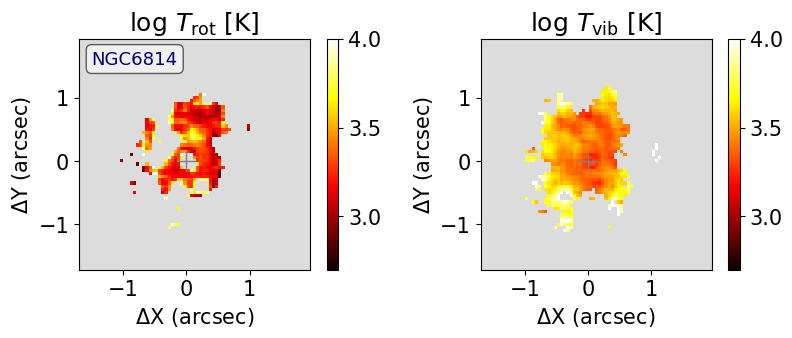}  \\
\includegraphics[width=0.3\textwidth]{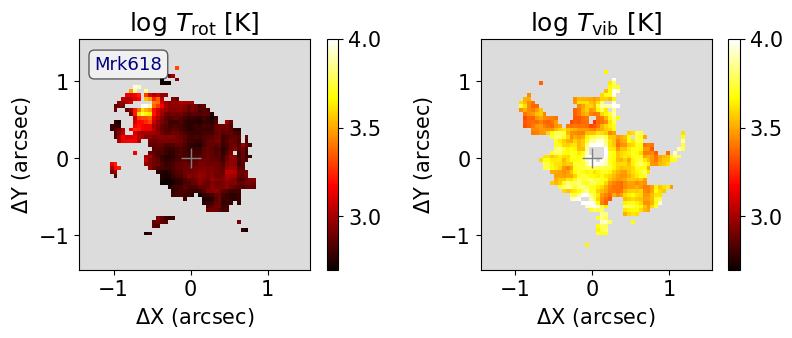}
&
\includegraphics[width=0.3\textwidth]{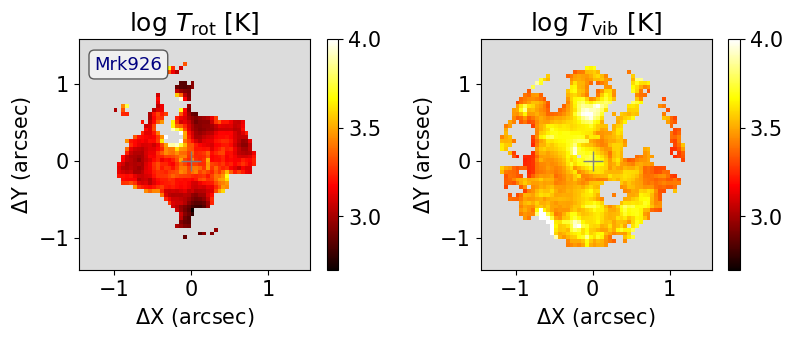} &
\includegraphics[width=0.3\textwidth]{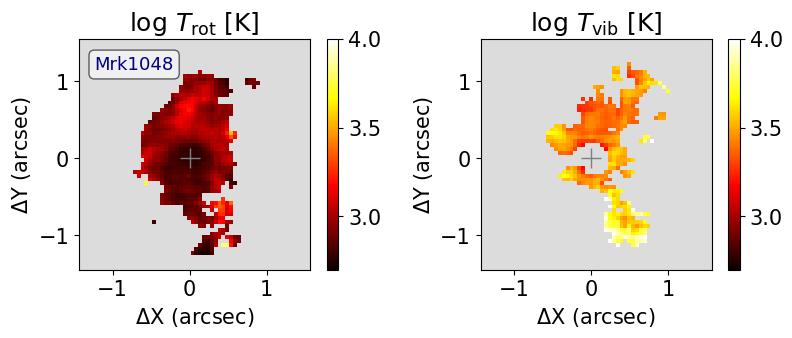}\\ 
\includegraphics[width=0.3\textwidth]{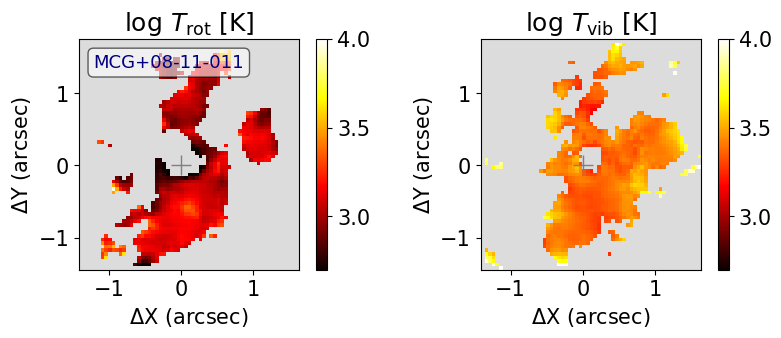}& & \\ 
 \end{tabular}
\caption{\small Same as \ref{fig:temperarutesT2}, but for type 1 AGN.}
    \label{fig:temperarutesT1}
\end{figure*}

\begin{figure*}
    \centering
\begin{tabular}{c c c c c c} 

\includegraphics[width=0.15\textwidth]{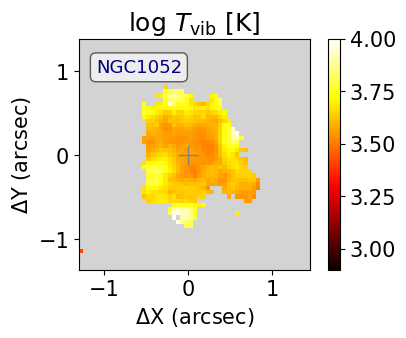}&
\includegraphics[width=0.15\textwidth]{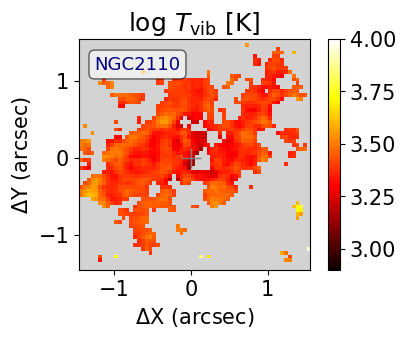}&
\includegraphics[width=0.15\textwidth]{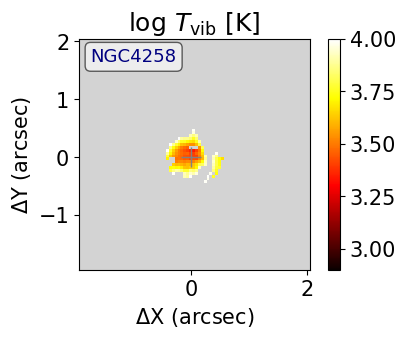}&
\includegraphics[width=0.15\textwidth]{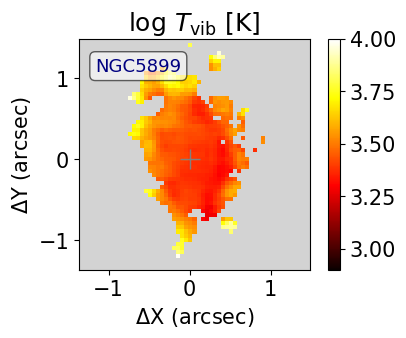}&
\includegraphics[width=0.15\textwidth]{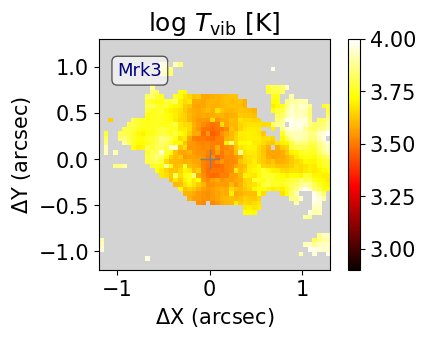}&
\includegraphics[width=0.15\textwidth]{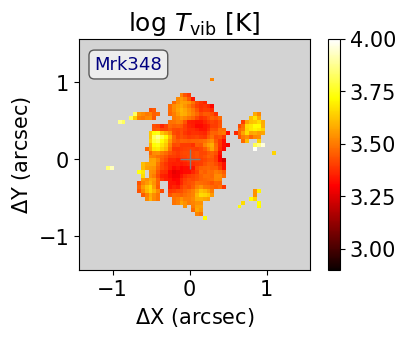} \\
\includegraphics[width=0.15\textwidth]{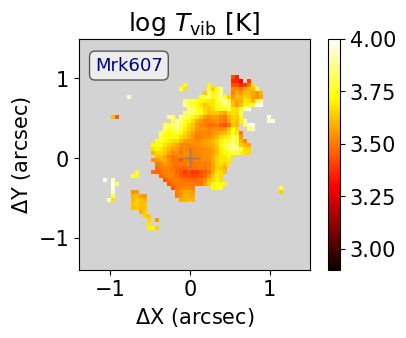}&
\includegraphics[width=0.15\textwidth]{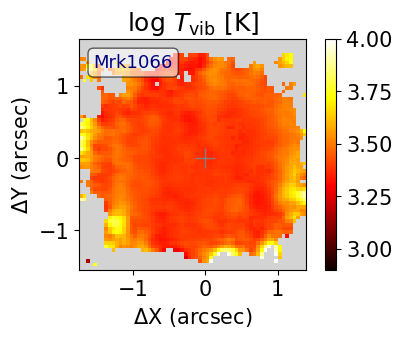}&
\includegraphics[width=0.15\textwidth]{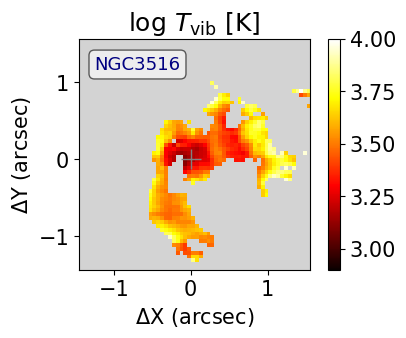}&
\includegraphics[width=0.15\textwidth]{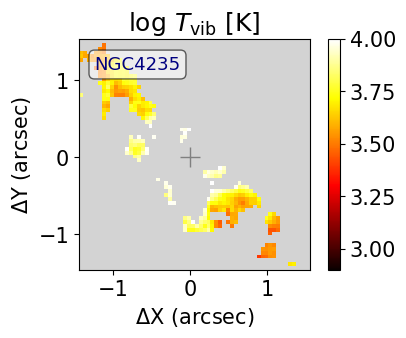}&
\includegraphics[width=0.15\textwidth]{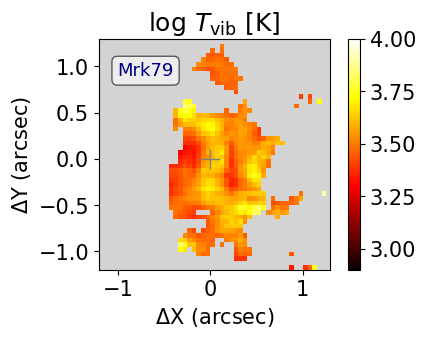}&
\includegraphics[width=0.15\textwidth]{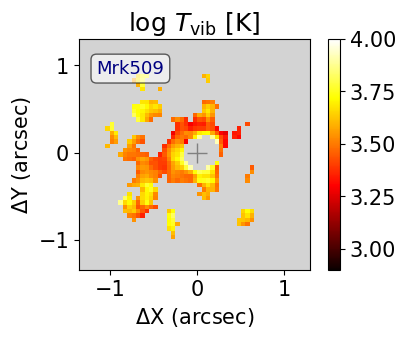}\\
\includegraphics[width=0.15\textwidth]{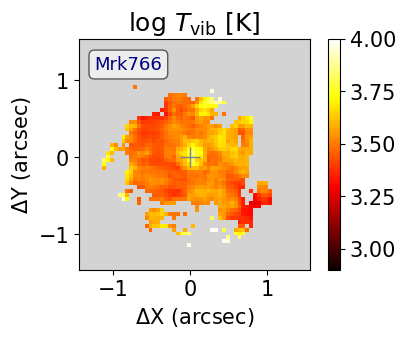}
\end{tabular}
\caption{\small H$_2$ vibrational for objects where the H$_2$\,1--0\,S(2) line is not detected.}
    \label{fig:temperarutesVib}
\end{figure*}


\bsp	
\label{lastpage}
\end{document}